\documentclass[]{ss_areviews}

\usepackage{natbib}
\bibpunct[]{(}{)}{;}{a}{}{,}
\usepackage{graphics,graphicx}
\usepackage{amssymb}

\newcommand{\ergps}{{\rm erg\ s^{-1}}}
\newcommand{\msun}{\ifmmode{\rm M}_\odot\else$\rm M_\odot$\fi}
\newcommand{\msunpy}{{\rm M_\odot\ year^{-1}}}
\newcommand{\pcm}{{\rm cm^{-3}}}
\newcommand{\nelec}{n_{\rm e}}
\newcommand{\nh}{n_{\rm H}}
\newcommand{\mh}{m_{\rm H}}
\newcommand{\tcool}{t_{\rm cool}}
\newcommand{\tflow}{t_{\rm flow}}
\newcommand{\kms}{{\rm km\ s^{-1}}}

\begin{document}

\bibliographystyle{plainnat}

\jname{Annual Reviews of Astronomy \& Astrophysics}
\jyear{2007}
\jvol{}
\ARinfo{}

\title{Heating Hot Atmospheres with Active Galactic Nuclei}

\markboth{McNamara \& Nulsen}{Heating Hot Atmospheres with AGN}

\author{B. R. McNamara
\affiliation{Department of Physics \& Astronomy, University of Waterloo,
Ontario, Canada,  Astrophysical Institute and Department of Physics and Astronomy, Ohio University, Athens, OH 45701, and
Harvard-Smithsonian Center for Astrophysics, 60 Garden
Street, Cambridge, MA 02138}
P. E. J. Nulsen
\affiliation{Harvard-Smithsonian Center for Astrophysics, 60 Garden
  Street, Cambridge, MA 02138}}


\begin{abstract}
High resolution X-ray spectroscopy of the hot gas in galaxy clusters
has shown that the gas is not cooling to low temperatures at the
predicted rates of hundreds to thousands of solar masses per year.
X-ray images have revealed giant cavities and shock fronts in the hot
gas that provide a direct and relatively reliable means of measuring
the energy injected into hot atmospheres by active galactic nuclei
(AGN). Average radio jet powers are near those required to offset
radiative losses and to suppress cooling in isolated giant elliptical
galaxies, and in larger systems up to the richest galaxy clusters.
This coincidence suggests that heating and cooling are coupled by
feedback, which suppresses star formation and the growth of luminous
galaxies. How jet energy is converted to heat and the degree to which  other heating mechanisms are
contributing, eg. thermal conduction, are not well understood.
Outburst energies require substantial late growth of
supermassive black holes. Unless all of the $\sim 10^{62}$ erg required to
suppress star formation is deposited in the cooling regions of
clusters, AGN outbursts must alter large-scale properties of the
intracluster medium.
\end{abstract}

\maketitle

\section{INTRODUCTION TO X-RAY CLUSTERS OF GALAXIES}

Clusters of galaxies are the largest gravitationally collapsed
concentrations of matter in the Universe.  With diameters of several
megaparsecs and masses up to $10^{15}\ \msun$, they are recognizable
in photographs as distinct concentrations of galaxies centered on one
or more brightest cluster members.  The space between the galaxies is
filled with a hot, dilute, plasma that emits X-rays and all is held in
place by the gravity of a dark matter halo.  The Perseus and Coma
clusters were among the first clusters to be identified as X-ray
sources by the Uhuru satellite in the early 1970s
\citep{gkg71,gkm71,fkg72}.  By the mid-1970s, at least 40 clusters of
galaxies were identified as extended and powerful X-ray sources
\citep{gs77}.  A thermal origin for the X-ray emission was confirmed
in the Perseus, Virgo, and Coma clusters with the detection of the
collisionally excited, 6 -- 7 keV Fe-K emission feature by the Ariel 5
\citep{mcd76} and OSO-8 observatories \citep{ssb77}.  Gas temperatures
typically lie in the range of 10 million to 100 million K,
corresponding to X-ray luminosities of $L_X \sim 10^{43}\ \ergps$ to
$10^{45}\ \ergps$.  Roughly 90\% of the baryons in clusters reside in
the hot plasma, while the rest are locked up in stars in galaxies
\citep{lms03}.  The hot gas forms a hydrostatic atmosphere, where the
temperature and density distributions reflect the gravitating
mass.  The atmosphere serves as a bank of baryons that failed to end
up in stars and galaxies, and as a repository for the heat exhaust and
detritus from stellar evolution and the growth of supermassive black
holes during galaxy and cluster formation.

This review focuses on the latter, emphasizing new results from the
Chandra and XMM-Newton X-ray Observatories showing that active
galactic nuclei (AGN) lying at the hearts of galaxy clusters are
pouring vast amounts of energy into the hot gas, some as prodigiously
as quasars.  The combination of high-resolution X-ray and radio
imaging is yielding reliable measurements of this energy, which is
apparently sufficient to suppress cooling flows and the substantial
growth at late times of giant elliptical (gE) and cD galaxies. Deep
Chandra images show that many clusters and gEs have multiple cavities,
giving the hot atmospheres a Swiss-cheese-like topology that reveals
the AGN outburst history independently of radio emission. These
spectacular images are giving new insight into the particle and
magnetic field content of radio sources, and are guiding the
development of new radio jet and galaxy formation models. We discuss
recent developments that link AGN outbursts to heating of the
intracluster gas, and we tie these developments to several outstanding
problems including the quenching of cooling flows, the exponential
decline in the numbers of bright galaxies, the relationship between
bulge mass and black hole mass in galaxies, and the possible
contribution of AGN to excess entropy (preheating) in the hot
atmospheres of groups and clusters.  We briefly discuss some of the
cosmological issues related to AGN heating in clusters.  General
reviews of clusters from an X-ray perspective were given recently by
\citet{mushotzky04} and \citet{a05}, a review of clusters as
cosmological probes was given by \citet{voit05}, and cold fronts and
shocks associated with cluster mergers are reviewed by \citet{mv07}.
We begin with a brief overview of the basic properties of clusters
emphasizing the hot intracluster medium (ICM) and the scaling
relations that describe it. 

\subsection{X-ray Emission from the Intracluster Plasma}

The intracluster plasma (which we refer to interchangeably as hot gas)
is composed primarily of ionized hydrogen and helium, mixed with
traces of heavier elements, at roughly 1/3 of solar abundances.  The
presence of the gas can be understood in the context of hierarchical
structure formation models.  The warm baryons were swept inward with
collapsing dark matter and subsequently heated to the virial
temperature of the halo by accretion shocks and adiabatic
compression.  Mean gas temperatures of several tens to one hundred
million Kelvin reflect the virial temperatures of halos, so that $T
\propto \sigma^2$, where $\sigma$ is the line-of-sight velocity
dispersion of the cluster galaxies.  Observed particle densities range
from $10^{-4}\ \pcm$ in the halos of clusters up to $10^{-2}\ \pcm$
and higher in the centers of some clusters.

The gas can generally be treated as an optically thin coronal plasma
in ionization equilibrium.  The electrons and ions interact through
Coulomb collisions and radiate mainly by thermal bremsstrahlung
emission in the X-ray band \citep[e.g.,][]{s88}.  At temperatures below
$3\times10^7\rm\ K$, X-ray emission from the gas is increasingly
dominated by the recombination lines of iron, oxygen, silicon, and
other elements \citep[e.g.,][]{s88}.  The emission from heavy elements,
particularly the iron K lines at 6 ­- 7 keV and the iron L lines below
1 keV in cooler plasmas, significantly alters the shape of the
spectrum. This and the exponential decline in emission at high
energies permit the temperature and metallicity of the hot gas to be
measured accurately with modern X-ray telescopes.

The X-ray telescopes used to study clusters have commonly employed
proportional counters or charge coupled devices (CCDs) as
detectors.  They are sensitive to photons with energies spanning the
range 0.1 ­- 10 keV, well-matched to thermal radiation from the
intracluster gas ($kT = 1$ keV for $T = 1.16\times10^7$ K).  Because
the emission processes are collisional, the power radiated per unit
volume is proportional to the square of the density.  X-ray surface
brightness can therefore be used to determine gas density.  With most
X-ray instruments, for gas temperatures $kT \simeq 2$ keV or greater,
the count rate from an optically thin thermal plasma depends almost
exclusively on its emission measure, $\int \nelec \nh \, dV$, where
$\nelec$ is the electron number density, $\nh$ is the equivalent
hydrogen number density, and the integral is taken throughout the
emitting volume.  As a result, gas densities can be determined quite
accurately, even when gas temperatures are poorly known.

The surface brightness of the hot gas declines with increasing radius,
approximately as $I_{\rm X} \propto r^{-3}$ at large distances.
Despite the rapid decline in surface brightness, the relatively low
X-ray background permits X-ray emission to be traced to very large
radii, making it an excellent probe of gas temperature, metallicity,
and mass throughout much of the volume of a cluster.

Surface brightness profiles have traditionally been characterized
using the isothermal ``$\beta$-profile''
\begin{equation}
I_{\rm X} \propto [1 + (r/r_{\rm c})^2]^{-3\beta + 1/2},
\end{equation}
where $r_{\rm c}$ is the core radius of the gas distribution, and the
parameter $\beta\simeq 2/3$ for relaxed, bright clusters.  As
introduced \citep[e.g.,][]{cf76,bff81,fj82}, $\beta$  is the ratio of the
energy per unit mass in galaxies to that in the gas.  However, this is
an approximation that further relies on both the gas and dark matter
being isothermal.  This model provides a reasonably good fit to the
surface brightness profiles of clusters at intermediate radii.  In the
central regions of some clusters, where the gas temperature declines
and the density rises rapidly, the fit is poorer.  At large radii,
beyond roughly $0.3 r_{200}$, the observed surface brightness profiles
steepen below the $\beta$-profile \citep{vkf06}.  Here, $r_{200}$ is
the radius within which the mean mass density of the cluster exceeds
200 times the critical density of the Universe.  For isothermal gas (or
gas with $kT > 2$ keV), the $\beta$-model corresponds to the electron
density profile
\begin{equation}
\nelec(r) = n_0 [1 + (r/r_{\rm c})^2]^{-3\beta/2},
\end{equation}
where $n_0$ is the central electron density.

More generally, gas density and temperature profiles can be determined
by ``deprojection'' \citep{fhc81}.  Under the often inadequate
assumption of spherical symmetry, the X-ray spectrum at any point in
the cluster is determined in terms of simple integrals of temperature
and density.  Typically, the gas is represented as a number of shells
of uniform density, temperature, and composition and its properties
are determined by fitting X-ray spectra extracted from corresponding
annular regions \citep{ettori00,pak04}.

\subsection{Gas Temperatures and Masses}

The gas temperature of a hot atmosphere is most sensitive to the
gravitating mass profile and, to a lesser degree, the history of
heating by gravitational and nongravitational processes
\citep{voit05,bab02}.  The connection between halo mass and gas temperature
is clearly evident in the correlation between galaxy velocity
dispersion and gas temperature \citep{es91}.  The scaling of this
relationship is $\sigma \propto T^{0.63}$ in clusters with
temperatures between 0.5 keV and 10 keV \citep[e.g.,][]{kwh03}, which
is close to the expected scaling $\sigma\propto T^{0.5}$.  Simple
theoretical models of gravitational collapse predict the present day
scaling relationships between virial mass, X-ray gas temperature, and
luminosity; $M \propto T^{3/2}$ and $L \propto T^2$ \citep{emn96}.
The observed scaling makes measured cluster temperatures and
luminosities valued proxies for the much less accessible masses of
dark matter halos.  Departures from these scaling relations reflect
physics beyond pure gravitational dynamics, including heating agents
such as supernova explosions and AGN, and additional pressure support
from magnetic fields and cosmic rays \citep{markevitch98,voit05}.

When the gas is spherically symmetric, in hydrostatic equilibrium, and
only the thermal gas pressure is significant, the run of temperature
and density with radius, in principle, permits the mass profile of
clusters to be computed from the equation of hydrostatic equilibrium
as
\begin{equation}
M(r) = - {kTr \over G\mu\mh} \left[ {d\log n_e\over d\log r} + {d\log
    T\over d\log r} \right],
\end{equation}
where $G$ is the gravitational constant, $\mh$ is the hydrogen mass,
and $\mu \simeq 0.61$ is the mean molecular weight of the ionized
plasma.  In practice, the logarithmic derivative of gas density is
often evaluated using a parametrized model, such as the $\beta$-model, 
for the gas density, with the parameters determined by fitting the
surface brightness profile.  Similar methods can be applied to finding
the logarithmic derivative of the temperature.

Mass determinations rely critically on the assumptions that the X-ray
emitting gas is hydrostatic and the sole source of pressure.  Other
possible sources of pressure, including magnetic fields \citep{gf04},
cosmic rays, and bulk motion of the gas, will cause masses to be
underestimated.  Velocity dispersion and weak lensing masses generally
agree with X-ray masses to within a few tens of percent, except
perhaps in systems undergoing major mergers.  Simulations have shown
that bulk motions cause the hydrostatic mass approximation to be
biased below the true value by 5\% to 20\% \citep{nvk07}.

\citet{vkf06} used Chandra observations extending to large radii to
measure the mass versus temperature relationship for relaxed clusters
with temperatures in the range of 0.7 -- 9 keV.  They found $M \propto
T^{1.5-1.6}$, which agrees with self-similar models \citep{vkf06}.
There are disagreements between researchers about the normalization
constant and temperature profile shapes, resulting in discrepancies 
in mass by a few tens of percent
\citep[e.g.,][]{mfs98,ib00,asf01,dm02,vkf06}.  These discrepancies are
often associated with the measurement of $dT/dr$ at large radii, where
systematic effects dominate.  However, some of the scatter may be real 
and perhaps associated with accretion, mergers, or AGN-related
activity.

The observed X-ray luminosity versus temperature relation for 3 -- 10
keV clusters scales as $L_{\rm X} \propto T^{2.6 - 2.8}$ which is
steeper than expected for self-similar scaling \citep[$\propto
  T^2$][]{mfs98,ae99}.  The steepening is less significant in hot,
relaxed clusters with gas temperatures above 5 keV \citep{asf01}, but
becomes pronounced in cooler clusters and groups.  This departure has
been attributed to excess entropy or preheating of the gas prior to
virialization.  Likely energy sources are the same supernovae that
enriched the gas with metals and AGN outflows from nascent
supermassive black holes. 

In the centers of clusters where radiative cooling is important, the
gas temperature declines by factors of two to three
\citep{asf01,vmm05}.  The temperature there reflects the reduction in
$GM(r)/r$ over the central galaxy.  It is in these so-called cooling
flow regions where AGN outbursts are seen to have their biggest
impact, as discussed below in detail. 

\subsection{Mass Partitioning}

A galaxy cluster's gravitational influence extends over hundreds of
cubic megaparsecs, which is enough in principle to capture a
representative sample of the dark matter and baryons in the Universe.
The form of the matter and its distribution throughout halos depends
on both the history of galaxy formation and the cosmological
parameters.  X-ray and optical/IR observations show that the bulk of
baryons reside in the hot ICM and that the fraction of baryons in
stars decreases with cluster mass (the cold gas content is
negligible). Only about 14\% of the baryons in a $10^{14}\ \msun$
cluster are in the stars, and this fraction decreases to 9\% in
$10^{13}\ \msun$ clusters, possibly owing to a declining efficiency of
galaxy formation with halo mass \citep{lms03}.

The gas mass fraction within $r_{500}$, the radius within which the
mean mass density of a cluster exceeds 500 times the critical density
of the Universe, depends weakly on gas temperature, rising to
$\simeq12\%$ in clusters hotter than $\sim4$ keV \citep{vkf06}.  Gas
mass fraction also varies with radius, and is affected by any process
that modifies the amount of star formation or the energy content of
the remaining gas, including radiative cooling and (pre-)heating
\citep{knv05}.  The baryon fraction in clusters is the sum of the gas
fraction and the stellar baryon fraction.  Based on near-infrared
luminosities of galaxies in clusters with masses of $\sim
3\times10^{14}\ \msun$ (within $r_{500}$), the mass fraction in stars
is approximately 1.64\%.  Therefore, the cluster baryon fraction
within this radius is $\simeq 14\%$.  This figure is close to the WMAP
microwave background 3-year measurement \citep{sbd07}, which gives a
Universal baryon fraction of 16\% to 19\%.  

If the true baryon
fraction is constant at a fixed radius in clusters and is constant
over cosmic time (redshift), the dependence of the measured baryon
fraction on luminosity distance can be a useful probe of the history
of expansion of the Universe.  Applying these assumptions to a study
of 26 X-ray clusters extending to $z = 0.9$ with the Chandra
observatory, \citet{ase04} found a high value for $\Omega_{\rm b}$
that is consistent with type 1a supernova and WMAP3 values.  It must
be emphasized that the assumptions that this technique relies on have
not been shown to be valid.  The baryon fraction is affected by the
history of radiative cooling, feedback, and star formation,
particularly in the cD, which are not well understood
\citep{knv05}. 

\subsection{Magnetic Fields}

Faraday rotation measurements of background radio galaxies and radio
galaxies within clusters have revealed that the intracluster gas is
threaded with magnetic fields \citep{ct02,k03,gf04,v04}.  The Coma
cluster's magnetic field was one of the first to be detected at a
level of a few microgauss \citep[e.g.,][]{k03}.  However, in the cores
of cooling flow clusters (Section 2) field strengths of tens of
microgauss have been inferred using Faraday rotation measures
\citep{ckb01}.  The ratio of magnetic pressure ($B^2 /8\pi$) to gas
pressure ($2\nelec kT$) is typically a few percent, so that magnetic
fields are dynamically unimportant in clusters, except occasionally in
the inner several kiloparsecs or so.  Magnetic fields may have been
deposited by radio galaxies and or quasars, or they may be primordial
fields that have been amplified over time by gas turbulence and/or
compression \citep{ct02}.

\subsection{Transport Coefficients}

Magnetic fields can modify the transport coefficients significantly
because the ratio of the Coulomb mean free path to the Larmor radius
of a thermal proton is $\simeq 10^8 B(kT)^{3/2} \nelec^{-1}$, where
the magnetic field, $B$, is in $\mu$G, the temperature, $kT$, is in
keV, and $\nelec$ is the electron density in $\pcm$.  The same ratio
for electrons is a factor of $\sqrt{m_{\rm p} /m_{\rm e}}$ larger, so
that electrons and ions are both tied rigidly to magnetic field lines
for typical values of the intracluster magnetic field
\citep[e.g.,][]{g06}.  One might then expect thermal conduction and
viscous forces to be controlled entirely by the structure of the
magnetic field \citep[e.g.,][]{t89}.  However, under similar conditions
in the solar wind, thermal conduction appears to be suppressed by only 
a modest factor and it has been argued that this should be true
generally for turbulent plasmas \citep[e.g.,][]{s88,rt89,nm01}.

Many empirical arguments favor greater suppression of thermal
conduction in clusters.  A number of researchers argue that thermal
conduction must be suppressed from the Braginskii value for a
nonmagnetized plasma \citep{s62,b65} by factors of up to 1000 or more
in order to have sharply defined ``cold fronts'' \citep{ef00,ksr02}.
\citet{vmm01} argue that cold fronts are a special case, because shear
can amplify magnetic field parallel to the front, effectively
suppressing heat flow across the front.  For example, they conclude
that a $\sim 10\ \mu$G field is required in the cold front in Abell
3667 to explain suppression of Kelvin-Helmholtz instability, although
\citet{ci04} point out that curvature suppresses growth of the
instability, significantly weakening this conclusion.  

As another
example, \citet{vmf01} found $\sim3$ kpc remnants of the interstellar
medium in NGC 4874 and NGC 4889 at the center of the Coma cluster.  In
order for these to survive thermal evaporation, they found that
thermal conduction must be suppressed there by a factor of 30 ­- 100
from the Braginskii value.  Again, these are special cases which may
not be representative of the general ICM.  However, \citet{mmv03}
argue that temperature variations in Abell 754 require thermal
conductivity to be suppressed by at least an order of magnitude.
Their result is determined for 100 kpc scale regions throughout the
cluster that are not associated with any special structures. 

There are fewer limits on the viscosity.  Based on the morphology of
H$\alpha$ filaments near the northwestern ghost cavity in the Perseus
Cluster, \citet{fsc03} argue that the flow there is laminar, hence
that the viscosity is suppressed by a factor of no more than 15.  This
interpretation is supported by numerical simulations \citep{rmf05}.
With electrons and protons both tied so tightly to the magnetic field 
lines, it is reasonable to expect similar levels of suppression for
thermal conductivity and viscosity.  Taken together, the observational
limits suggest suppression of thermal conduction by somewhat more than
expected from theory and observations of the solar wind and, perhaps,
a similar level of suppression of viscosity. The constraints are
showing signs of converging, but there is no consensus. Unsatisfactory
though it is, transport coefficients in the hot ICM are not well
determined. 

\subsection{Metal Abundances}

The intracluster plasma is enriched with heavy elements to an average
metallicity of roughly 1/3 of the solar value \citep{arb92}, which is
roughly the universal average mean metal abundance
\citep{renzini04}. Elements heavier than helium are created by stellar
evolution, particularly by core collapse supernova explosions of
massive stars (SNe II), and by thermonuclear detonations of accreting
white dwarf stars (SNe Ia), which also play a significant role in
their dispersal.  Core collapse supernovae presumably seeded the
Universe with metals during the early stages of galaxy formation,
while SNe Ia, which are associated with the late stages of stellar
evolution of less massive stars, dominate metal production over longer
time spans.  Clusters of galaxies are close to being ``closed boxes''
and thus retain the memory of metal enrichment through star formation
and stellar evolution.  SNe Ia are major producers of iron, whereas
SNe II produce high yields of the alpha elements (Si, S, Ne, Mg).
X-ray measurements of the relative abundance of the metallic species
are thus able to constrain the history of star formation.

The iron mass in the hot gas in clusters correlates with the
luminosity of the elliptical and lenticular galaxy population but not
with the luminosity of spirals.  Moreover, alpha element abundances
relative to iron in hot clusters are inconsistent with those in the
Milky Way \citep{blh05}.  Therefore, most of the heavy elements were
created by stars bound to the early-type galaxies \citep{arb92}.  The
iron mass in the gas exceeds the iron mass in stars in galaxies by at
least a factor of two \citep{arb92,renzini04}, implying that cluster
galaxies have ejected most of the metals they produced over cosmic
time.  If the iron in the ICM was shed by the stars, then the ratio of
the iron mass to total stellar mass in clusters is a factor of four
larger than expected for a population of stars like those in the Milky
Way \citep{loew06}, implying that star formation as it proceeds in the
Milky Way would have difficulty producing the observed levels of iron
in clusters.  It is unlikely that SNe Ia could have supplied most of
the iron unless the supernova rate was much higher in the past.
Furthermore, SNe II underproduce iron by a factor of five if galaxies
formed with a Salpeter initial mass function (IMF).  It appears that
most of the iron was produced by SNe II in a rapid phase of early star
formation that proceeded with an IMF heavily weighted toward massive
stars relative to the IMF operating in present day spiral galaxies
\citep{renzini04,loew06}.  No evidence has been found for strong changes
in abundance with lookback time out to $z \sim 0.5$, which is
consistent with early enrichment scenarios \citep{ml97}.

Abundance gradients corresponding to average metallicity increases of
factors of two or more are routinely found in the central $\sim100$
kpc regions of cD clusters
\citep{efm97,ime97,fdp00,dm01,ib01}. High-resolution Chandra studies
have shown that the gradients are strongest near the cD and that they
sometimes track the stellar isophotes \citep{wmm04}.  For example, in
Hydra A, \citep{dnm01} found that the iron abundance rises from 30\%
of the solar value at 100 kpc to about 60\% in the nucleus of the cD,
and that the silicon abundance rises from roughly half the solar value
to the solar value in the nucleus.  They attribute the gradient to
constant SNe Ia production from the central galaxy over the past
several gigayears.  A similar situation is found in M87 \citep{fmb02},
where most of the iron seems to have originated from SNe Ia, while
only 10\% was injected from SNe II.  Intriguingly, \citet{tkd04} found
that SNe II enrichment from massive star formation may be important in
the cores of some clusters.  Star formation is often observed in cD
galaxies centered in cooling flows, which could contribute 
significantly to the gradients.

\section{CLASSICAL COOLING FLOWS}

A cooling flow cluster is characterized by bright X-ray emission from
cool, dense gas in the central region of the cluster \citep[see ][ for
a more comprehensive review of the basic operating principles of a
cooling flow]{fab94}.  Within the cooling radius, where the cooling
time of the gas is less than the time since the last major heating
event, the surface brightness of the gas near a central cD galaxy
often rises dramatically above a $\beta$-model, by factors of up to
100, corresponding to a rise in gas density by factors of 10 or
more. The X-ray luminosity within the cooling region reaches values of
$10^{45}\ \ergps$ in the extreme.  In many cases it is more than 10\%
of the cluster's total luminosity. If this luminosity is uncompensated
by heating, the gas will radiate away its thermal and gravitational
energy on a timescale of $\tcool = p/[(\gamma - 1) \nelec \nh
  \Lambda(T )] < 10^9$ year \citep{silk76,cb77,fn77,mb78}, where $p$
is the gas pressure, $\Lambda(T)$ is the cooling function, and
$\gamma$ is the ratio of specific heats of the gas.  As the gas
radiates, its entropy decreases and it is compressed by the
surrounding gas, causing it to flow inward.  The cooling time decreases
as the gas density increases and, eventually, the gas temperature
drops rapidly to $<10^4$ K, so that cooled gas condenses onto the
central galaxy. The condensing gas is replenished by hot gas lying
above, leading to a steady, long-lived, pressure-driven inward flow 
of gas at a rate of up to $1000\ \msunpy$ \citep{fab94}.

Observed cooling times are significantly longer than free-fall times,
so that the gas remains very nearly hydrostatic as it cools. The flow
is then governed by cooling, making the flow time $\tflow = r/v$,
where $v$ is the radial speed of inflow, approximately equal to the
cooling time.  Counterintuitively, the heat lost to radiation does not
necessarily make the gas temperature decrease.  The inexorable entropy
decrease is offset by adiabatic compression as gas flows inward, so
that, typically, the temperature of the cooling gas follows the
underlying gravitational potential, i.e., $kT/(\mu \mh)$ is a multiple  
of order unity of the ``local virial temperature,'' $GM(r)/r$
\citep[e.g., solutions of ][ with $k = 0$]{nul86}.  Cooling gas would
therefore remain approximately isothermal in an underlying isothermal
potential.  Because the virial temperature of cluster central galaxies
(revealed by their stellar velocity dispersions) is typically lower
than that of the surrounding cluster, the gas temperature declines
inward in a classical cooling flow.  Rather than being a direct
manifestation of cooling, this temperature drop reveals the flattening
of the underlying gravitational potential.  (If gas cools enough to
flow inward, this feature will persist in more up-to-date cooling flow
models.)  The Mach number of the flow increases inward, until it
approaches unity.  Up to that point, linear growth of the thermal
instability is suppressed by buoyant motions \citep{bs89}, but beyond
it cooling is too fast and the gas is expected to form a rain of
thermally unstable clouds that cool rapidly to low temperature. 

The classical cooling flow is approximately steady within the region
where the cooling time is shorter than its age. The power radiated
from the steady flow equals the sum of the enthalpy carried into it
and the gravitational energy dissipated within it. To a first
approximation, the gravitational energy can be ignored, so that the
luminosity $L_{\rm X} \simeq  \dot M (5 kT)/(2\mu \mh) \simeq 1.3
\times 10^{44} T_5 \dot M_22\ \ergps$, where the temperature of the
gas entering the flow is given by $T_5 = (kT/5\rm\ keV)$ and the
cooling rate is given by $M_2 = \dot M/(100\rm\ \msunpy)$.  This
expression is exact for steady, isobaric cooling. 

This single-phase cooling model predicts central spikes in surface
brightness that are clearly stronger than observed, a discrepancy that
prompted the introduction of inhomogenous cooling flow models. Such
models postulate a broad spectrum of gas temperature and density at
each radius \citep{nul86}.  Both Rayleigh-Taylor and shear
instabilities can disrupt an overdense cloud if the distance it must
move to reach its convective equilibrium position exceeds its size
(alternatively, if its fractional overdensity is greater than its size
divided by the distance to the cluster center).  Such nonlinear
overdensities therefore tend to be shredded finely, slowing their
motion relative to the bulk flow and enabling them to be pinned to the
flow by small stresses (e.g., magnetic stresses).  Tying overdensities
to the bulk flow then permits the growth of thermal instabilities
throughout the core of the cluster, leading to widespread deposition
of cooled parcels of gas and making the mass flow rate, $\dot M$, a
strong function of the radius.  This solution avoids the need for
strong X-ray and UV surface brightness spikes centered on the cD's
nucleus by distributing the hundreds of solar masses of gas and star
formation deposited each year throughout the cooling region of the
cluster. 

This cooling flow model has been under siege for years, primarily
because of its failure to predict the observed amount and spatial
distribution of star formation, line emission, and other expected
products of cooling, which are generally observed only in the inner
few tens of kiloparsecs. The spatial distribution of cool gas and star
formation is more consistent with the single-phase model, but at
levels that fall orders of magnitude below the predictions. This
failure implies that the gas is not condensing at the predicted rates,
and that radiation losses are either being replenished, or the gas is
condensing into an unseen state. Sensitive searches for the repository
in optical, infrared, and radio bands have severely restricted the latter
possibility, if not ruled it out entirely.

\subsection{The Modern View of Cooling Flows}

The strongest spectroscopic signatures of cooling gas are X-ray
emission lines below 1 keV of various charge states of Fe L
\citep{bmc02}.  Early observations of M87 and a few other bright
clusters made with the Einstein Observatory's Focal Plane Crystal
Spectrometer [FPCS]\citep[][]{ccj82,cmd88} apparently detected these Fe L
lines at the $4\sigma$ to $6\sigma$ level. Remarkably, the line
strengths from gas at temperatures of a few million to a few tens of
million degrees agreed with predictions of the classical cooling flow
model. Unfortunately, the FPCS, with an effective area of only one
square centimeter, was unable to achieve detections of higher
statistical significance. Nevertheless, for two decades these
concordant results were bolstered by signatures of cooling at other
wavelengths: filaments of H$\alpha$ emission from warm, ionized gas
\citep{heckman81}; star formation \citep{jfn87,mo89}; and pools of
cold atomic and molecular gas \citep[][]{edge01} provided support
for a classical cooling flow model, albeit at lower rates.

The situation changed dramatically when far more sensitive XMM-Newton
Reflection Grating Spectrometer (RGS) observations failed to confirm
the picket fence of lines, including the Fe L features, at $\sim 1$
keV from cooling gas in spectra of the clusters Abell 1835
\citep{ppk01} and Abell 1795 \citep{tkp01}.  The prominent and useful
Fe XVII line at 12 \AA\ is either weak or absent in cooling flow 
clusters \citep{pkp03}, although there is a hint of Fe XVII emission
in the Abell 2597 cluster at a level that is consistent with a cooling
rate of $\sim100\ \msunpy$ \citep{mf05}.  Ultraviolet line emission
also suggests cooling rates about an order of magnitude smaller than
previous estimates \citep{ocd01,bfm06}.

The relationship between cooling rate and line power that is central
to this issue was first noted by Cowie (1981).  In terms of the
entropy, S, the energy equation of cooling gas is $\rho T dS/dt =
-\nelec \nh \Lambda(T)$, where $\rho$ is the gas density, $t$ is
the time and other terms are as above. Rearranging this to describe a
fluid element in gas cooling steadily at the rate $\dot M$ gives the
mass of the element as $dM = \dot M \, dt = - \dot M \rho T d
S/[\nelec \nh \Lambda] = \rho \, dV$, where $dV$ is its volume, so
that the emission measure of the element is $\nelec \nh\,  dV = - \dot
M T dS/\Lambda$ (where $dS < 0$ for cooling gas).  In terms of the
pressure and temperature, $dS = [\gamma \, dT/{(\gamma - 1)T} -
  dp/p]k/(\mu \mh )$.  As the gas cools and flows inward, its pressure
generally increases, so that $-dS$ is minimized for isobaric cooling
($d p = 0$ and $dT < 0$; late stages of cooling can be isochoric, 
reducing the integrand below by a factor of $\gamma$ at sufficiently
low temperatures).  If the contribution of a line (or group of lines)
to the cooling function is $\Lambda_{\rm line} (T )$, then the total
power radiated in that line by gas cooling isobarically in a steady
flow is 
\begin{displaymath}
P_{\rm line} = \int \nelec \nh \Lambda_{\rm line}(T)\, dV
= \dot M {\gamma\over\gamma - 1} {k \over \mu \mh} \int_0^{T_{\rm i}}
{\Lambda_{\rm line}(T) \over \Lambda(T)} \, dT,
\end{displaymath}
where $T_{\rm i}$ is the temperature that the gas cools from (the
sense of integration has been reversed here). If the gas pressure
increases as the gas cools, the power radiated in the line would be
greater and this expression would overestimate the cooling rate. 

The failure to detect the low energy X-ray lines at the expected
levels indicates that the canonical cooling rates were overestimated
by an order of magnitude or more, although the picture is not so
simple \citep{pf06}.  In the classical cooling flow model, thermally
unstable gas clouds are expected to be approximately isobaric, so that
the constant pressure cooling model should be reasonably accurate,
especially for lines emitted mainly by gas that is cooler than
average. As outlined above, the constant pressure cooling model makes
specific predictions of the strengths of lines from low temperature
cooling gas \citep[cf.][]{bmc02,pf06}.  These predictions are
inconsistent with the line strengths observed by \citet{pkp03}, which
require the emission measure to decrease more rapidly with temperature
than expected for the constant pressure cooling flow model.  This 
behavior cannot be explained by merely reducing the cooling rate.

The absence of multiphase gas is confirmed in moderate spectral
resolution CCD data from Chandra, XMM-Newton, and ASCA.  In general,
the deprojected gas profiles in cooling flows can be adequately
modeled by a single temperature plasma at each radius
\citep{mwn00,dnm01,mp01}, except perhaps near the nucleus of the cD,
where star formation is frequently observed. 

It may be possible for nonradiative cooling to proceed at or near the
canonical rates without revealing X-ray line emission.  Processes such
as gas-phase mixing, differential photoelectric absorption, and
rapidly cooling, unresolved, high metallicity inclusions in the hot
gas \citep{fmn01} are possible.  If so, the problem of identifying the
permanent repository for the cooling material would remain.

\section{RADIO LOBE RELATED X-RAY STRUCTURE AND CAVITIES IN CLUSTER
  CORES}

Disturbances in the hot gas near NGC 1275 were first noted in an early
Einstein Observatory image of the Perseus cluster \citep{bff81,fhc81}.
A decade later, Rosat's 5 arcsec High Resolution Imager (HRI) associated 
the disturbances with two cavities filled with radio emission
emanating from the nucleus of NGC 1275 \citep{bvf93}.  Similar
disturbances were later noted in HRI images of other bright cD
clusters \citep[e.g.,][]{cph94,hs98,oe98,rlb00}, but limitations in
Rosat's spatial and spectral resolution stalled any further progress
on the nature and import of these disturbances until the launches of
Chandra and XMM-Newton in 1999. 

We now know that nearly three dozen cD clusters and a similar number
of gE galaxies and groups harbor cavities or bubbles in their X-ray
halos \citep{fse00,mwn00,mwn01,scd01,hcr02,mkp02}.  Cavity systems are
difficult to detect, so this is surely a lower limit to their
numbers. Like the radio lobes that created them, cavities are usually
found in pairs of approximately elliptical X-ray surface brightness
depressions, 20\% to 40\% below the level of the surrounding gas. This
is the expected decrement toward a spheroidal empty cavity embedded in
a $\beta$-model atmosphere \citep{fse00,mwn00,bsm01,bsm03,ndm02}.
Cavity systems in clusters vary enormously in size, from diameters
smaller than 1 kpc like those in M87 \citep{fnh05,fcj07} to diameters
approaching 200 kpc in the MS0735.6+7421 and Hydra A clusters
\citep{mnw05,nmw05,wmn07}.  A correlation exists between radio
luminosity and cavity power \citep[Section 5.1; e.g.,][]{brm04,df06},
but with a large scatter that is poorly understood.

One of Chandra's early surprises was the discovery of cavities devoid
of bright 1.4 GHz radio emission. They were dubbed ghost cavities, and
were interpreted as aging radio relics that had broken free from the
jets and had risen $20-30$ kpc into the atmosphere of the cluster
\citep{mwn01,fcb02}.  We now know they are filled with low-frequency
radio emission and may be connected by tunnels back to the nucleus
\citep{csb05,wmn07}.  Some clusters have multiple pairs of cavities,
apparently produced by multiple radio outbursts or quasicontinuous
outflows.

The work required to inflate the cavities against the surrounding
pressure is roughly $pV \sim 10^{55}$ erg in gEs \citep[e.g.,][]{fj01}
and upward of $pV = 10^{61}$ erg in rich clusters
\citep[e.g.,][]{rmn06}.  The total energy needed to create a cavity is
the sum of its internal (thermal) energy, $E$, and the work required 
to inflate it, i.e., its enthalpy, $H = E + pV$.  This is several
times $pV$. The displaced gas mass is several $10^{10}\ \msun$ in an
average cluster system such as Abell 2052 \citep[e.g.,][]{bsm01} but
can exceed $10^{12}\ \msun$ in powerful outbursts such as those in 
MS0735.6+7421 and Hydra A.  The cavities in these systems occupy
between 5\% and 10\% of the volume within 300 kpc giving the hot ICM a
Swiss cheese-like topology \citep{wmn07}.  The bright rims or shells
surrounding many cavities are cooler than the ambient gas
\citep{fse00,mwn00,bsm01,bsm03,ndm02} and thus are not active shocks as
anticipated in early models \citep[e.g.,][]{hrb98}. Evidently, the
cavities are close to being in pressure balance with the surrounding
gas.  A nearly empty cavity will rise into the cluster atmosphere like
a buoyant weather balloon, traveling at a speed approaching the local
free-fall velocity.  The cool rims are probably composed of displaced
gas dragged outward from the center by the buoyant cavities
\citep{bsm01,cbk01,rhb01}.

Cavity systems are often surrounded by belts \citep{swa02}, arms
\citep{ywm02,fnh05,fcj07}, filaments, sheets \citep{fst06}, and
fragile tendrils of gas maintained against thermal evaporation,
perhaps, by magnetic fields threaded along their lengths
\citep{nb04,fcj07}.  This structure is usually composed of cooler gas
and is associated with H emission that may be tracing circulation
driven by rising radio lobes and cavities \citep{fsc03}.
``Swirls'' of cool X-ray gas were found in the central regions of the
Perseus \citep{fst06} and Abell 2029 \citep{cbs04} clusters, which may
be related to merger activity or circulation flows generated by radio
sources \citep[e.g.,][]{mb03,hby06}. At present, this wealth of
structure is not well understood.

\begin{figure}
\centerline{\includegraphics[height=9truecm,angle=270]{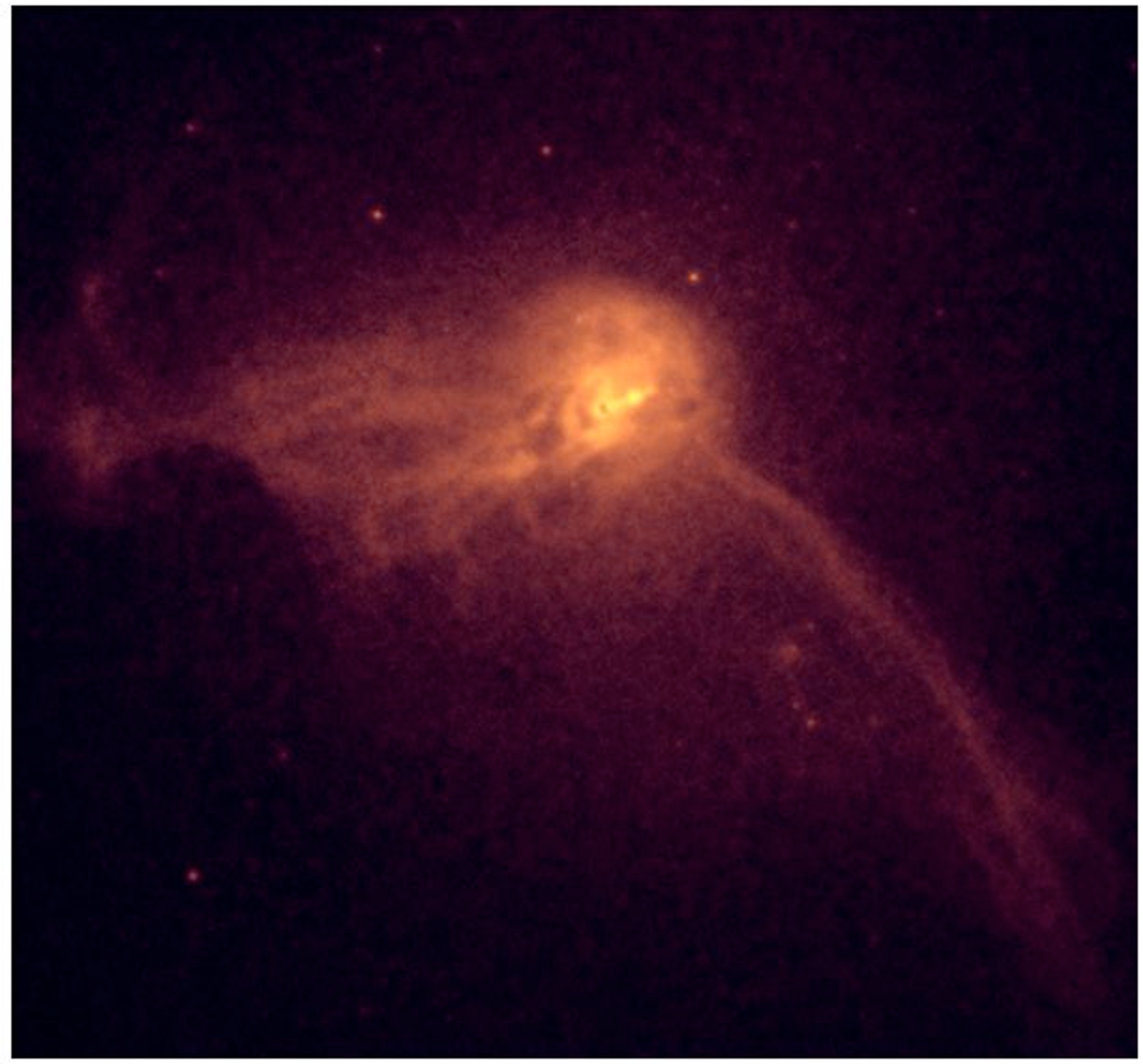}
\includegraphics[height=7truecm,angle=270]{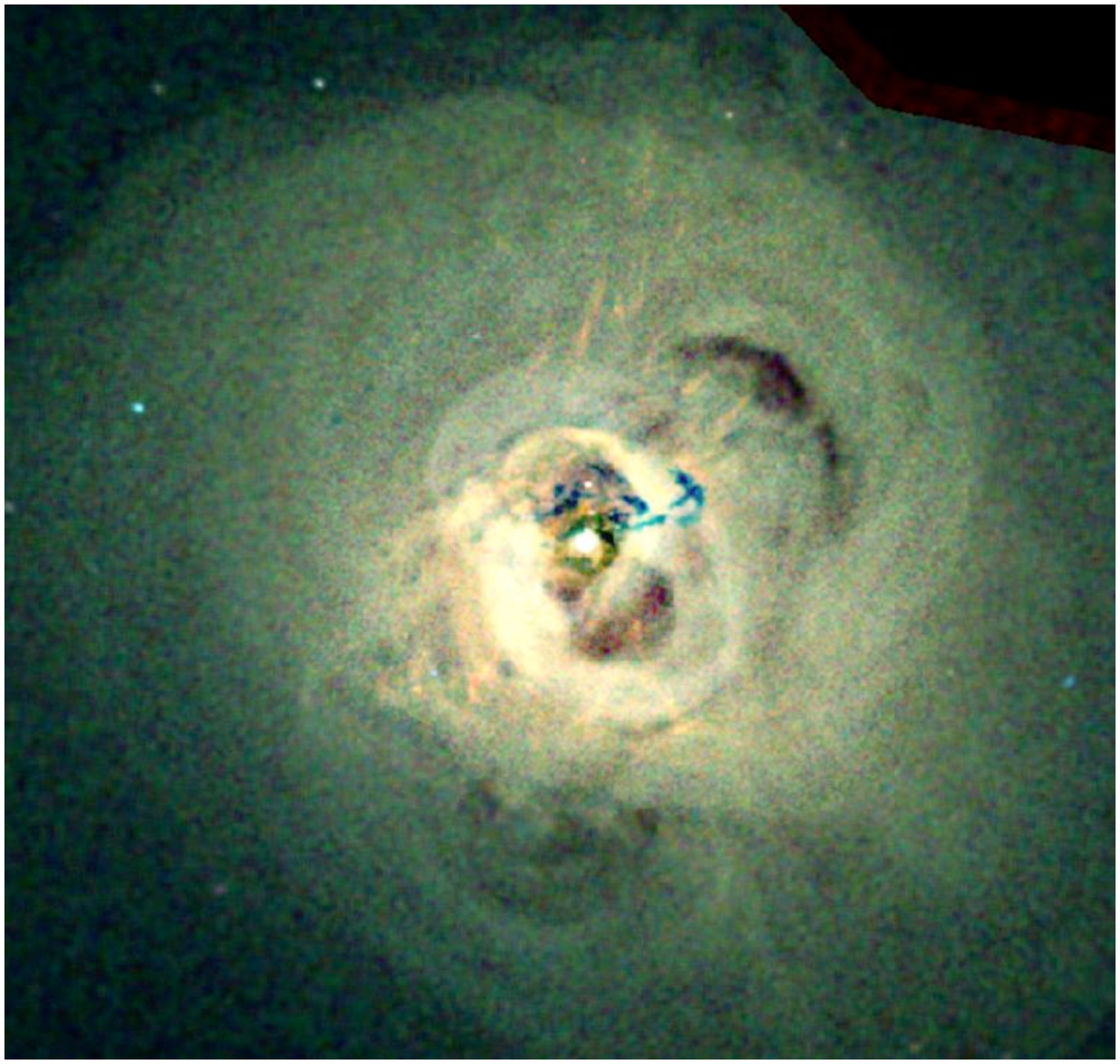}}
\caption{{\it Left.} Chandra X-ray image of M87 showing structure in
  the hot gas associated with the AGN outburst \citep{fcj07}. Several
  small cavities and filaments of uplifted gas are visible. X-ray
  emission from the radio jet is seen at the center.  {\it Right.}
  Chandra X-ray image of the Perseus cluster exposed for 900,000
  s from Fabian et al. (2006). Two inner cavities containing the active radio lobes and two
  outer ghost cavities are seen. The sound waves (weak shocks) are
  visible as a series of circular ripples surrounding the
  cavities. The central blue structure is absorption by a foreground
  galaxy.} \label{fig:pretty}
\end{figure}

Owing to their proximity and high surface brightnesses, the M87 and
Perseus clusters are spectacular exemplars of these structures (Figure
\ref{fig:pretty}).  Keep in mind, however, that neither is outstanding
in its power output. The AGN (cavity) power in Perseus is about 25
times larger than that of M87, and Perseus itself is feeble compared
to Cygnus A and MS0735.6+7421, which are roughly 215 and 270 times,
respectively, more energetic than Perseus \citep{rmn06}. Some 18 of
the 33 cavity systems studied by \citet{rmn06} exceed the AGN power
output of Perseus, suggesting the Perseus and Virgo clusters serve as
useful benchmarks for average to low power outbursts.

\subsection{Shock Fronts}

\begin{figure}
\centerline{\includegraphics[height=20truecm]{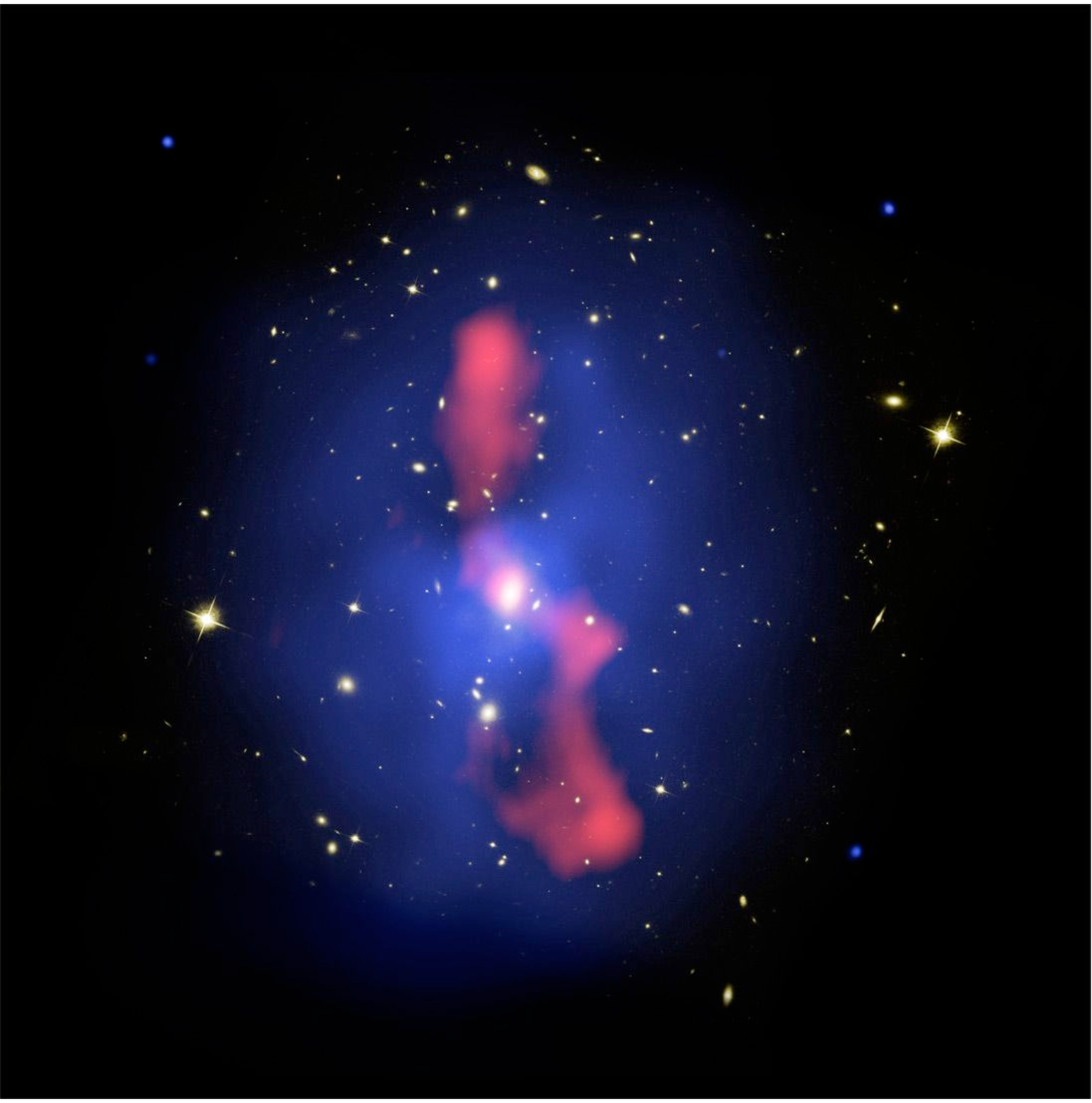}}
\caption{Hubble Space Telescope visual image of the MS0735.6+7421
  cluster superposed with the Chandra X-ray image in blue and a radio
  image from the Very Large Array at a frequency of 330 MHz in
  red. The X-ray image shows an enormous pair of cavities, each
  roughly 200 kpc in diameter that are filled with radio emission.
  The radio jets have been inflating the cavities for $10^8$ years
  with an average power of $< 2 \times 10^{46}\ \ergps$.  The
  displaced gas mass is $< 10^{12}\ \msun$.  The cavities and radio
  source are bounded by a weak shock front. The cavities are well
  outside the central galaxy and cooling region of the cluster. The
  supermassive black hole grew by at least $< 3 \times 10^8\ \msun$
  during the outburst.} \label{fig:ms07}
\end{figure}

With the notable exceptions of the Mach 8 shock in Centaurus A
\citep{kvf03}, the Mach 4 shock in NGC 3801 \citep{ckh07}, and
perhas surrounding the overpressured quasar 3C186 \citep{siem05} there is
little evidence for strong shocks surrounding the radio-lobe cavities
studied with Chandra.  However, evidence of weaker, more remote shocks
is accumulating.  \citet {jfv02} reported gas features near NGC 4636
consistent with a Mach 1.7 shock.  \citet{swa02} saw evidence of
shocks around the radio source Cygnus A.  More recently, Chandra
images of clusters have revealed roughly elliptical breaks in surface
brightness enveloping the inner radio-lobes of M87 \citep{fnh05} and 
the enormous cavity systems in MS0735+7421 (Figure \ref{fig:ms07}),
Hydra A, and Hercules A \citep{mnw05,nhm05,nmw05}.  These structures
resemble the classical cocoon shocks thought to envelope powerful
radio sources as they advance into the surrounding medium. Modeled as
a spherical shock from a nuclear outburst, the surface brightness
discontinuities can be reproduced by conventional shock waves with
Mach numbers lying between 1.2 and 1.7
\citep{mnw05,nhm05,nmw05}. Though the shocks are only mildly
supersonic (weak) they can encompass huge volumes, $200-400$ kpc in
radius and require energy deposition upward of $10^{61}$ erg into the
intracluster gas. The shock energy corresponds to several $pV$ per
cavity, comparable to the free energy of the cavities themselves \citep[see
also][]{wmn07}.  Thus the cavity enthalpy provides only a lower limit
to the total outburst power. The ages of these outbursts, estimated
from shock models (approximately the shock radius divided by the shock
speed), are reassuringly similar to buoyant rise times of their
cavities, although significant differences found, for example, in
Hydra A are related to the outburst history and the partitioning of
energy between enthalpy and shock heating (Section 3.5).

Detecting the expected gas temperature rises behind these shock fronts
has proved difficult. Radiative losses are negligible on timescales of
interest, so that shocks in the ICM should conserve
energy. Furthermore, the magnetic field in the bulk of the ICM is
dynamically insignificant and does not affect shocks significantly for
pure magnetohydrodynamic shocks \citep[e.g.,][]{ndm02}.  Some shock
energy can be absorbed by particle acceleration, but this is expected
to be small for weak shocks.  Under these conditions, the ICM should
be well approximated as an ideal gas with a constant ratio of specific
heats, $\gamma = 5/3$, leading to the well-defined relationship, 
\begin{equation}
{T_2\over T_1} = {(\gamma + 1) \rho_2 / \rho_1 - (\gamma - 1) \over
[(\gamma + 1) - (\gamma - 1)\rho_2 / \rho_1] \rho_2 / \rho_1},
\end{equation}
between the shock temperature jump, $T_2 /T_1$, and density jump,
$\rho_2/\rho_1$.  For example, for the Mach 1.2 shocks in Hydra A and
M87 the temperature jumps by 20\% behind the shock. However, the
temperature declines rapidly behind the shock because of adiabatic
expansion. When projected onto the sky, emission from the shocked gas
is also diluted by emission from the surrounding unshocked gas. As a
result, the projected emission-weighted temperature has a peak rise of
only 5\%. Given that thousands of photons are required to measure
temperatures accurately, few such temperature rises have been
detected. Examples of detections include the 14 kpc 
ring in M87 \citep{fcj07}, MS0735.6+7421 \citep{mnw05}, Hercules A
(PEJ Nulsen, in preparation), Centaurus A \citep{kvf03}, and NGC 4552
\citep{mnj06}. 

For MS0735.6+7421 and Hercules A, the mean jet power released into the
ICM over $\sim 10^8$ years is more than $10^{46}\ \ergps$, which is
comparable to a powerful quasar.  The bulk of this energy is deposited
beyond the cooling region of those clusters.  Despite this quasar-like
power, the cD hosts share few quasar characteristics (e.g., broad
nuclear emission lines), although this issue needs further study. The
energy deposited in the inner 1 Mpc of MS0735.6+7421 corresponds to a
few tenths of a keV per particle, which is a significant fraction of
the $\sim 1$ keV required to supply the excess entropy (preheating) in
clusters \citep{wfn00,rrn04,vd05}.  Thus several outbursts of this
magnitude during the life of a cluster, particularly in the early
stages of its development, could preheat it (see Section 1.2). No
clear distinguishing characteristics have yet been noted between the
cD galaxies hosting powerful outbursts and cDs in other cooling
flows. Thus occasional powerful outbursts, if they occur in all
systems, could dominate the energy output from smaller, more frequent
outbursts integrated over cluster ages. 

\citet{vd05} have identified several clusters with strongly boosted
central entropy profiles and relatively weak radio sources that
otherwise show no evidence of recent AGN activity. They suggested that
the central entropy boosts were imprinted by shocks generated by
powerful AGN outbursts that occurred in the past, although the AGN are
dormant at present. Furthermore, Voit \& Donahue found that the entropy
profiles are generally consistent with shock heating in the inner few 
tens of kpc, but the mode of heating switches to cavities and sound
waves further out.  An additional consequence of repeated powerful
outburts is the substantial growth at late times of the black holes at
the core of the AGN, which is discussed further in Section 7.2.

Luminosity boosts, gas clumping, and gas outflows caused by AGN
outbursts could, in principle, affect measurements of cosmologically
important quantities such as gas mass fractions, the luminosity
function, and cluster masses. However, these effects are likely to be
subtle and are just beginning to be explored
\citep[e.g.,][]{knv05,gmn07}. 

\subsection{Ripples and Sound Waves}

The best known example of weak shocks or sound waves is that in the
Perseus cluster.  It is seen as a spectacular system of ripples and
other disturbances in an 890 ksec Chandra image \citep{fst06}.  The
half dozen ripples lying beyond the inner cavity system (Figure
\ref{fig:pretty}) are separated by roughly 11 kpc and are visible out
to a radius  of 50 kpc \citep{fsa03}.  The ripples appear to be
pressure disturbances with amplitudes of 5\% to 10\%, or sound waves
(weak shocks) propagating outward with a period of $\sim 10^7$ year
\citep{fsa03,fst06}.  Apparently the gas within 25 kpc surrounding the
cavity system is overpressured by about 30\%, implying that the entire 
inner halo is expanding.  Cavity $pV$ work alone would then
underestimate the AGN power in Perseus.  Temperature jumps across the
ripples and the band of high pressure have not been found, despite the
ample number (70 million) of detected photons available.
\citet{fst06} have suggested that thermal conduction is suppressing 
the temperature jumps by creating isothermal shocks.  This phenomenon
is poorly understood.

\subsection{Detectability of Cavity Systems}

Studies of cavity populations should in principle yield information on
the AGN duty cycle, the energy per AGN outburst, and outburst ages,
once a reliable dynamical model has been established and the
complicating issues of cavity detectability, stability, and lifetime
are understood. In a genuinely random sample of galaxy clusters, radio
jets and their associated cavities are expected to emerge from the
cD's nucleus at random orientations with respect to the plane of the
sky. The decrement in surface brightness of a cavity relative to the
surrounding cluster, i.e., its detectability, will be a strong
function of its size and distance from the cluster center, as well as
its aspect with respect to the plane of the sky. In the simplest
approximation, a small bubble of radius $r$ on the plane of the sky at
a distance $R$ from the cluster center produces a count deficit
(cavity) scaling as $r^3 (1 + R^2 /a^2)^{-3\beta}$, where $\beta$ and
$a$ are the $\beta$-model parameters for the cluster. The cluster
count from over the bubble scales as $r^2 (1 + R^2 /a^2)^{-3\beta +
  1/2}$ and the noise in this scales as its square root. Thus the
signal-to-noise ratio scales as $r^2 (1 + R^2 /a^2)^{-3\beta/2-1/4}$,
implying that cavities are easiest to detect when they are large and
are located in the bright central regions of the cluster. This simple
argument does not take projection into account, but should apply
roughly to bubbles within $45^\circ$ of the plane of the sky as seen
from the cluster center. Bubbles far from the plane of the sky are
difficult to detect at any radius. 

The detectability of a cavity system as a function of its age and
nuclear distance (time) was modeled by \citet{eh02}, who considered
spherical bubbles of adiabatic, relativistic plasma embedded in an
isothermal cluster with a core radius $r_{\rm c} = 20$ kpc.  As a
cavity rises, it expands adiabatically from its initial volume, $V_0$,
to a modestly larger volume, $V_1$, at its observed location as $V_1 = 
V_0 (p_0/p_1)^{1/\Gamma}$, where $p_0$ and $p_1$ are the ambient
pressures at the respective locations, and $\Gamma$ is the ratio of  
specific heats within the cavity.  For a spherical cavity, the ratio
of its radius, $r$, to the distance, $R$, from the cluster center
therefore evolves as 
\begin{equation}
{r \over R} = {r_0\over R_0} \left(R\over R_0\right)^{-1} \left[ p(R)
  \over p(R_0)\right]^{-1/(3\Gamma)},
\end{equation}
where $p(R)$ is the pressure of the ICM at distance $R$ and subscripts
``0'' denote initial values.  En{\ss}lin \& Heinz found that the contrast
of a cavity launched from the center of the cluster along the plane of
the sky is a slowly decreasing linear function of distance, until the
cavity vanishes into the background at large distances. The
detectability of cavities rising at oblique angles with respect to the
plane of the sky declines according to a power law with distance. From
figure 3 of \citet{eh02}, we find that the detectability declines
$\propto R^{-1}$ and $\propto R^{-2.5}$ for cavities launched at
$45^\circ$ and $90^\circ$, respectively, from the plane of the
sky. These estimates do not include the effects of cavity rims, which
enhance cavity contrast, nor of cavity disruption (Section 3.6), that
works in the opposite direction. Nevertheless, they imply that once
cavities have ventured distances of several times their initial radii,
their chances of detection decline rapidly. 

\subsection{Cavity Statistics}

The statistical properties of cluster cavities drawn from the Chandra
archive, together with their attendant radio sources, have been
studied in some detail
\citep[e.g.,][]{brm04,df04,df06,dft05,rmn06}. These studies are hobbled
by unknown selection effects in the Chandra archive. Nevertheless,
they suggest a high incidence of detectable cavity systems in
clusters, groups, and galaxies, spanning a large range of gas
temperature. The detection frequencies found in the samples of
\citet{brm04} and \citet{rmn06} are 20\% (16/80) and 25\% (33/130),
respectively.  Using the flux-limited sample of the brightest 55
clusters observed by the Rosat observatory \citep{pfe98},
\citet{dft05} found between 12 and 15 clusters with cavity systems,
giving an overall detection rate of between 22\% and 27\%.  Their
detection rate rose to 70\% (12/17) in strong cooling flows
\citep{dft05}. The detection rate is close to 25\% (27/109) for a
sample of nearby gEs with significant hot atmospheres \citep{njf07}.
While there is considerable overlap between the three cluster samples,
the gE sample is largely independent (Section 4.3). 

\begin{figure}
\centerline{\includegraphics[height=10truecm]{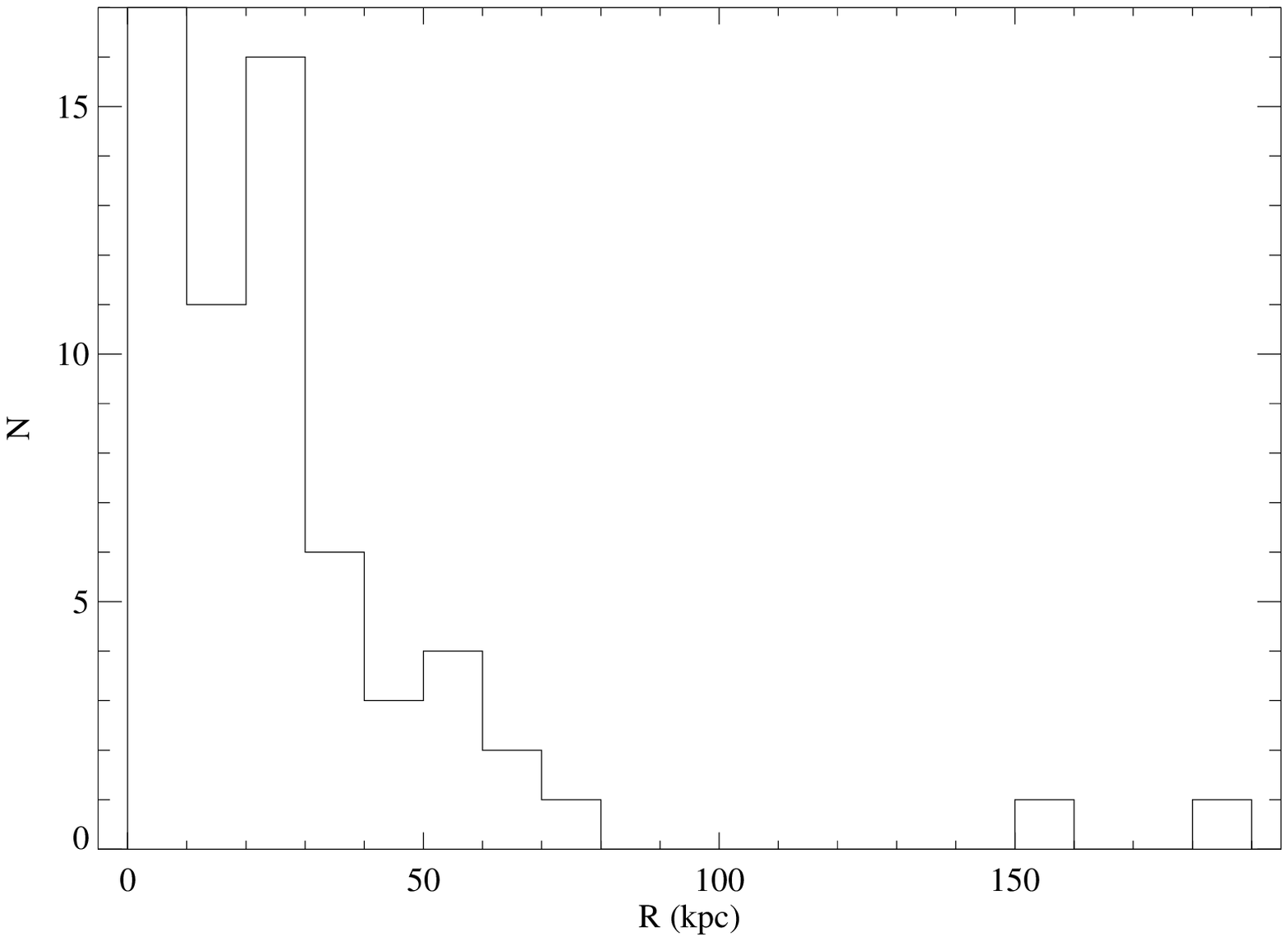}}
\caption{Distribution of cavity distance from the nucleus of the host
  cD galaxy. } \label{fig:rdist}
\end{figure}

The \citet{rmn06} sample is the most extensive cluster sample
available, and thus provides a good representation of average cavity
properties in clusters. The distribution of projected distances
between the nucleus of the host cD and cavity centroid in Figure
\ref{fig:rdist} shows that the detection rate peaks in the inner 30
kpc or so and declines rapidly at larger distances. Only the rarest
and most powerful outbursts produce detectable cavities beyond $\sim
100$ kpc. Within 100 kpc the detection frequency 
declines formally as $\sim R^{-1.3}$, but is consistent with $\sim
R^{-1}$, the expected rate of decline for cavities launched on random
trajectories (Section 3.3). 

\begin{figure}
\centerline{\includegraphics[height=10truecm]{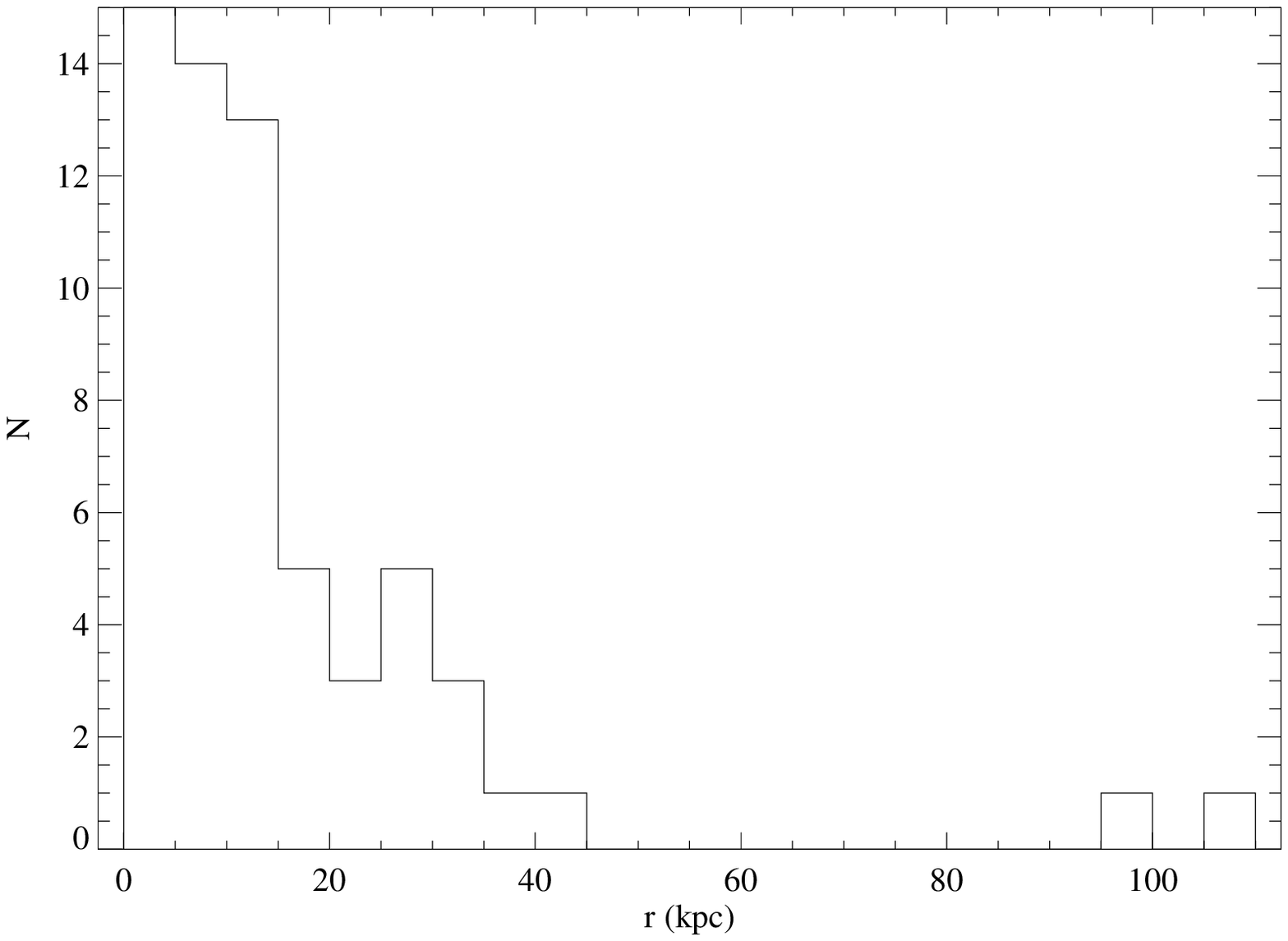}}
\caption{Distribution of cavity sizes.  An ellipse was fitted to each
  cavity and its is size expressed as $\sqrt{ab}$, where $a$ and $b$
  are the semiaxes.} \label{fig:szdist}
\end{figure}

\begin{figure}
\centerline{\includegraphics[height=10truecm]{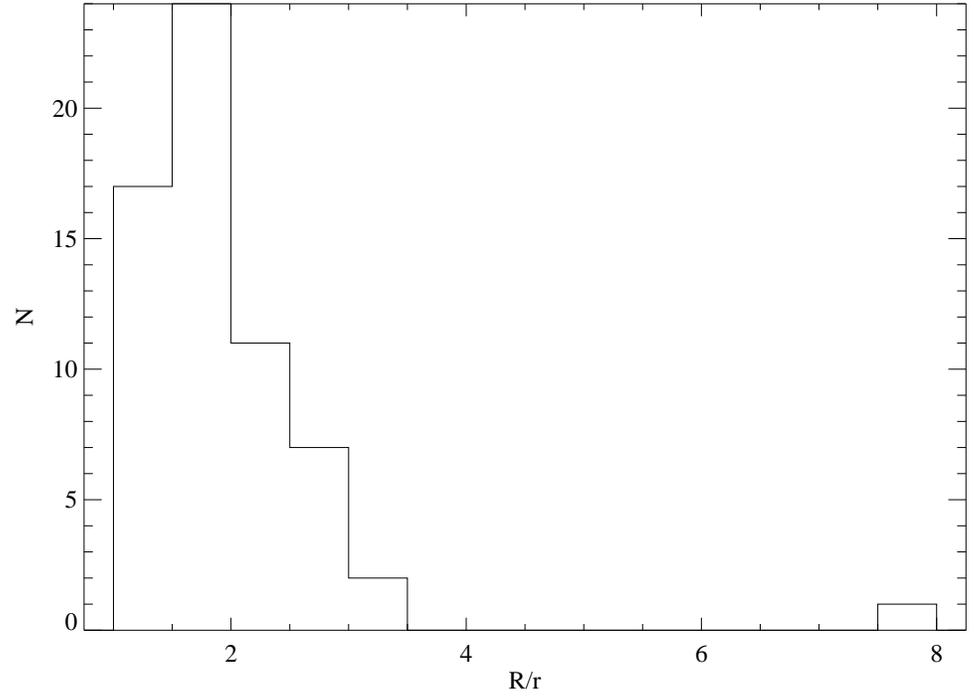}}
\caption{Distribution of the ratio of cavity size to nuclear distance.
  See Figures 3 and 4. } 
\label{fig:ratdist}
\end{figure}

\begin{figure}
\centerline{\includegraphics[height=10truecm]{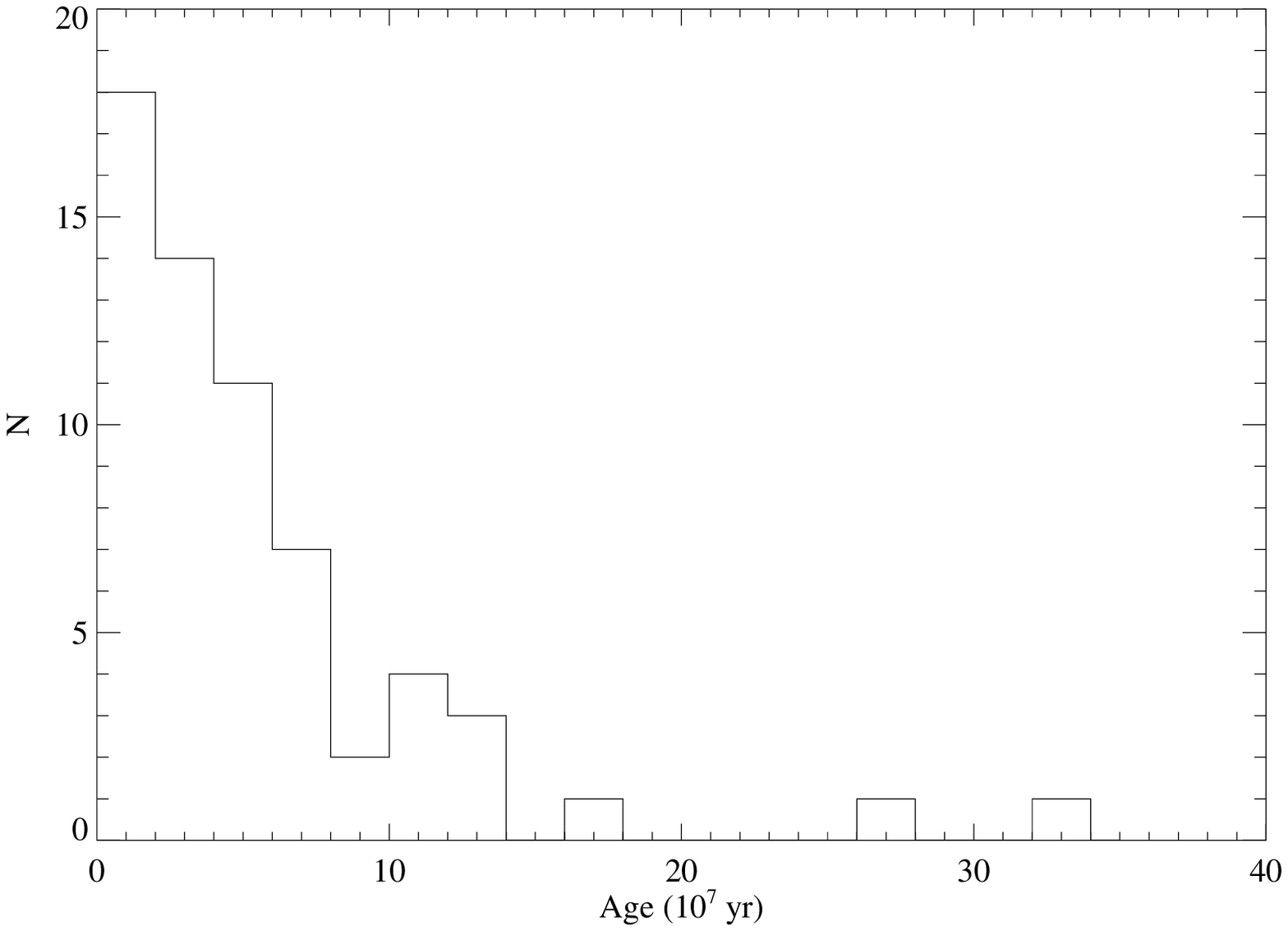}}
\caption{Distribution of cavity buoyancy ages. }  
\label{fig:agedist}
\end{figure}

The distribution of observed cavity sizes (Figure \ref{fig:szdist})
shows a typical value of 10 -- 15 kpc in radius with no preferred
size.  The distribution of the ratio of projected nuclear distance to
radius, shown in Figure \ref{fig:ratdist}, peaks at $R/r \simeq 2$ and
falls off rapidly beyond.  Evidently, cavities travel roughly their
own diameters before they disintegrate or become too difficult to
detect. Other than the most powerful systems, cavities are found
within the light of the central galaxy. The distribution of buoyancy
ages (Figure \ref{fig:agedist}) shows a typical age of $\sim 10^7$
years but some are greater than $\sim 10^8$ year. 

The difficulty of detecting cavities coupled with the highly variable
image quality (depth) of the Chandra archive suggest that the existing
inventory of cavities is incomplete. Most, if not all, cooling flow
clusters probably harbor cavity systems or have done so in the recent
past. 

\subsection{Cavity Kinematics and Ages}

During its initial stages, the tip of a radio jet advances
supersonically into the surrounding medium
\citep[e.g.,][]{scheuer74,gn73,br74,hrb98,eh02}.  As the ram pressure
of the jet declines with respect to the ambient pressure, it
decelerates, eventually shuts down, and the radio lobes quickly reach
pressure balance with the surrounding hot ICM \citep{b04}.  During the
initial, supersonic stage, the cavity created by the tip of a jet can
be long and narrow. Only ICM near the advancing jet tip is subjected
to the strongest shocks, and relatively little of the X-ray emitting
gas within the region encompassed by the jet is shocked. A narrow
cavity produces a small deficit in the X-ray emission, making it
difficult to detect. Enhanced X-ray emission from the relatively small
volume of shocked gas may not stand out when superimposed on the
general cluster emission. In order to form the roughly spherical
cavities that are observed, the expansion of a cavity has to ``catch
up'' with the jet tip, so that the cavity displaces a much greater
proportion of the X-ray emitting gas within the extent of the jets.
This may come about when the pressure within a radio lobe becomes
comparable to the ram pressure of the jet, so that the lobe expands
into the surrounding gas at a speed similar to the tip of the jet, or
because the jet wanders enough to carve out a large cavity
\citep[e.g.,][]{hby06}.  At formation, such a cavity can have an age
that is appreciably shorter than its sound crossing time and a lot
less than its buoyant rise time.

A cavity is buoyed outward with a force, $F_{\rm b} = Vg(\rho_{\rm a}
- \rho_{\rm b})$, where $V$ is its volume, $g$ is the local
gravitational acceleration, $\rho_{\rm a}$ is the ambient gas density,
and $\rho_{\rm b}$ is the density of the cavity. As the influence of
the jet wanes, the cavity will start to rise outward under control of
the bouyant force. Its terminal speed, determined by balancing the
buoyant force against the drag force, is then $v_{\rm t} \simeq
\sqrt{2gV/(SC)} \simeq (4v_{\rm K} /3) \sqrt{2r/R}$, where $S$ is the
bubble's cross section, and $C$ is the drag coefficient \citep{cbk01}.
The second form is for a spherical cavity of radius, $r$, at a
distance $R$ from the cluster center, with $C = 0.75$ \citep{cbk01}.
The Kepler speed, $v_{\rm K} = \sqrt{g R}$, is comparable to the local
sound speed, so that the terminal speed is almost invariably
subsonic. In practice, the volume, $V$, is determined from X-ray
measurements of the projected size of a cavity. Because the cavities
generally lie within the body of the central galaxy, $g$ can be
estimated using the local stellar velocity dispersion, under the
approximation that the galaxy is an isothermal sphere, as $g \simeq 2
\sigma^2 / R$.  Alternatively, the gravitating mass
distribution can be determined on the assumption that the surrounding
gas is hydrostatic, as outlined in Section 1.2, then used to calculate
the gravitational acceleration. 

Three estimates are commonly used for cavity ages: the buoyant rise
time, the refill time, and the sound crossing time.
From above, the time taken for a bubble to rise at its buoyant
terminal speed from the center of the cluster to its present location,
i.e., its buoyant rise time, is approximately $t_{\rm buoy} \simeq
R/v_{\rm t} \simeq R \sqrt{SC/(2g V)}$.  This is a
reasonable estimate for the age of a cavity at late times, long after
it has detached from the jet that formed it.  The ``refill time'' is
the  time required for gas to refill the displaced volume of the
cavity as the bubble rises, i.e., the time taken
for a cavity to rise buoyantly through its own diameter. If a cavity
is formed rapidly and the jet then shuts down, this is an upper limit
to the time taken by the cavity to move away from where it formed,
hence to its age. In the notation above, $t_{\rm r} = 2 \sqrt{r/g}$.
As discussed,  the early expansion of a cavity is likely to be
supersonic, whereas motion of cavities under the control of buoyancy
is almost invariably subsonic.  Employing the simple compromise
assumption, that a bubble is launched from the nucleus and travels at 
approximately the sound speed, it follows that the time it takes to
rise to its projected position is then the sound crossing time,
$t_{\rm c} = R/c_{\rm s}$.  The sound speed is given by $c_{\rm s} =
\sqrt{kT/(\mu \mh)} \simeq 1100 T_5^{1/2}\ \kms$, where the ratio of
specific heats, $\gamma \simeq 5/3$, and $\mu \simeq 0.62$.
Generally, $t_{\rm c} < t_{\rm buoy} < t_{\rm r}$ for observed
cavities. 

Which timescale best approximates the true age depends on several
unknown factors.  Viscous stresses \citep{rmf05} and magnetic stresses
\citep{deyoung03} can reduce the terminal speed, so that the estimate
given here may be high, causing ages to be underestimated and cavity
powers to be overestimated. However, equating the age to the buoyant
rise time presumes that cavity dynamics are controlled entirely by
buoyancy. There are at least a few instances where this is clearly not
so. 

Other unknown factors include the trajectory with respect to the plane
of the sky and the effective location at launch.  Projection causes
rise times to be underestimated, but generally by less than a factor
of two.  If cavity dynamics are controlled by the jet, all of the age
estimates above are too long and powers would likewise be
underestimated.  Although adiabatic cavities expand as they move
outward, the ratio $r/R$ is invariably a decreasing function of $R$,
so that the relatively large observed values of $r/R$ (Figure
\ref{fig:ratdist}) add weight to the argument that most observed
cavities have been caught early, before, or soon after becoming
buoyancy dominated.  Thus, outburst ages may generally be 
overestimated rather than underestimated.

For example, consider the outburst in Hydra A, for which shock models give a
robust age estimate of $\simeq 1.4 \times 10^8$ year \citep{nmw05}.
The buoyant rise time for its large northern cavity \citep[cavity E
  of][]{wmn07} is $5.2 \times 10^8$ year, some three to four times
longer. This discrepancy would be expected if the dynamics of the
cavities in Hydra A are still under the control of the jet. The same
applies to most systems with shocks closely surrounding the cavities,
and it is likely that shocks will be detected in deeper exposures of
many more cavity systems. Because the shocks are relatively weak in
most systems, their cavities cannot be advancing much faster than the
speed of sound.  Thus we get a crude estimate of cavity age from the
sound crossing time.  We note that this still overestimates the age of
the northern cavity of Hydra A by a factor $\simeq 2$.

\subsection{Stability of Cavities}

A significant issue is cavity disruption, which, in effect, can make
cavities nonadiabatic.  Buoyant cavities are prone to Rayleigh-Taylor
and shear instabilities \citep[e.g.,][]{bk01,sbs02}, which can disrupt
them, mixing all or part of their contents with the surrounding
gas. The fate of a cavity's energy then depends on its
constituents. Highly energetic particles can diffuse over large
distances without depositing much of their energy as heat
\citep[e.g.,][]{bm88}.  Lower energy particles (such as nonrelativistic
protons) will deposit energy locally.  The fate of magnetic energy is
unclear. Leakage of particles and fields may account for the radio
mini-halo and extended nonthermal X-ray emission in the Perseus
cluster \citep{sfd05}.

Abell 2052 is an example of a cavity system in the throes of
disruption. The rim surrounding its northern cavity appears to be
breaking apart where radio plasma is leaking out \citep{bsm01}.  M87
is the nearest cavity system in a cluster and one of the best
studied. Its "bud" cavity is strongly suggestive of instability
\citep{fnh05}.  However, the numerous small cavities in M87
\citep{fcj07} suggest that cavities can fragment without being
destroyed. Furthermore, the highly irregular shapes of the cavities in
the simulations that demonstrate strong instability are at odds with
the apparently regular shapes of observed cavities, also suggesting
that real cavities are more stable than the simulated ones.

Magnetic draping \citep{l06} may help to explain this. A buoyant
cavity can entrain magnetic field from the surrounding gas, stretching
it and naturally creating a dynamically significant magnetic field
around itself. Magnetohydrodynamic simulations suggest that this
effect, together with internal cavity fields, can help to stabilize 
cavities \citep{deyoung03,rdr04,jd05}. Growth rates of instabilities
are also affected by the viscosity \citep{rmf05,kpp05}, which is
poorly known (Section 1.5). As noted by \citet{ps06}, while a cavity
is being inflated by a jet, its surface is often decelerating. Early
in the lifetime of the cavity, the deceleration can exceed the local
acceleration due to gravity, preventing Rayleigh-Taylor instability.
As discussed in Section 5.1, cavities that survive rising through
several pressure scale heights, whether whole or as fragments, will
liberate a significant part of their enthalpy as thermal energy in
their wakes.

\subsection{Radio Lobe Composition Inferred from X-Ray Observations
of Cavities}

Extragalactic radio sources are essentially bipolar outflows of
magnetic field and relativistic particles ejected from an AGN
\citep{b56,br74,bbr84,deyoung01,hk06}.  Their structure includes a
core associated with the AGN, oppositely collimated jets emanating
from the core, and lobes that bloom from the jet terminals. Jets are
narrow, collimated conduits that transmit mass, momentum, energy, and
electromagnetic field from the nucleus to the lobes, which in turn
transmit much of the energy to the surrounding medium. Radio sources
emit synchrotron radiation throughout the radio (and sometimes optical
and X-ray) spectrum, from relativistic electrons gyrating along
magnetic field lines. Synchrotron radiation reveals only the existence
of relativistic electrons and magnetic fields, not their momentum flux
and power \citep{bbr84,hk06}.  Charge neutralizing particles, such as
protons for example, could carry most of the momentum without
betraying their existence through the emission of radiation. We will
focus on the lower synchrotron power FR I radio sources found in
clusters, rather than the higher synchrotron power FR IIs associated
with powerful radio galaxies and quasars. Cavities and the associated
shock fronts function essentially as gauges of the total
energy output of jets, and they allow the contents of radio jets and
lobes to be studied with greater certainty than was previously
possible. These studies are revealing that even the jets of faint
synchrotron sources can carry powers comparable to the luminosities of
powerful quasars. 

The energy content of radio lobes is at least the sum of energy in
particles, $E_{\rm p}$, and magnetic field, $E_B$;
\begin{equation}
E_{\rm tot} = E_B + E_{\rm p}
= \Phi V {B^2 \over 8\pi} + (1 + k) A L_{\rm syn} B^{-3/2},
\end{equation}
where $V$ is the lobe volume, and $\Phi$ is the volume filling factor
of the magnetic field, $B$.  The energy in electrons is related to the
radio power per unit frequency, $L_{\rm syn}$, as $E_{\rm e} = A
L_{\rm syn} B^{-3/2}$.  The unknown factor, $k$, accounts for unseen
particle species \citep{df04,gf04,deyoung06}, so that the total
particle energy can be expressed as $E_{\rm p} = (1 + k) E_{\rm e}$.
Corresponding approximately to the minimum energy condition
\citep{gf04,deyoung06}, it is customary to assume energy equipartition
between magnetic field and particles, yielding
\begin{equation}
B_{\rm eq} = \left[4 \pi (1 + k) A L_{\rm syn} \over \Phi
V\right]^{2/7}. 
\end{equation}

For electron Lorentz factors of a few thousand, synchrotron radiation
at radio frequencies requires $\sim \mu$G magnetic field strengths. If
the magnetic field strength greatly exceeds the equipartition value
the magnetic field would control the jet dynamics. If the field
strength is substantially below the equipartition value, the particle 
pressure controls the dynamics. No models of magnetic field-dominated
jets have been found to be stable over decades of scale, from the
Schwarschild radius to the lobes, and thus it seems unlikely that the
energy density in magnetic fields greatly exceeds that of particles
\citep{deyoung06}. 

The internal energy of the cavity, $E_{\rm tot}$, as well as its
filling factor, $\Phi$, and magnetic field strength, $B$, can be
constrained by X-ray observations. As discussed in Section 5.1, the
internal energy ranges from $pV$ for a magnetically dominated cavity 
to $3pV$ for a lobe dominated by relativistic particles (this is the
enthalpy minus the work of inflation, $pV$). The gas pressure
surrounding a cavity provides a measure of the energy density in field
and particles required to support it against collapse.  Several
studies have shown that at equipartition, lobe pressures are
approximately an order of magnitude smaller than the surrounding gas
pressure \citep{bsm01,fcb02,deyoung06}.  One way around this is to
suppose that the filling factor of the radio plasma, $\Phi$, is less
than unity. In that case, if the radio plasma is in equipartition at
the external gas pressure, $p$, its filling factor would be $\Phi =
(p_{\rm eq} / p)^{7/4}$, where $p_{\rm eq}$ is the equipartition
pressure determined assuming a filling factor of unity.  Thus, the
filling factor of the radio plasma might be only percents and total
lobe energies reduced by similar factors. 

However, this possibility
has been ruled out by X-ray observations of cavities and shocks. X-ray
count deficits over lobes are typically consistent with the X-ray
emitting gas being completely excluded from radio lobes.  While it is
difficult to place stringent limits on the amout of X-ray emitting gas
within lobes, cavities would not be evident in X-ray images unless the
lobes displace most of the X-ray emitting gas. Furthermore, the
expanding radio lobes are the pistons that drive shocks into the
surrounding gas. To obtain significant shocks, the work done by an
expanding lobe, $\int p\, dV$, must be comparable to $p_{\rm s} V_{\rm
  s}$, where $p_{\rm s}$ is the preshock pressure and $V_{\rm s}$ is
the volume encompassed by the shock. For known shocks, this generally
requires the lobes to displace most of the gas within the volume they
occupy, i.e., it demands filling factors, $\Phi$, close to unity. The
pressure support could then be supplied by some combination of thermal
or relativistic particles. 

\citet{deyoung06} modeled radio jets as pipe-like conduits of energy
collimated by the surrounding gas pressure. The energy flux was
inferred from the $pV$ work and buoyancy ages, based on X-ray
observations alone, while the lengths and cross sections of the jets
were taken from high-resolution radio observations. \citet{deyoung06}
found that the energy in $pV$ work alone is so large that the jets
would decollimate unless most of the energy and momentum are carried
by cold, heavy particles (e.g., protons) that do not contribute
significantly to the internal isotropic pressure of the jet (but see Dunn, 
Fabian \& Celotti (2006) for a discussion of electron-positron jets). 
De Young's results are consistent with the high ratio of relativistic
particle energy to electron energy ($k$) found by Dunn, Fabian, \& Taylor (2005).
Despite
the many unknowns concerning, for example, jet stability, confinement,
and acceleration mechanisms, De Young's analysis suggests that
electrons are unable to supply the observed jet power alone and must
be aided by heavy particles, or perhaps Poynting flux. 

The fluid supporting the cavities could be dominated by a hot thermal
plasma, or cosmic ray pressure \citep{mb07}. Constraints on such a
fluid can be placed by asking what combination of gas temperature and
density would be required to provide pressure support without
violating the X-ray surface brightness constraints.  This technically
challenging measurement has yielded constraints for a few systems of
$kT \gtrsim 15$ -- 20 keV \citep{ndm02,bsm03}.

Sunyaev-Zeldovich (SZ) measurements in the submillimeter band in
principle provide a novel means to discriminate between a thermal and
nonthermal cavity fluid \citep{pes05}.  The SZ decrement is sensitive
only to thermal gas.  Thus the presence or absence of a decrement
toward the cavities themselves would provide a test.  The effect is
subtle and the observations are difficult to perform using existing
instrumentation, but should be possible in the future with the Atacama 
Large Millimeter Array \citep[ALMA;][]{pes05}. These measurements
coupled to deeper X-ray observations have the potential to place
valuable new constraints on the composition of radio jets. In the
longer term, SZ effect can also be used to detect shock fronts
\citep{cl06}.  Likewise, GLAST may place interesting constraints on
the cosmic ray content of cavities through observations of the pion
decay continuum produced in cavity walls \citep{mb07}. 

\subsection{Radiative Efficiency of Radio Sources}

\begin{figure}
\centerline{\includegraphics[height=10truecm]{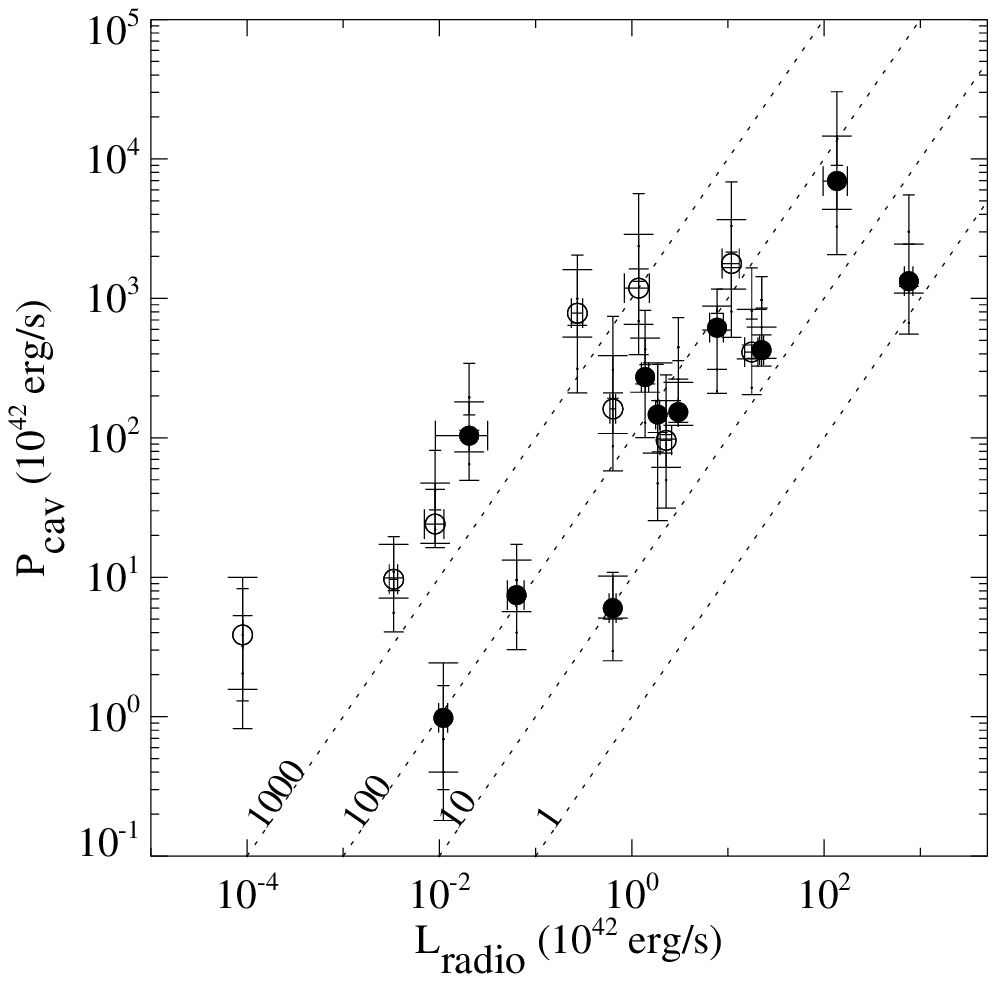}}
\caption{Total radio luminosity (10 MHz -- 10 GHz) plotted against jet  
power ($4pV/t_{\rm bouy}$) taken from Laura B\^{\i}zan's PhD thesis
  (2007). The diagonal lines represent ratios of constant jet power to
  radio synchrotron power.  Jet power correlates with synchrotron
  power but with a large scatter in their ratio.  Radio sources in
  cooling flows are dominated by mechanical power. The radio
  measurements were made with the Very Large Array telescope.}
\label{fig:pcavlrad}
\end{figure}

The plot of jet power determined from X-ray cavity data versus total
synchrotron power (core plus lobes) in Figure \ref{fig:pcavlrad} shows
a trend, as one might expect. The median ratio of jet (cavity) power to synchrotron power is $\sim
100$ \citep{brm04}.  However, the mean is much larger $\sim 2800$
owing to the large scatter. When considering radio flux from the lobes
alone, the average ratio rises dramatically to $\sim 4700$ (L. Birzan,
private communication). Figure \ref{fig:pcavlrad} clearly demonstrates that most of the jet
power is deposited into the surrounding medium and that a negligible
fraction is radiated away.  Although this result is not terribly
surprising in a qualitative sense, we now have a quantitative
measurement of this ratio that, interestingly, is on the high side of
theoretical expectations \citep[e.g.,][]{deyoung01}.  Furthermore, the
large scatter in the relationship, ranging between unity and several
thousand, shows that synchrotron luminosity is a poor measure of true
jet power. Factors contributing to the scatter include variations in 
age, field strength, and jet composition, but how these variables
combine to create the scatter is not understood. Several radio faint
cD galaxies with jet powers that equal or exceed the energy output of
powerful quasars and radio sources like Cygnus A have been identified
through X-ray observations, yet they would not have been identified 
as such using optical and radio observations alone.  X-ray observations have
revealed that under some circumstances black holes produce powerful
mechanical outflows with little accompanying radiation.

It is worth noting that adopting a $pV$-based standard for AGN
energy output blurs the canonical separation between high-power FR II radio
sources and lower-power FR I radio sources that are typically found
in the centers of clusters. Cygnus A is the best known FR
II radio source, and is the most powerful radio source in the 3C
catalog within z = 1 \citep[cf.][]{ywm02}.  Seven objects in the
\citet{rmn06} sample of cavity systems exceed Cygnus A in mechanical
power, yet none of them are considered to
be powerful radio sources. The most distant object in the sample lies
at $z \sim 0.5$, but most lie within $z \sim 0.2$.

\subsection{Simulations of Mixing by Radio Lobes}

The rising abundance gradients found in the cores of cD clusters are
presumably established over a few gigayears by stellar evolution in cD
galaxies. The gradients should in principle then be sensitive to
erasure by mixing induced by merger and AGN activity over that time
frame, thus providing interesting constraints on both the outburst and
merger history, as well as jet entrainment models. 

AGN-induced mixing has been explored recently by several groups using
2D and 3D hydrodynamical simulations
\citep{bruggen02,obb04,hka06,rbr06}. The studies generally assumed an
initial metallicity gradient added to a $\beta$-like model atmosphere
with gas particles tagged by metallicity. Moderate jet powers of $1.4
\times 10^{41}\ \ergps$ to $6 \times 10^{43}\ \ergps$, and quasar-like
jets exceeding $10^{46}\ \ergps$ lasting 100 Myr to 3 Gyr have been
explored.  \citet{rbr06} tailored their simulation to the conditions
in Perseus and included a prescription for metal injection from the
central galaxy. 

These studies found that lower power jets allowed to run for 100 Myr
or so produced only modest dredging that reduced existing metallicity
gradients by at most a few tens of percent. However, \citet{hka06}
found  that quasar-like outbursts are able to reduce metallicity
gradients to roughly 10\% of their initial values by transporting the
metals outward, primarily through convection and entrainment behind
the cavities. Short duration (powerful) jets producing big cavities,
and wide jets were most effective. Simulations generally produce
anisotropic abundance distributions aligned along the jet axis, a
prediction that can be tested using available X-ray and radio
observations. However, these studies considered jets launched at
constant aspect angles into static atmospheres. An isotropic
distribution of metals could be preserved through turbulent mixing and
gas circulation \citep[e.g.,][]{mb03,hby06} or by launching bubbles on
random trajectories, perhaps through jet precession, as has been
observed in some clusters \citep[e.g.,][]{gfs06}.

\subsection{Observations of Outflows and Mixing}

The outward mixing of metal-enriched gas has been invoked to explain
why central metallicity peaks are broader than the light profiles of
cD galaxies \citep[e.g.,][]{dnm01,dm01,rcb05}. Moreover, there are
several striking examples of plumes and shells of metal-enriched
plasma and cold clouds that were apparently dragged outward by radio
jets and lobes advancing into the ICM.  A metal-enriched shell of gas
was found nearly 90 kpc from the nucleus of NGC 1275 in the Perseus
cluster \citep{sfa04}, and a plume of cool, X-ray emitting gas extends
several tens of kpc along the radio axis of Hydra A \citep{ndm02}.  A
metal-rich shell or cap of gas was found near the edge of the southern
cavitiy, 34 kpc from the nucleus of the central galaxy in the HCG 62
group \citep{gxg07}.  However, the absence of a similar feature
associated with the northern cavity and other circumstantial evidence 
for merger activity lead the researchers to propose the gas may have
been stripped from an interloping galaxy. Arcs and shells of H$\alpha$
emission surround the cavities and radio sources in several clusters
\citep{bsm01,fsc03,csf05,hcj06}, and cold molecular gas surrounds the
inner cavities in the Perseus cluster \citep{sce06}. The cooling time
of the hot gas near the cavities is too long for the gas to have
cooled locally, so it was probably dragged there from below.

\section{STABILIZING COOLING FLOWS BY FEEDBACK}

\subsection{Heating and Cooling Rates in Clusters}

The failure to find large quantities of cooling gas with the expected
properties of a cooling flow \citep{pkp03} implies that more than 90\%
of the energy radiated away is being replenished.  Only a few percent
of the gas associated with the cooling flow forms stars and even less
accretes onto the central supermassive black hole. For an AGN to be a
viable agent, it must be powerful, persistent, an efficient heater,
and it must distribute the heat throughout the cooling region. If jets
are underpowered, the remaining issues are moot.

As discussed in Section 2, in terms of the classical cooling flow
model, the power radiated from the core of a cluster can be expressed
as $L_{\rm X} \simeq 1.3 \times 10^{44} T_5 \dot M_2\ \ergps$, where
the classical cooling  is given by $\dot M_2 = \dot M /
(100\ \msunpy)$.  This radiated power is equivalent to $\sim 10\%$ of
the gravitational binding energy released by only $\simeq
0.02\ \msunpy$ of accretion onto a black hole. The fact that AGN
outbursts are so frequently associated with cool and cooling X-ray
atmospheres suggests that the rates of cooling and AGN heating may be
thermostatically controlled. Three lines of evidence draw us to this
conclusion. First, central cooling times are as short as $\sim 3
\times 10^8$ year, $\sim 1/30$ of the Hubble time, in many clusters
and they are even shorter in elliptical galaxies \citep{vf04}.  The
$\sim 100\ \msunpy$ of cooling gas expected in the classical cooling
flow model is rarely observed, implying that the energy radiated is
being replenished on a shorter timescale than the small cooling times
found within $\sim 10$ kpc of the nucleus \citep{nmw05}.

\begin{figure}
\centerline{\includegraphics[height=9truecm]{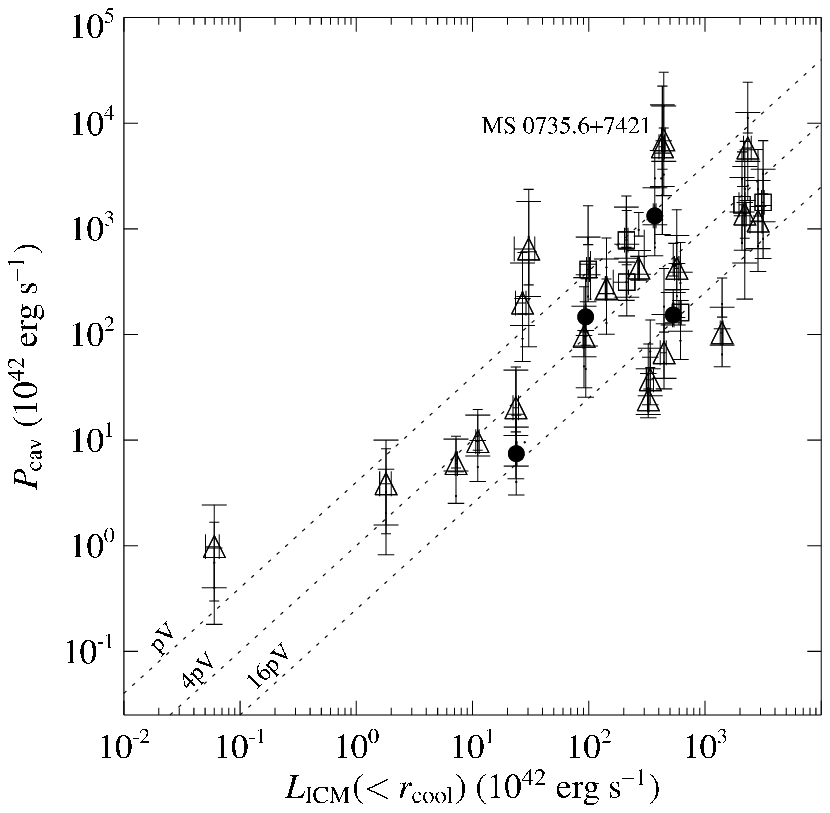}}
\caption{Cavity power of the central AGN plotted against the X-ray
  luminosity of the intracluster medium (ICM) within the cooling
  radius, after correcting for mass deposition \citep{rmn06}.  The
  symbols and wide error bars denote values of cavity power calculated
  using the buoyancy timescale. Short and medium width error bars
  denote the limits of the cavity power calculated using the sound
  speed and refill timescales, respectively. Diagonal lines denote
  equality between heating and cooling rates assuming $pV$, $4pV$, and
  $16pV$ of energy per cavity, respectively.} \label{fig:pcavlx}
\end{figure}

If the heating rate were not tied to the radiative cooling rate, then
it must exceed it to ensure that very little gas cools to low
temperatures in the majority of cooling flows.  This would only be
possible if the heating rate exceeded typical cooling rates, giving
rise to net heating of the ICM. Net heating would then drive up the
central entropy, and in the long-term, the central cooling times would
approach the ages $\sim H_0^{-1}$ of most clusters, in contradiction
of observations. Thus, it is difficult to maintain a significant
proportion of clusters, groups, and galaxies with short central
cooling times, unless heating rates are tied to cooling rates by some
feedback loop. Second, the trend between cavity-based power estimates
and the X-ray luminosity of cooling gas
shown in Figure \ref{fig:pcavlx}, strongly implicates AGN: they
apparently know about the cooling gas and vice versa. Third, the
entropy profiles of the gas in cooling flows fall inward in roughly
power-law fashion \citep{pjk05}, but often flatten in the cores
\citep{dhc06} implying entropy (energy) input at a level of $\sim
10\ \rm keV\ cm^2$ \citep{mbb04,vd05}, which is consistent with the
observed level of AGN energy input. The flattening is seen in both
radio active and radio quiet clusters. It suggests the operation of an
intermittent heating mechanism that maintains a roughly steady power
when averaged over times comparable to the central cooling time. The
coincident rates (70\%) of cavities \citep{dft05} and radio emission
\citep{burns90} in cooling flow clusters further implicates AGN,
although heat conduction or other agents may augment AGN heating
\citep{rb02,bm03}.

A quantitative comparison between X-ray (cooling) luminosity and jet
power (heating) averaged over the lifetime of a cavity $pV /t_{\rm
  bub}$ is shown in the heating versus cooling diagram (Figure
\ref{fig:pcavlx}) taken from \citet{rmn06}. In this diagram, the
cooling  luminosity is the radiated power that must be replenished by
heating. This quantity is found by subtracting from the total X-ray
luminosity within the cooling radius the luminosity from gas that
could be condensing out without violating observations.  This
correction amounts to less than 10\% of the total X-ray luminosity. The
diagonal lines in Figure \ref{fig:pcavlx} represent equality between
heating and cooling assuming $pV$, $4pV$, and $16pV$ of heat input per
cavity. Cavities filled with a nonrelativistic monoatomic gas ("hot
bubbles") would supply approximately $2.5pV$ per cavity, whereas
cavities filled with relativistic gas would supply roughly $4pV$ per
cavity (Section 5.1). The true effective energy per cavity could be
substantially greater than $4pV$ if much of the outburst energy has
been dissipated by shocks (Section 5.4), or if other cavity systems 
exist below the threshold of detectability. Thus, the data points
should be treated as lower limits.

More than half of systems with detectable cavities liberate enough
energy to balance or exceed radiation losses at the present
time. Other systems including those without detectable cavities do
not. Either this implies that AGN require help from other energy
sources in order to suppress cooling, or it reflects the elusive
nature of X-ray cavities and the transience and variable power output
of AGN. If AGN outbursts are transient, then the time averaged AGN
heating power needs to match the power radiated. Additional energy
supplied by AGN associated with the broader cluster galaxy population
\citep{nsb06} will heat the gas beyond the cooling region but is
apparently inconsequential within the cooling region itself
\citep[e.g.,][]{bvk07}. 

The trend in Figure \ref{fig:pcavlx} does not take account of the 30\%
of cooling flow clusters that lack identifiable cavities \citep{df06},
which would in principle populate the lower right-hand side of the
diagram. It is not known whether such objects have had AGN activity in
the recent past, but the associated cavities and shock fronts are not
visible in existing data, or whether other heating mechanisms are at
work.  Given that deeper observations have invariably revealed more
structure \citep[e.g.,][]{fcj07}, there is good reason to
suspect that the fraction of clusters for which the AGN power is
sufficient to balance radiative losses will grow beyond half as deeper
Chandra images become available.

Regardless of this, the existing data already suggest that AGN heating
can balance cooling.  For the sample of \citet{rmn06}, the mean
cooling power is $6.45 \times 10^{44}\ \ergps$, while the mean cavity
power is $1.01 \times 10^{45}\ \ergps$ (using $4pV$ per cavity and the
bouyant lifetimes). Assuming that cooling flow clusters without
evident cavities have similar cooling powers and zero cavity power, we
can correct the ratio of mean cavity power to cooling power for these
with a factor of 0.7 \citep{df06}, giving 1.1 for this ratio. Thus,
within the substantial uncertainties, it is plausible that
time averaged AGN heating powers balance radiative cooling. A more
accurate assessment of the ratio of heating to cooling must await a
study of a complete, unbiased sample of cooling flow clusters.

If we assume for the moment that all cooling flows are suppressed by
AGN, the scatter in Figure \ref{fig:pcavlx} may be a consequence of
variable AGN power output. \citet{nb05} modeled AGN power output as a
Gaussian process with a log-normal distribution at a fixed cooling
luminosity, and an observationally motivated outburst timescale of
$10^8$ years. Their model implies that in any given system there is a
good chance of finding smaller than average jet powers, because much
of the power is generated by less frequent but more powerful
outbursts. In this context, the objects falling below the $2pV$ --
$4pV$ lines may be in a lower than average outburst state and thus may
be in or moving toward a cooling cycle. In the \citet{nb05} model, the
powerful, rare outbursts are experienced by all
systems. These outbursts may be responsible for the high entropy
pedestals observed in radio quiet cD galaxies \citep{dhc06,ppk06}.

How effective AGN heating is over the lives of clusters depends on how
and how much their power output varies over time. Although AGN power
output is a strong function of halo gas mass, power outputs vary
widely at a given mass. The systems with the largest cavities
represent the extreme in power output. We do not know whether they are
unique to some clusters or whether all systems occasionally experience
them. These rare but powerful outbursts can easily dominate a cD
galaxy's AGN power output over the age of a cluster. 

To summarize, under the assumptions outlined above, AGN are
powerful enough to supress cooling in many and perhaps all cooling
flow systems. However, this conclusion depends on how well the X-ray
method traces true jet power and how efficiently cavity enthalpy and
shock energy is converted to heat. 

\subsection{Heating and Cooling in Elliptical Galaxies and Groups}

\begin{figure}
\centerline{\includegraphics[height=9truecm,angle=270]{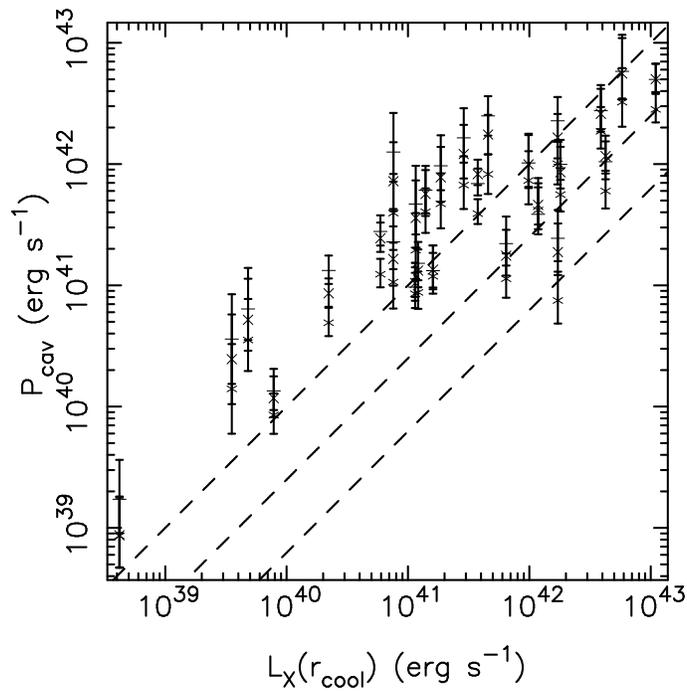}}
\caption{Cavity power versus cooling power for nearby giant elliptical
  galaxies \citep{njf07}. Cooling power is the X-ray luminosity from
  within the projected radius where the cooling time is $7.7 \times
  10^9$ years. Cavity powers are determined using an energy of pV per
  cavity and a range of cavity age estimates (see Figure 8). The
  dashed lines show equality for cavity energies of $pV$, $4pV$, and
  $16pV$, top to bottom.  All but one system lie above the $4pV$ line,
  indicating that radiative losses can bebalanced by AGN power.} 
\label{fig:egals}
\end{figure}

The cooling time of the hot gas in the centers of some gE galaxies is
less than $\sim 10^8$ year, which is shorter than found in the cores
of clusters. As in clusters, the hot gas there is expected to cool and
form stars in the host gEs, yet it fails to do so at the expected
rates of $\sim 1\ \msunpy$ \citep{mb03}. Like clusters, many gEs have
cavities or other disturbances in the hot gas implicating AGN as
significant heating agents \citep{mb03,ds07}.  \citet{jfc07} have
assembled Chandra X-ray observations of a nearly complete sample of
roughly 160 nearby gE galaxies, 109 of which show significant diffuse 
emission from hot gas. Of those, 27 have significant AGN
cavities. \citet{njf07} evaluated the relative rates of AGN heating
and radiative cooling for the 27 ellipticals with detectable cavities,
giving the results shown in Figure \ref{fig:egals}. Apart from minor
differences, this figure can be viewed as extending Figure
\ref{fig:pcavlx} to less massive halos.  Although heating matches or
exceeds cooling in roughly half of the cluster systems in Figure
\ref{fig:pcavlx}, AGN power exceeds radiative losses in all but one or
two of the nearby gEs shown in Figure \ref{fig:egals}.  Roughly one
quarter of Jones's gEs with significant emission from hot gas have
detectable cavities, a detection rate that is similar to the overall
rate of detection in clusters, but is smaller than the 70\% detection
rate in cluster cooling flows \citep{dft05}.  Based on central cooling
times, all 109 of the gEs with significant hot atmospheres are cooling
flows. Assuming $1pV$ per cavity, the total heating power of the 27
nearby gEs harboring cavity systems is $\sim 2 \times 10^{43}\ \ergps$.
This is to be compared to the total cooling power of $\sim
10^{44}\ \ergps$ for all 109 nearby gEs with significant emission from
hot gas. Allowing for a cavity enthalpy of $4pV$ for relativistic gas
and the significant boost to this owing to shock energy, overall
heating and cooling powers for the nearby gE sample match reasonably
well. Although there is still significant systematic uncertainty, it
seems that intermittent AGN outbursts are a plausible mechanism for
preventing X-ray emitting gas from cooling and forming stars in nearby
gE galaxies. A similar result was found indirectly for gEs culled from 
the Sloan survey by \citet{bkh06}, who determine jet power by assuming
a one to one correspondence with radio power, calibrated using cavity
data from \citet{brm04}.

Because the binding energy per particle is lower in groups, the same
nongravitational heating energy per particle will have a greater
effect in groups than in clusters.  Rosat studies of poor
clusters and groups reported an ``entropy floor'' or pedestal in the
gas entropy profile \citep[e.g.,][]{pcn99}.  It was argued that the
entropy floor caused the steepening in the luminosity versus
temperature relation, which was more pronounced at group
masses. Entropy floors can be produced by a number of processes,
including AGN outflows, supernova driven galactic winds, and by the
effects of cooling \citep{voit05}. With higher resolution data from
Chandra the situation has become murkier. Entropy floors are no longer
seen, but there is an apparent excess of entropy at larger radii,
beyond the inner cooling region \citep{psf03}. \citet{mushotzky04} has
argued that the steeper than self similar power law scaling of
luminosity versus temperature for rich clusters extends to groups, but
with a larger scatter at the group scale. 

Unfortunately, AGN heating is not as well studied in groups as in
clusters. We know that the gas in many groups is being disturbed by
radio sources emanating from their central galaxies \citep{chb05}, for
example, as seen in HCG 62 \citep{miy06,nmf07}. The cavity power of
the lobes is sufficient to quench the cooling flow 
but the overall significance of AGN heating in
groups is unclear \citep{dn06}.

The effects of AGN heating on intragroup gas has been explored by
separating X-ray bright systems with central galaxies into radio-loud
and radio-quiet bins. Analyzing an X-ray-selected sample of groups,
\citet{chb05} found that in 63\% of groups with a dominant central
elliptical galaxy the dominant elliptical harbors an active radio
source.  Many of these radio sources are interacting with the hot gas
filling the groups. \citet{chb05} found that the radio loud groups are
slightly hotter on average at a given X-ray luminosity than radio
quiet groups, which they attribute to AGN heating. They point out that 
AGN heating at the level they find extending over $5 \times 10^9$ year
would be able to supply the $\sim 1$ keV per particle of excess energy
required to preheat groups \citep[e.g.][]{vd05}.

However, in a Chandra study of 15 groups, \citet{jph07} found only
modest  steepening of the gas temperature in radio-loud groups
compared to radio-quiet ones, and steeper entropy gradients in groups
with brighter central galaxies and presumably more massive nuclear
black holes. It is unclear whether these trends implicate AGN heating
or heating by the galaxies themselves. Jetha et al.\ found power law
entropy profiles extending to small radii, and only small differences
in the gas profiles of radio-loud and radio-quiet groups. They found
that AGN heating may have a more significant effect in quenching
cooling than for preheating. 

\subsection{Uncertainties in Estimating Jet Power}

Model assumptions about the time and energy dependence of outbursts
are a significant source of uncertainty in the calculation of mean
cavity (jet) power. If AGN inject energy in a series of short,
isolated outbursts, the interval between outbursts characterizes
$t_{\rm bub}$. This interval is impossible to estimate for individual
systems in the absence of multiple generations of cavities, such as
those in Perseus \citep{fse00} and Abell 2597.  Multiple
cavity systems are also required to estimate the mean energy per
outburst. The ghost cavities in Perseus and Abell 2597 imply an
outburst every 60 - 100 Myr or so, and in both clusters the earlier
outbursts were stronger than the current ones, indicating variability
in outburst strength. The ripples in Perseus indicate
that some outbursts may occur on shorter timescales than the cavity
ages, implying the outburst period is also variable.

If outbursts are more nearly continuous, i.e., the jets remain
active for a substantial fraction of the outburst cycle, then the
current jet power is a more useful measure of the average value. In
general, the off-time must be included to measure a population
average. The large systems (Hydra A, MS0735.6+7421, Hercules A) 
appear to be operating in this mode \citep[e.g.,][]{wmn07}. Most
observational treatments implicitly assume a more nearly
continuous distribution of outbursts.

Sources of uncertainty in the measurements of cavity energy include
the volume estimates from the projected cavity sizes and shapes and
the unknown composition of the cavity plasma, which combined give an
uncertainty of at least a factor of several.  In addition, the energy
per outburst is probably underestimated owing to adiabatic losses,
cavity disruption, undetected cavities, and the omission of shock
energy \citep[e.g.,][]{nsb06,bbo07}. These effects are offset to some
degree by the unknown fraction of outburst energy that is converted to
heat.

\section{HEATING MECHANISMS}

\subsection{Cavity Heating}

The energy required to create X-ray cavities around radio lobes is the
sum of the $pV$ work required to displace the X-ray emitting gas and
the thermal energy of the contents of the lobe, i.e., the enthalpy,
\begin{displaymath}
H = E + pV = {\Gamma \over \Gamma - 1} p V,
\end{displaymath}
where $p$ is the pressure in the lobe and $V$ is its volume. The
second form applies if the lobe is filled with an ideal gas with
constant ratio of specific heats, $\Gamma$. If the lobe is dominated
by relativistic particles, $\Gamma = 4/3$ and $H = 4pV$, whereas if it
is dominated by nonrelativistic gas, $\Gamma = 5/3$ and $H = 2.5pV$.
Lobes may also be dominated by magnetic field, in which case $H =
2pV$. Other possibilities lie between these extremes, so that,
although the equation of state ($\Gamma$) for lobes is not known, lobe
enthalpy is likely to fall in the range $2pV$ -ÃÂ­ $4pV$.

\begin{figure}
\centerline{\includegraphics[height=8truecm]{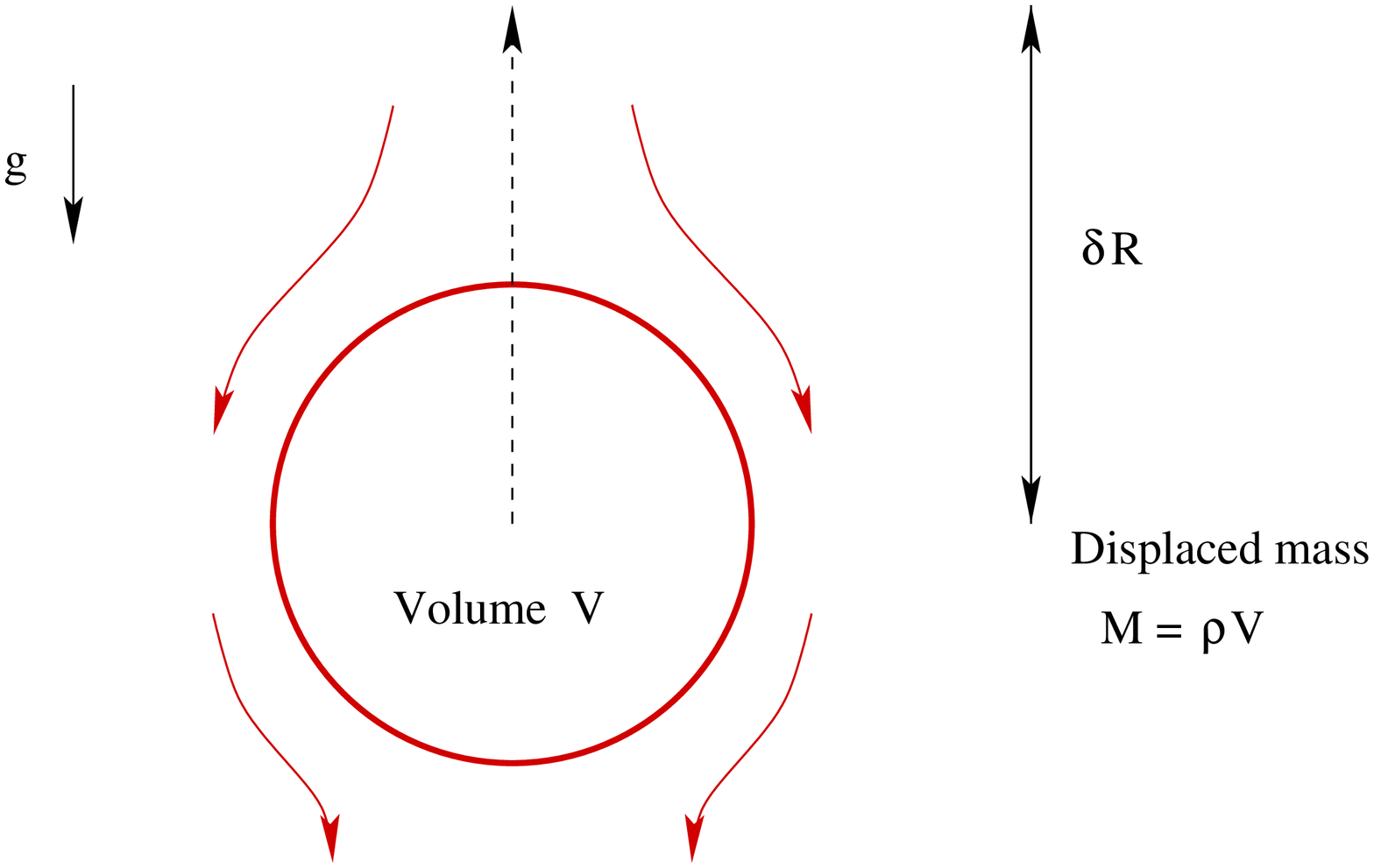}}
\caption{Buoyant cavity. As the cavity rises, gas falls inward to fill
  the space, turning gravitational potential energy into kinetic
  energy in the cavity's wake.} \label{fig:bubble}
\end{figure}

Simulations show that buoyant cavities can heat the surrounding gas as
they rise through a cluster atmosphere \citep{bk02,rhb02}.
\citet{csf02} argued that essentially all of the enthalpy of a rising 
cavity can be thermalized in its wake and a simple mechanism for this
is outlined in \citep{brm04}.  As a buoyant cavity rises, some X-ray
emitting gas must move inward to fill the space it vacates (Figure
\ref{fig:bubble}), so that gravitational potential energy is turned
into kinetic energy in the ICM. The potential energy released as a
cavity rises a distance $\delta R$ is
\begin{equation}
\delta U = M g \,\delta R = V \rho g \, \delta R = -V {dp \over dR} \,
\delta R = - V \, \delta p,
\label{eq:cavdissip}
\end{equation}
where $M = \rho V$ is the mass of gas displaced by the cavity, $V$ is
its volume, $\rho$ is the density of the surrounding gas and $g$ is
the acceleration due to gravity. The third equality relies on the
surrounding gas being close to hydrostatic equilibrium, so that $\rho
g = -dp/dR$.  The last equality expresses the result in terms of the
change in pressure of the surrounding gas over the distance $\delta
R$, $(dp/dR)\, \delta R = \delta p$. Because the rising cavity moves
subsonically \citep{cbk01}, its pressure remains close to that of its
surroundings and we may regard this as the change in its pressure. The 
first law of thermodynamics, $d E = Td S - pd V$, expressed in terms
of the enthalpy is $dH = TdS + Vdp$. Entropy remains constant for an
adiabatic cavity (radiative losses from the cavity are negligible), so
that this gives $\delta H = V \, \delta p$. Thus, Equation
\ref{eq:cavdissip} shows that the kinetic energy created in the wake of
the rising cavity is equal to the enthalpy lost by the cavity as it
rises. 

Regardless of the viscosity, we should expect this kinetic
energy to be dissipated, creating heat locally in the wake of the
cavity. If the viscosity is high, i.e., the Reynolds number is low,
the motion is damped viscously in a laminar wake that is comparable in
size to the cavity. If the Reynolds number is high, the wake is
turbulent and kinetic energy is damped on the turn-over timescale of
the largest eddies (also comparable to the size of the cavity). The
turbulent cascade maintains the dissipation rate by propagating energy
to sufficiently small scales for viscous dissipation to keep pace
\citep[in fact, the turbulent spectrum is fully characterized by the 
dissipation rate;][]{ll87}. In both cases, kinetic energy is damped
before diffusing far from the axis on which the cavity rises. This
leads to the important result that the enthalpy lost by a buoyantly
rising cavity is thermalized locally in its wake, almost regardless of
the physical properties of the cavity and the surrounding gas.

Heating by cavity enthalpy is the physical basis of the
``effervescent'' heating model \citep{b01,rb02,rrn04}. Chandra
observations have established not only that cavities are common, but
also that many clusters contain multiple cavities, though not all
coincident with active radio lobes \citep{fse00}. The numerous
cavities seen in deep observations of M87 \citep{fcj07} also fit
this model. The enthalpy of an adiabatic cavity depends on the
surrounding pressure as $H_{\rm b} = H_{\rm b, 0} (p/ p_0)^{(\Gamma
  -1)/\Gamma}$, where $H_{\rm b, 0}$ is the initial enthalpy of the
cavity and $p_0$ is its initial pressure. If the mean power injected
by an AGN as cavity enthalpy is $L_{\rm b}$, then the mean heating
rate per unit volume averaged over the sphere at radius $R$ due to 
liberated cavity enthalpy is
\begin{equation}
\Pi_{\rm b} = - {L_{\rm b} \over 4 \pi R^2} {d \over dR} \left(p \over
p_0\right)^{(\Gamma - 1)/\Gamma},
\label{eq:cavheat}
\end{equation}
where $p_0$ is the pressure at the radius, $R_0$, where the cavity is
formed. The value of $R_0$ depends on outburst details, but it is
always nonzero. Very little cavity enthalpy is thermalized within
$R_0$. With minor modifications, this is the effervescent heating rate
used by \citet{b01} and successors. Using this model, \citet{rb02}
argued that cavity enthalpy needs to be supplemented by thermal
conduction from the surrounding cluster in order to prevent
cooling. However, \citet{rrn04} find that this conclusion depends on
the assumed total rate of AGN heating.  Note that these are all 1D
models, employing mixing length theory to model convection. Such
models probably omit significant features of the full, 3D flow.

As well as providing better models for the convection associated with
anisotropic cavity heating, 3D simulations allow more realistic
treatments of cavity stability and mixing \citep{qbb01}.
\citet{dbt04} made 3D hydrodynamic simulations for the longer-term
effects of cavity heating by injecting energy to simulate the
formation of radio lobes at random positions with a Gaussian
distribution around cluster centers. Although they found that the
resulting heat input can prevent catastrophic cooling, their model
clusters do not produce cool cores as observed. They argue that
additional preheating, taking place earlier in the collapse hierarchy,
could resolve this disagreement. AGN heating rates in this model are
fixed, not determined by feedback, and \citet{dbt04} conclude that
prevention of catastrophic cooling is not sensitive to AGN power. A
significant channel of heating for their model is the mixing of gas
from the simulated cavities with the general ICM.  Relativistic
particles can be slow to transfer energy to gas \citep[e.g.,][]{bm88}
and the fate of magnetic fields is poorly understood, making it
unclear how effectively radio plasma mixes with and heats the ICM in
practice. 

\subsection{Heating by Weak Shocks}

Although a substantial energy, comparable to the cavity enthalpy, is
required to drive the weak shocks seen in association with AGN
outbursts in some clusters \citep{fnh05,mnw05}, much of the shock
energy ends up as additional potential energy in the gas. This helps
to delay cooling by reducing gas density and increasing the total
energy that must be dissipated. However, the key requirement on the
process that prevents the gas from cooling is to make up for entropy
lost by radiation from the gas. Shock heating probably plays a
significant role in this, especially close to the AGN \citep{frt05}. A
fundamental distinction between sound waves and shocks is the entropy
created by dissipation at shock fronts.  For weak shocks, the entropy
jump per unit mass, $\Delta S$, is proportional to the cube of the 
shock strength \citep[e.g.,][; measured here by the fractional
  pressure increase, $\delta p/ p$, where $p$ is the preshock pressure
  and $\delta p$ is the pressure increase across the
  shock]{ll87,dnm01}. The equivalent heat input per unit mass is $T\,
\Delta S$, where $T$ is the gas temperature. To lowest nonzero order,
this equivalent heat input amounts to a mean heating rate per unit
volume due to repeated weak shocks of 
\begin{equation}
\Pi_{\rm s} = {(\gamma + 1) \over 12 \gamma^2} {\omega p \over 2 \pi}
\left(\delta p \over p\right)^3,
\label{eq:shock}
\end{equation}
where $\gamma$ is the ratio of specific heats for the ICM and the
interval between outbursts is expressed as $2\pi /\omega$ (compare the
sound heating rate, Equation \ref{eq:sound}). 

The repeated weak shocks around M87 seen by Forman and colleagues
\citep{fcj07} demonstrate the possible significance of weak shock
heating. Based on its X-ray surface brightness profile, the innermost
shock, 0.8 arcmin (4 kpc) from the AGN, has a Mach number of 1.4, 
so that its equivalent heat input is only 2\% of the gas thermal
energy. However, there is another comparable shock at about twice the
radius, suggesting repeated outbursts every $\sim 2.5 \times 10^6$
year, while the cooling time of the gas is $\sim 2.5 \times 10^8$ year. 
Over the cooling time, this leaves ample time for multiple weak shocks
to make up for the energy radiated by the gas \citep{njf07}. In M87,
weak shocks are capable  of preventing gas near the AGN from cooling.

When a shock overruns a cavity, the high sound speed of the cavity
causes the shock to propagate faster through the cavity than around
it. This results in the formation of a vortex ring around the cavity
after it has been overrun \citep{cbk01,eb02}. \citet{hc05} note that
this process, an example of Richtmyer-Meshkov instability, can
increase the fraction of shock energy that is thermalized in the ICM,
especially for weak shocks, because it converts shock energy into
localized kinetic energy that can then be dissipated as heat. If the
ICM contains many small cavities, Heinz \& Churazov find that the
attenuation length due to the Richtmyer-Meshkov instability is
inversely proportional to the fraction of the ICM occupied by the
cavities. For example, if cavities in the Perseus cluster have a
filling factor of $\sim10\%$, most of the energy of the weak shocks
would be dissipated within the central 100 kpc. The significance of
this process is hard to assess, because small bubbles are not
generally accessible to observation. If, for example, bubbles occupy 
$\sim 10\%$ of the volume of the ICM, then they also contain $\sim
10\%$ of its thermal energy.  Because the bubbles rise at their
terminal speeds, $v_{\rm t} \simeq v_{\rm K} \sqrt{r/R}$
\citep{cbk01}, they must be replaced continually. The power required
to maintain a large bubble filling factor is substantial, unless the
bubbles are very small, and a large part of that power would be
dissipated as bubble heating (Section 5.1). 

\subsection{Heating by Sound Damping}

Repeated weak shocks may also be regarded as a superposition of sound
waves. \citet{fsa03} showed that viscous damping of sound waves
generated by repeated AGN outbursts may represent a significant source
of heating. The heating power per unit volume due to dissipation of a
sound wave can be expressed as \citep{ll87,frt05}
\begin{equation}
\Pi_{\rm d} = \left[{2\nu \over 3} + {(\gamma - 1)^2 \kappa T \over
    2\gamma p}\right] {\omega^2 \rho \over \gamma^2} \left(\delta p
\over p\right)^2,
\label{eq:sound}
\end{equation}
where $\rho$, $T$, $p$ and $\gamma$ are the density, temperature,
pressure, and ratio of specific heats of the gas, respectively,
$\kappa$ is the thermal conductivity, $\nu$ is the kinematic viscosity
($\nu = \mu/\rho$, where $\mu$ is the viscosity), $\omega$ is the
angular frequency, and $\delta p$ is the pressure  amplitude of the sound
wave. This expression includes both viscous and conductive
dissipation. Both terms in the leading coefficient have the form of a
mean free path times a thermal speed \citep{s62}. For an unmagnetized
plasma, the mean free paths of the electrons and protons are the
same.  For the kinematic viscosity, the thermal speed is that of the
protons, whereas in the conductive term it is that of electrons.  Thus
conductive dissipation would be greater in the absence of a magnetic
field. 

Heating by weak shocks is a separate mechanism from heating by sound
dissipation. In numerical simulations, shocks are controlled by
viscous stresses \citep{rbb04a,rbb04b,brh05}, so that these two
processes are lumped together.  Dissipation of sound depends on the
transport coefficients, but shock heating does not. Therefore,
uncertainty in the transport coefficients (Section 1.5) translates
directly into uncertainty in the heating rate due to sound
dissipation. If the transport coefficients are suppressed by no more
than an order of magnitude, about the range suggested by observations,
analytical estimates \citep{fsa03} and simulations
\citep{rbb04a,rbb04b,brh05} agree that sound dissipation plays a
significant role in converting AGN energy into heat in the ICM. The
two mechanisms have distinct dependencies on the parameters (Equations
\ref{eq:shock} and \ref{eq:sound}), leading to distinct spatial
distributions of heating. 

Only the fundamental frequency should be used in applying the sound
dissipation rate (Equation \ref{eq:sound}). The thickness of a weak
shock front can be expressed as $w \simeq \lambda /(M - 1)$, where
$\lambda$ is the effective particle mean free path and $M$ is the shock 
Mach number \citep[e.g.,][]{ll87}.  When the width of the shock front
becomes comparable to or larger than the wavelength, the disturbance
transitions from a shock wave to a sound wave. At this transition the
sound dissipation rate, which depends on transport coefficients, and
the shock dissipation rate, which does not, cross over. When the
thickness of a shock is considerably smaller than the distance between 
successive shocks, the shock front contributes substantial power in
high harmonics when decomposed into a Fourier series. The fundamental
frequency is determined by the shock repetition rate.  Because the
dissipation rate varies as $\omega^2$ (Equation \ref{eq:sound}) the 
linear treatment suggests that the high frequency components will
dissipate quickly, producing significant heating. However, the entropy
created at the shock front is determined entirely by the
Rankine-Hugoniot jump conditions, regardless of the transport
coefficients. Nonlinear effects adjust shock thickness to make the
dissipation rate match that required by the jump conditions. Although
shock thickness is determined by the transport coefficients, the
dissipation rate for the high frequency terms is not. This dissipation
is accounted for as the shock heating rate (Equation \ref{eq:shock}). 
Only the dissipation rate of the component at the fundamental
frequency depends directly on the transport coefficients and only it
should be regarded as subject to sound damping.

In the absence of magnetic fields, the kinematic viscosity, $\nu$, and
the coefficient of the conductive dissipation rate, $\kappa T/ p$ in
Equation \ref{eq:sound} both scale approximately with temperature as
$T^{5/2}$. If this scaling applies in reality, the sound dissipation
rate is sensitive to the ICM temperature, which would reduce its
significance in cool systems.  However, because the sound dissipation
rate is also sensitive to frequency, higher outburst rates in cool
systems could offset the reduction in transport
coefficients. \citet{fsa03} find that outbursts occur every $\sim
10^7$ year in the Perseus cluster, where the ICM temperature is $\sim
4$ keV, whereas the period of outbursts in M87 is $\sim 2.5 \times 10^6$
year for an ICM temperature 1$-$2 keV \citep{njf07}. This hints that
the feedback process may be able to adjust the frequency as well as
the power of AGN outbursts.

\subsection{Cavity Enthalpy versus Shock Energy}

The division of jet energy between cavity enthalpy and shocks is
affected significantly by the history of an outburst. Here, shock
energy is used loosely to mean the total work done on the ICM by an
inflating cavity, most of which ends up as thermal and gravitational
potential energy in the hot gas. In fact, because cavity expansion
lifts gas outward, it invariably reduces the gas pressure after a
shock has passed. As a result, the gas expands and its total thermal
energy is generally decreased after the outburst, especially when
shocks are weak. Thus, the ultimate repository for most of the shock 
energy is probably potential energy in the ICM.

At one extreme, consider a jet that dumps its energy explosively into
the gas in a single brief event. Initially this would create a tiny
cavity with volume, $V_{\rm i}$, and pressure, $p_{\rm i}$, much
greater than the pressure of the surrounding gas. The thermal energy
of the cavity, $E_{\rm i} = p_{\rm  i} V_{\rm i} /(\Gamma - 1)$, would
equal the energy deposited by the jet  (radiative losses are assumed
negligible throughout this process). This cavity would expand
explosively, driving a strong shock into the ICM, until its pressure
reached that of its surroundings, $p_{\rm f} \ll p_{\rm i}$. Because
the expansion is adiabatic, the final energy of the cavity is
\begin{equation}
E_{\rm f} = {1\over \Gamma - 1} p_{\rm f} V_{\rm f} = {1\over\Gamma -
  1} p_{\rm f} V_{\rm i} \left(p_{\rm i} \over p_{\rm f}
\right)^{1/\Gamma}
= E_{\rm i} \left(p_{\rm f}\over p_{\rm i}\right)^{(\Gamma -
    1)/\Gamma},
\end{equation}
which can be much smaller than its initial thermal energy. The balance
of the initial energy is the work done by the expanding cavity on the
ICM, i.e., the shock energy.  This extreme case illustrates that, in
principle, there is no upper limit on the fraction of the energy from
an AGN outburst that ends up in the ICM. At present, the known
cavities in clusters are all weakly overpressured. However, the
southwestern lobe of Cen A has a pressure that is two orders of
magnitude greater than that of the surrounding, unshocked gas
\citep{kvf03}, showing that the explosive extreme can be approached
for systems in poorer environments. 

At the opposite extreme, consider a cavity that is inflated
gently. Its pressure would remain close to that of the surrounding ICM
throughout the expansion. To the extent that the pressure also remains
constant during cavity inflation, the ratio of the work done by the
expanding cavity to its final thermal energy would then be $\Gamma -
1$. In practice, the pressure will generally decline as a cavity
inflates and expands outward into lower pressure gas. This causes
additional adiabatic energy loss, boosting the ratio of the work done
to its final thermal energy. 

The main conclusion here is that work done by an expanding cavity is
rarely less than its thermal energy and may be considerably
greater. This is broadly consistent with findings for observed
clusters, that the cavity enthalpy is comparable to the energy
required to drive the shocks \citep{mnw05,nmw05}.  This argument also
shows that the relative energies contain information about the history
of outbursts. 

\subsection{Distribution of AGN Heating within a Cluster}

Ignoring all dissipation, we can use conservation of energy to
estimate the radial dependence of the amplitude of sound waves and
weak shocks, 
\begin{equation}
{\delta p\over p} \propto \rho^{-1/2} T^{-3/4} R^{-1},
\label{eq:econs}
\end{equation}
where $\rho(R)$ and $T(R)$ are the density and temperature of the ICM,
and $R$ is the radius. With this scaling, the ratio of weak shock
(Equation \ref{eq:shock}) to sound heating rate (Equation
\ref{eq:sound}) almost invariably decreases with radius
\citep{frt05}. As might be  expected, this arises because the shock
heating rate is more sensitive to $\delta p/p$, which decreases with
the radius. The radial temperature gradient that typifies cool cores
\citep[e.g.,][]{asf01} probably decreases the ratio further, as the
transport coefficients increase with radius.

Because Equation \ref{eq:econs} assumes that energy is conserved, it
only applies after the expanding radio lobes stop driving the pressure
disturbance, i.e., after the disturbance separates from the
cavities. Using it to extrapolate to small $R$ would overestimate the
relative strength of the disturbance there. The assumption of
spherical symmetry, which is adequate at large distances from the
radio lobes, also fails at radii smaller than the lobe separation,
where the off-center energy deposition must be taken into account.

The ratio of cavity heating (Equation \ref{eq:cavheat}) to shock
heating rate (Equation \ref{eq:shock}) is proportional to $v_{\rm K}^2
\rho^{3/2-1/\Gamma} T^{5/4-1/\Gamma} \propto  \rho^{3/2-1/\Gamma}
T^{9/4-1/\Gamma}$, where the Kepler speed is $v_{\rm K} = \sqrt{g R}$
and $g$ is the acceleration due to gravity, and the ICM is assumed to 
be hydrostatic. The second form relies on the further approximation,
$v_{\rm K}^2 \propto T$. The relatively gentle temperature rise with
radius in cool cores is insufficient to offset the density decrease,
so that this is generally a decreasing function of the radius (for
reasonable values of $\Gamma$). Thus, cavity heating is more centrally
concentrated than weak shock heating. As noted above (Section 5.1),
although it is formally the most centrally concentrated of these three
heating processes, cavity heating is ineffective inside the radius
where the cavities are formed. Thus, weak shock heating is likely to 
be the most significant heating process closest to the AGN. Cavity
heating probably takes over this role immediately outside the region
where the radio lobes are formed \citep{vd05}.

If AGN outbursts deposit comparable amounts of energy in shocks and
cavity enthalpy, then the reasoning of this section suggests that the
dominant mode of AGN heating changes with radius. Closest to the AGN,
weak shocks (or, possibly, sound dissipation) are likely to be most
significant. Note that weak shock heating can plausibly stop the
innermost gas from cooling in M87 (Section 5.2). The total rate of 
shock heating may not be large \citep{frt05}, but because the gas
closest to the nuclear black hole is the most likely to be accreted,
the heating process at work on that gas plays a critical role in any
AGN feedback cycle. Cavity heating may well take over beyond the
radius where the radio lobes are formed. On larger scales, sound
damping may become the dominant AGN heating process. On even larger
scales, thermal conduction can play the dominant role in the hotter
clusters. In short, it appears likely that no single AGN heating
process is the most significant. It may also be that AGN heating does
not act alone to prevent copious gas deposition and star formation. 

\subsection{Energy Injection by Radio Jets}

In order to understand the process of AGN feedback, it is necessary to
understand how AGN outbursts are fueled and triggered (Section 7.3),
as well as the spatial distribution and form of the energy deposited
by jets \citep[e.g.,][]{ob04}.  Observations of shocks and cavities
created by AGN outbursts have motivated the development of
increasingly sophisticated simulations of the interaction of jets with 
the ICM.  \citet{vr06} found that even a variable hydrodynamic jet
flowing into a static atmosphere is incapable of tranferring a
significant fraction of its energy to the atmopshere.  In order to
avoid channeling jet energy beyond the halo, some additional physics
is required. 

\begin{figure}
\centerline{\includegraphics[height=10truecm]{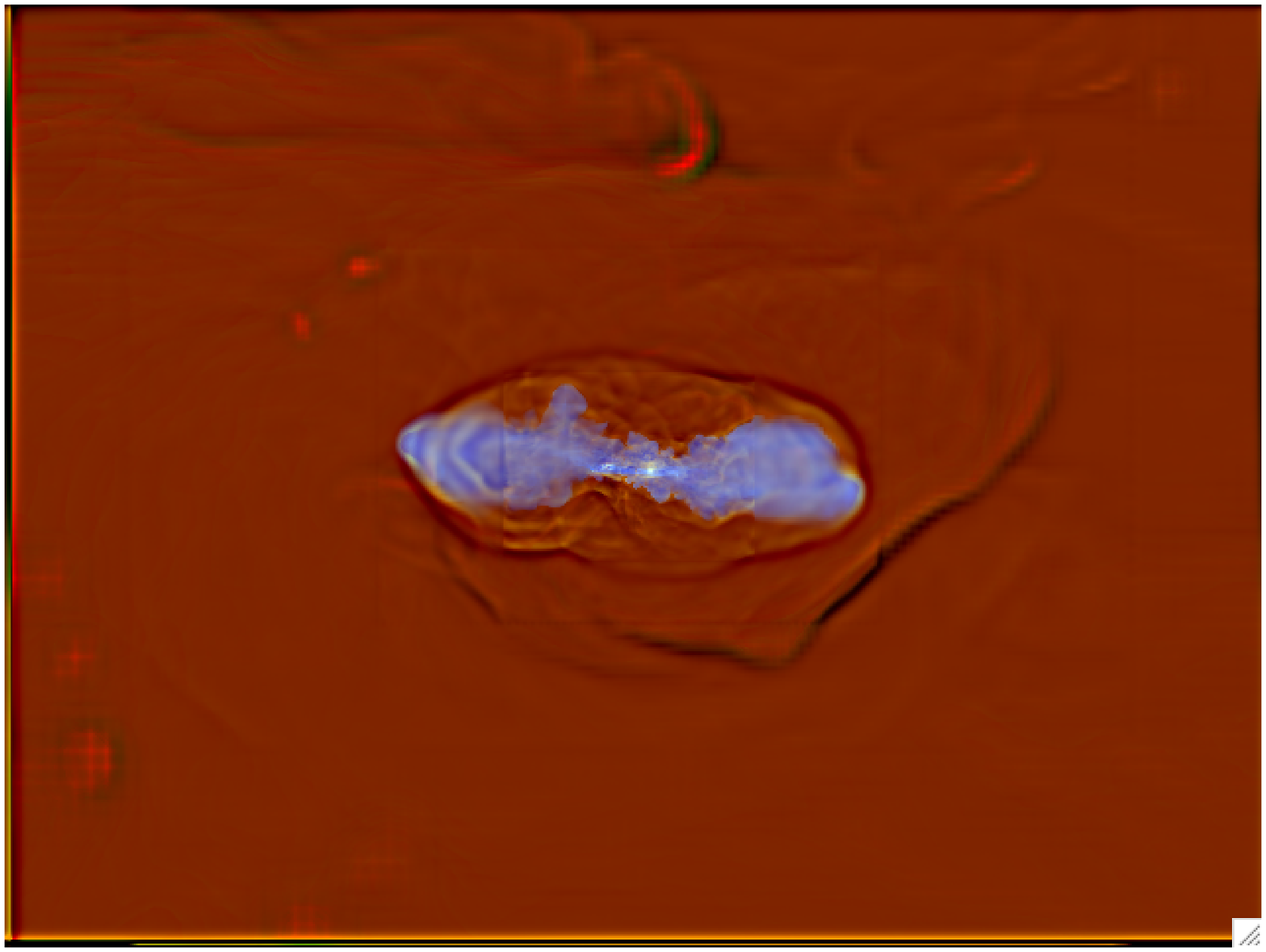}}
\caption{Simulation of jets interacting with the intracluster medium
  (ICM) from \citet{hby06}.  Radio synchrotron emission is in blue and
  X-ray emission in red. This simulation produces a realistic cocoon
  shock, cavities, and ripples. The strong gas feature to the lower
  right is an unrelated merger shock. Motion of ICM owing to ongoing
  cluster growth and wobbling of the Cygnus A-like jet both have a
  significant impact on the outcome of the model.} \label{fig:simulate}
\end{figure}

A recent jet simulation by \citet{hby06} is situated in a cluster
atmosphere drawn from a realistic simulation of heirarchical structure
formation. The simulated atmosphere includes the ongoing effects of
cosmological evolution, such as merger shocks and turbulence. The jet
also wobbles, simulating a dentist drill effect that is thought to
result from instabilities along the jet \citep{scheuer74}. Rather than
punching through the atmosphere, the jet deposits much of its energy
into gas close to the AGN. This produces a fairly realistic simulation
of Cygnus A (Figure \ref{fig:simulate}). Simulations of this
sophistication need to be coupled with a realistic model for fueling
and triggering outbursts in order to fully test feedback models. 

Although radio jets appear to be the main route by which AGN energy is
conveyed to the hot gas in nearby galaxies and clusters, AGN outbursts
produce uncollimated winds and intense radiation that can also heat
the gas. Powerful, uncollimated nuclear winds should couple strongly
to the surrounding gas, making them an effective means of
heating. However, they do not appear to be a major route for AGN
energy injection in nearby systems
\citep[e.g.,][]{kne07}. Photoionization and inverse Compton heating
can also couple a small fraction of the energy radiated in an AGN
outburst to the hot ISM \citep{co01,soc05}. Among other things, the
significance of this process depends on the relative values of the gas
temperature and the Compton temperature of the AGN radiation
\citep[$\sim 2$ keV;][]{soc05}, making it most effective in lower
temperature systems during the quasar era.  Again, there is no
evidence of such heating in nearby systems.

\section{HEATING WITHOUT FEEDBACK}

\subsection{Conduction}

Whether or not thermal conduction is an effective heating agent has
been a controversial issue for decades. Plenty of thermal energy is
available in the outskirts of clusters to heat the cores. The
controversy concerns whether and how it can be effectively
tapped. Assuming the magnetic fields threading clusters are tangled
(Section 1.4), the conductivity is suppressed below the classical
Braginskii value, inhibiting inward heat flux (Section 1.5). The
magnetic field topology is not observable, so thermal conduction is
poorly understood. Apart from the difficulties of maintaining local
equilibrium between conductive heating and radiation \citep[the fine-tuning
problem;][]{sfn84,mrb87,bd88}, there are several clusters where heat
conduction cannot balance cooling throughout the core, even at the
full Braginskii rate \citep{zn03,vf04,wmm04}.

On the other hand, if the thermal conductivity is only mildly
suppressed (Section 1.5), conduction may be effective in the outer
reaches of the cooling regions of clusters
\citep{nm01,zn03,vf04,wmm04}, which can substantially reduce the power
required from AGN heating. The heat flux depends strongly on
temperature, as $\sim T^{5/2}$, so it is less important in lower
temperature systems, including gE galaxies. 

\citet{djs04} investigated the effect of including thermal conduction
in numerical simulations of galaxy cluster formation. With the thermal
conductivity set to one third of the Braginskii value, they found that
although the temperature structure of the ICM is significantly
modified, the fraction of baryons that cools and  turns into stars is
little affected. This supports the argument that thermal conduction 
acting alone does not explain why star formation is inhibited in
cluster cores. 

\subsection{Other Heating Mechanisms}

Many mechanisms have been proposed to prevent gas from cooling to low
temperatures at the centers of cooling flows, but very few of these
involve feedback \citep[e.g.,][]{fab94}.  As discussed in Section 4.1,
without feedback it is difficult to account for the many clusters with
short central cooling times. Nevertheless, such mechanisms could
significantly reduce the demands on AGN heating. Most other mechanisms 
that have been proposed to stop the gas from cooling rely on the
energy available from mergers or, more generally, cluster binding
energy \citep[cf.][]{b03}. Preventing the core gas from cooling in
a system like the Perseus cluster requires $\sim 10^{62}$ erg over a
Hubble time. Because a major merger releases $\sim 10^{64}$ ergs, only
a small part of that energy needs to be tapped in order to prevent gas
from cooling and forming stars. Set against this, cool cores occupy a
small fraction, $\sim 0.001$, of the total cluster volume, presenting
a relatively small target for undirected heating. Furthermore, the
stably stratified cluster atmosphere, which resists the inward
propagation of turbulence, combined with high central pressure
(density), tends to shield the central gas from disturbances in the
more tenuous gas that inhabits the bulk of a cluster. 

Simulations show that major mergers can disrupt some cluster cooling
flows although, even for head-on mergers, the effect need not be long
lasting \citep[e.g.,][]{glr02}. Higher merger rates may explain the
reduced fraction of bright clusters with strong central X-ray peaks at
$z > 0.5$ \citep{vbf07}. The processes that make cooling flow clusters
different from clusters with long central cooling times are still the
subject of debate \citep[e.g.,][]{mbb04,omb06}, but it seems likely
that merger history plays some role. Thermal conduction should make it
very difficult to re-establish cooling cores in hot clusters, unless
it is suppressed by a large factor (Section 1.5).

Simulations of the growth of structure consistently find that the ICM
should be turbulent, with typical turbulent velocities $\sim
100\ \kms$ \citep[e.g.,][]{knv05}. Dissipation of turbulence could be
a significant source of heating. Outflows from central AGN may create
a similar level of turbulence, as suggested by observations of the
Perseus cluster \citep{cfj04}, so that the source of turbulence in
cool cores is unclear.  \citet{dc05} used a semiempirical model to
argue that a combination of thermal conduction, turbulent dissipation,
and turbulent diffusion could balance radiative losses for turbulent
velocities in the range 100 -­ 300 $\kms$. However, their model does
not explain why the turbulence should have the spatial distribution or
power required to balance radiative losses locally.

\section{FEEDBACK AND GALAXY FORMATION}

\subsection{Star Formation in cD Galaxies}

Brightest clusters galaxies (which we refer to for convenience as cD
galaxies), with masses upward of $\sim 10^{12}\ \msun$ and with halos
extending hundreds of kiloparsecs into the surrounding cluster, are
the largest and most luminous galaxies in the Universe \citep{s88}.
cDs are similar in appearance to gE galaxies, but there are
significant differences. Their central surface brightnesses tend to be
lower, their velocity dispersions rise less steeply with increasing
luminosity, and they often possess stellar envelopes that lie above
the $R^{1/4}$ law profiles that characterize gE galaxies. Their
locations at the centers of clusters suggest they grew to such
enormous sizes by swallowing stars and gas from neighboring galaxies
through mergers and stripping \citep{go72,m85}. This process is
augmented at late times by the accretion of intracluster gas
\citep{cb77,fn77}.

\begin{figure}
\centerline{\includegraphics[height=4.5truecm]{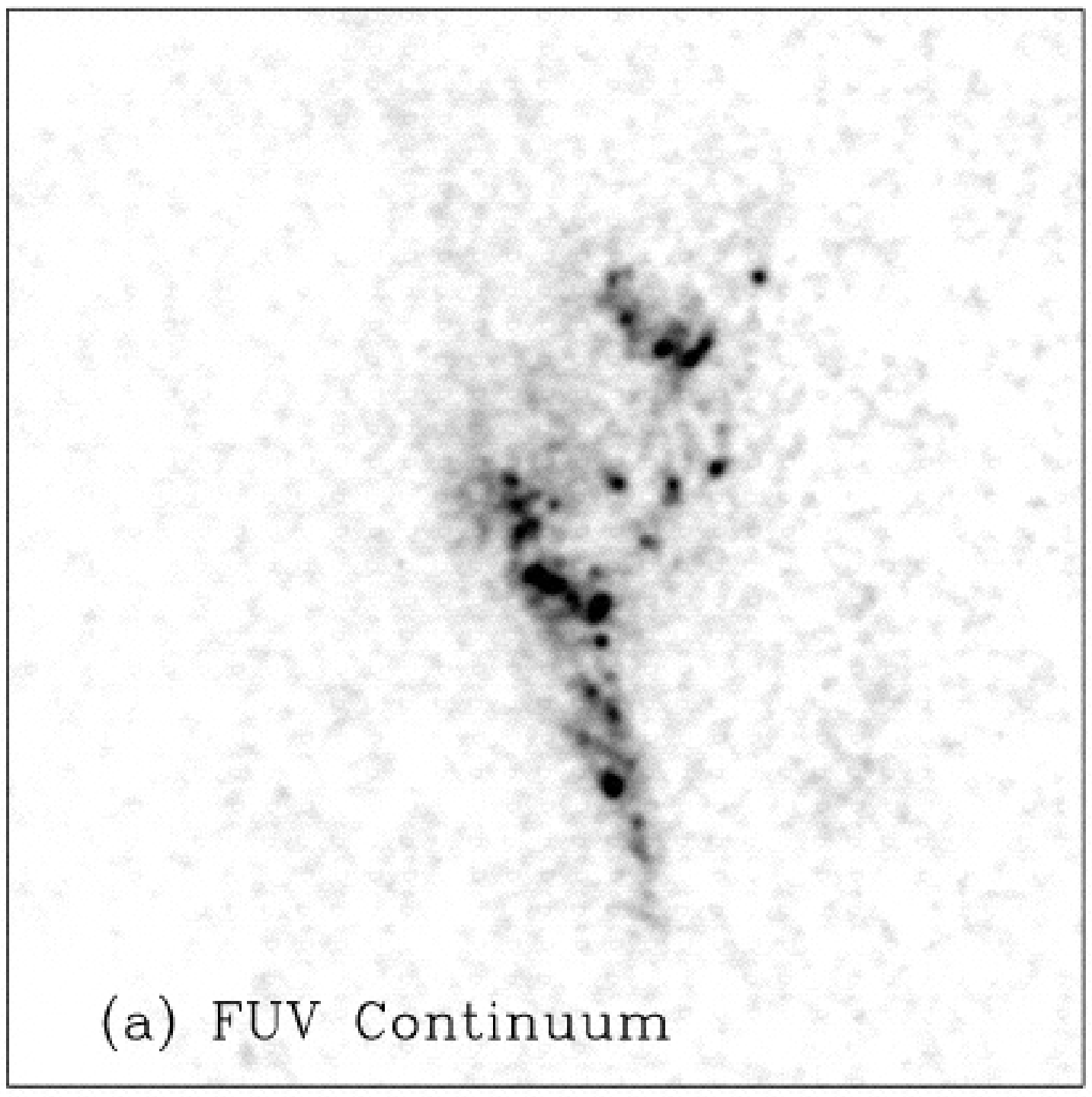}
\includegraphics[height=4.5truecm]{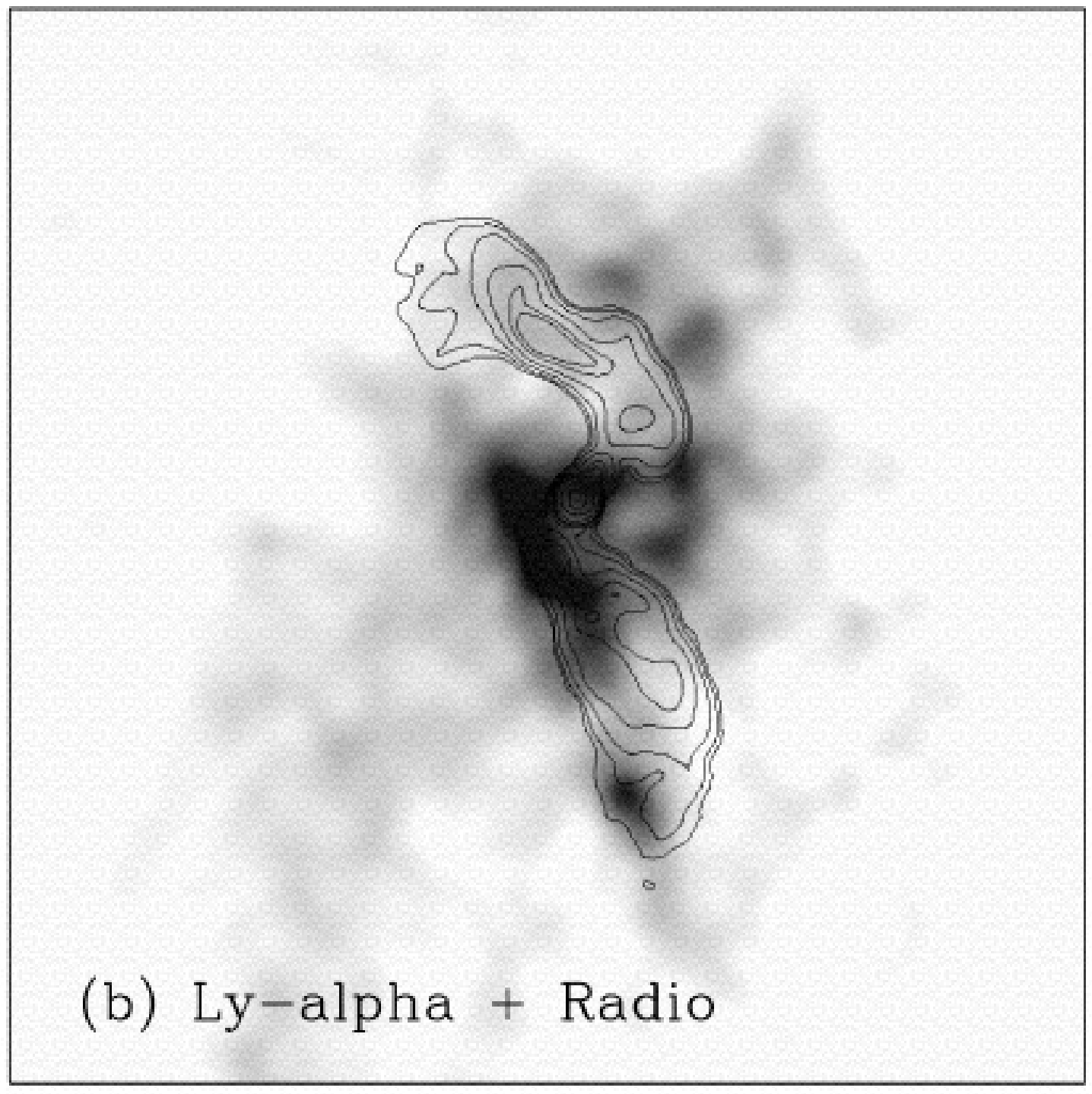}
\includegraphics[height=4.5truecm]{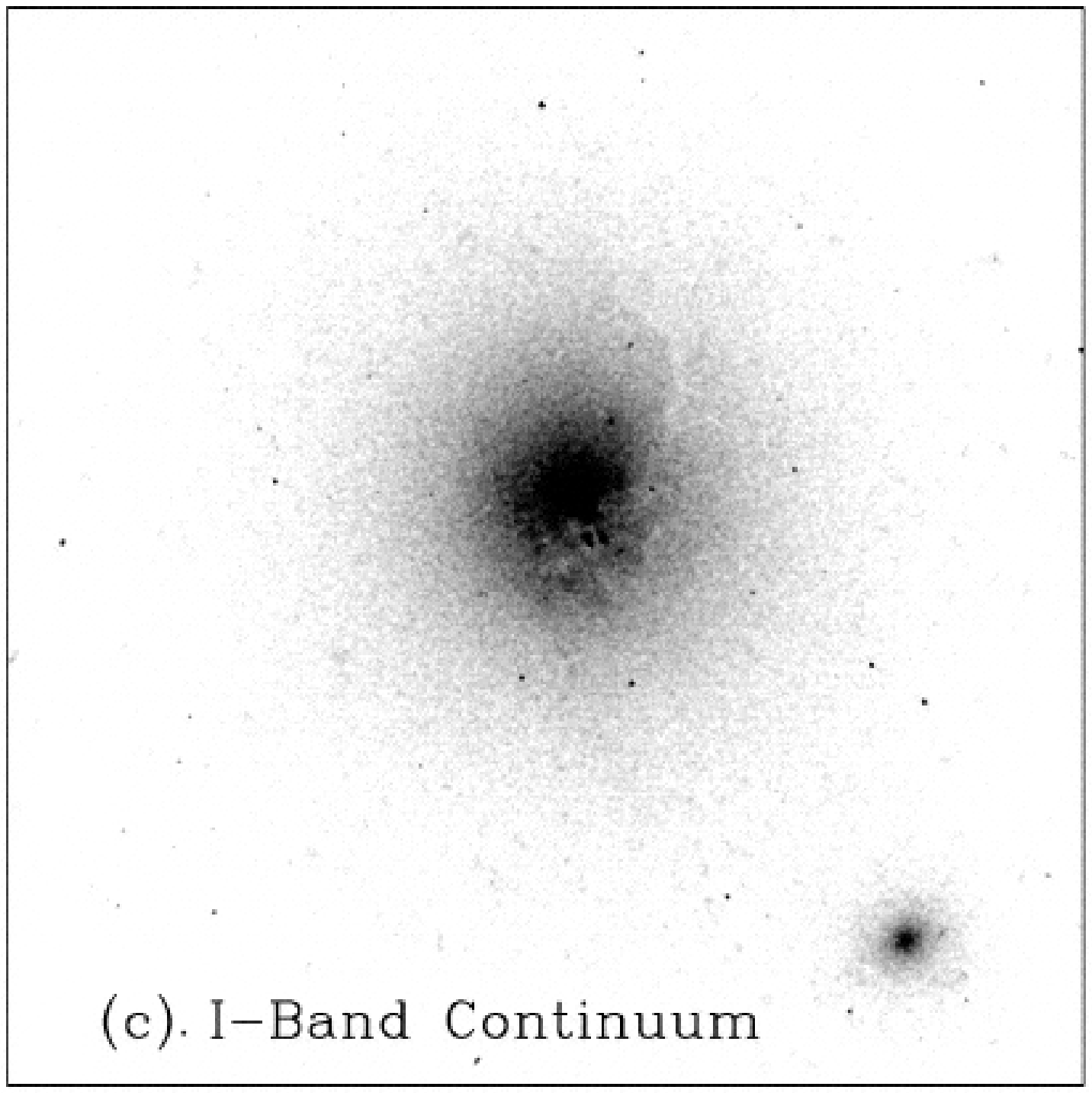}}
\caption{Hubble Space Telescope images of the central $19 \times 19$
  arcsec ($24 \times 24$ kpc) of the cD in Abell 1795 from
  \citet{obm04}: {\it Left:}  Far-UV continuum, {\it middle:} Ly
  emission with radio contours superposed, and {\it right:} I-band
  continuum. Note the bright knots of star formation and Ly emission
  that are seen preferentially along the radio lobes. A dust lane is
  evident in the I-band image. The star formation rate in this system
  is 10 -- 30 $\msunpy$.} \label{fig:sf}
\end{figure}

The stellar populations of most cDs are metal rich and dormant, and
they rarely show optically active nuclei. Only 10\% of optically
selected cDs in non-cooling flow clusters harbor nuclear emission line
fluxes above an equivalent width of a few angstroms
\citep{bkh06,ehb07}. In comparison, 20\% of galaxies lying outside of
cluster cores \citep{bkh06} have detectable line emission, suggesting
that the gas processes driving emission line activity in most galaxies
are suppressed in cluster cores. The detection rate and strength of
nebular emission in cD galaxies rises dramatically, to at least 45\%,
in cooling flow clusters \citep{cae99,ehb07}.  These emission line
systems often extend tens of kiloparsecs from the nucleus
\citep{hbb89}. They appear to be powered  by a combination of shock
heating, stellar photoionization, and irradiation by the surrounding
X-ray gas, but rarely by photoionization from an AGN. At least half 
of these systems are experiencing star formation
\citep[e.g.,][]{jfn87,mo89}, and the nebular clouds and sites of star
formation are usually embedded in pools of $10^9$ -- $10^{11}\ \msun$
of atomic and molecular gas \citep[e.g.,][]{edge01}. The ratio of
young stars to galaxy mass is typically only $\sim 10^{-3}$.  The
corresponding star formation rates are typically a few $\msunpy$, but
in extreme  cases they approach or exceed $100\ \msunpy$ 
\citep[e.g.,][]{cae99,mrb06}. 
Young stellar population masses of $10^8~\msun$ to $10^{10}~\msun$ are found
in these systems.  The largest starbursts rival those
observed during the most rapid period of galaxy growth at $z = 2$ -- 3
\citep{jgc05}. Many cDs experiencing star formation are also
experiencing powerful AGN outbursts. An example of star formation in
the Abell 1795 cD is shown in figure \ref{fig:sf} from \citet{obm04}.

There are two schools of thought regarding the origin of this star
formation, its accompanying cold gas and nebular emission: accretion
through mergers \citep[e.g.,][]{hfs92}, or cooling flows
\citep{fab94}. Mergers are an appealing mechanism because they are the
principal route to the growth of structure in the context of the cold
dark matter cosmogony, and they are an established and effective
mechanism for triggering star formation in galaxies. Moreover, they
provided a plausible alternative to classical cooling flows, that
predict excessive star formation rates in cD galaxies
\citep{fab94}. Just how effective mergers are at stimulating star 
formation in cluster cores in general and cDs in particular is
unclear. Gas rich donor galaxies rarely inhabit the cores of clusters,
perhaps because their gas becomes vulnerable to stripping long before
they are able to reach the center. Furthermore, galaxies in the cores
of clusters are less likely to host star formation and nuclear
activity than galaxies in other environments \citep{bvk07}, which
augurs against merger induced star formation. When star formation is
observed in cDs, with rare exceptions, it is centered on a cooling
flow. Most importantly, star formation rates approach or agree with
the upper limits on cooling rates from XMM-Newton and Chandra
\citep{mwm04,hm05,rmn06,sce06}.  The velocity structure of the cold
gas fueling star formation is also consistent with having recently
condensed from a static atmosphere \citep{jbb05,sce06}.

The emerging consistency between star formation and cooling rates is
noteworthy in the broader context of galaxy formation. The mechanism
suppressing cooling flows and regulating the growth of cD and gE
galaxies may also be responsible for the exponential decline in the
luminosity function of bright galaxies, and perhaps the relationship
between bulge luminosity and black holes \citep[e.g.,][]{bbf03}.  The
possibility that these two fundamentally important properties of
galaxies may be explained using a single mechanism has led to the
development of new galaxy formation models that incorporate AGN
feedback. Cooling flows should be relatively clean examples of galaxy
and supermassive black hole growth through the accretion of cooling
gas in large halos. 

\subsection{Galaxy Formation Models}

Based on the Millennium Run dark matter simulation, \citet{csw06}
modeled the formation and evolution of galaxies and their supermassive
black holes in a concordance CDM cosmology. Their model combined
numerical and semianalytic techniques, with the goal of understanding
the effects of AGN feedback on galaxies of different masses. In this
model, the AGN is powered by continuous accretion, scaling in
proportion to the supermassive black hole and hot gas masses divided
by the Hubble time. In keeping with earlier work, they follow galaxies
through a ``quasar mode'' of halo merging and gas accretion that leads
to rapid bulge and black hole growth at $z = 2$ -- 3. This is followed
by a more quiescent ``radio mode'' that suppresses cooling flows at
late times. The suppression becomes increasingly effective in more
massive halos and is able to stop cooling entirely in galaxy halos
with virial temperatures $T > 3 \times 10^6$ K (i.e., groups and
clusters) from $z = 1$ to the present. cD galaxies continue to grow
slowly at late times by dissipationless ``dry mergers'' lacking star
formation. Radio mode feedback develops in a static hot halo driven by
black hole accretion at a strongly sub-Eddington rate that adds
negligibly to the black hole mass at late times. \citet{csw06}
consider feeding the nucleus with cold cloud and Bondi accretion from
the hot halos, but lacking a feedback prescription, the outcome of
their model is insensitive to the mode of accretion. 

Using smooth particle hydrodynamic simulations, \citet{ss06} focused
on the formation and evolution of the cD and the ICM surrounding it in
the presence and absence of AGN feedback. Like \citet{csw06},
\citet{ss06} distinguish between major galaxy and black hole growth
(BHAR model) at early times and more gentle AGN accretion and radio
bubble feedback (Magorrian mode) at late times. In both the
\citet{csw06} and \citet{ss06} models, the correlation between black
hole mass and bulge mass is imprinted during the quasar era and
changes little at late times, when the suppression of cooling flows
becomes important. This model assumes bubble injection repeats every
$10^8$ year with the energy per bubble scaling with the halo mass to
the 4/3 power in the BHAR phase, and scaling with the black hole
accretion rate during the Magorrian phase. In the absence of bubble
heating, massive galaxies grow to unrealistically large sizes. The
introduction of hot buoyant bubbles reduces cooling flows
substantially, but cooling and star formation does not cease entirely
at redshifts below $z = 1$. This behavior is qualitatively consistent
with observations of real clusters. The \citet{ss06} model
successfully produces entropy pedestals in the cores of clusters
\citep{vd05,dhc06}. However, their temperature and density profiles
are flatter than those in real cooling flow clusters with active AGN.

These and other studies demonstrate that periodic AGN outbursts with realistic
energies and duty cycles are able to suppress cooling flows and to
recover the observed exponential turnover in the luminosity function
of bright galaxies. They may also play a role in the development of
the black hole bulge mass relation. Despite these impressive results,
a great deal of the picture is missing in these simulations, including
the ``microphysics'' of heating and of course the ``macrophysics'' of
star formation, both of which are poorly understood. These models also
lack a working feedback prescription that reproduces the observed
levels of cooling, AGN activity, and star formation in gEs and cDs at
late times. They also fail to reproduce the observed temperature and
density profiles of cooling atmospheres. A complete model for AGN
heating should include a physically realistic treatment of the
behavior of radio jets and their interaction with the ICM.

It is worth noting an important difference between the mode of energy
output from AGN during the quasar epoch and present day cooling flows
that is not explicitly captured in these models. Quasars are thought
to be triggered by cold gas funneled onto the nucleus during halo
mergers in the early Universe, leading to the rapid buildup of bulges
and supermassive black holes \citep[e.g.,][]{b99,sdh05}. Only a small
fraction of the $\sim10^{45-46}\ \ergps$ of power radiated by this
process must couple to the gas in order to drive an outflow capable of 
suppressing star formation and black hole growth. Roughly the same
driving power $\sim 10^{43-45}\ \ergps$, is required to offset a
massive cooling flow. The enthalpy and shock energy released by
cooling flow AGN, even in cDs with relatively weak radio sources, can
rival the power output of a quasar. However, the bulk of the power
emerging from an FR I source in a cooling flow is mechanically coupled
to the gas through shocks and cavity enthalpy, and is not released as
radiation. Apparently, as the specific accretion rate throttles down
from the Eddington regime during the quasar era to the strongly
sub-Eddington regime at late times, the power output switches from
being radiation dominated to being mechanically dominated. This shift
is qualitatively consistent with the observed behavior of low-mass
X-ray binaries \citep{css05}.

\subsection{Supermassive Black Hole Growth as a Consequence of
  Feedback}

The demanding power requirements of the long-term suppression of
cluster cooling flows by AGN should produce black holes with masses
exceeding $10^9\ \msun$ \citep{fr04}. Expressing the cooling
luminosity in terms of the classical cooling rate, $L_{\rm cool} = 2.5
\dot MkT/(\mu \mh)$, multiplying by the age of the cooling flow, $t$,
and equating to the energy released by the black hole, $\eta M_{\rm h}
c^2$, gives the minimum black hole mass required to stop cooling as
\begin{displaymath}
M_{\rm h} = {1.5\over \eta} {s^2\over c^2} \dot M t \simeq
2.25\times10^{-4} \dot M t,
\end{displaymath}
for gas cooling from 5 keV and $\eta = 0.1$. Cooling proceeding at a
rate of $\dot M = 100\ \msunpy$ since $z = 1$, which corresponds to a
lookback time of $t = 7.7 \times 10^9$ year in a concordance cosmology,
would form a $1.7 \times 10^8\ \msun$ black hole. The power output
required to quench a $1000\ \msunpy$ cooling flow (e.g., Abell 1835) 
would produce a $1.7 \times 10^9\ \msun$ black hole, rivaling the
largest known black hole masses. Evidently, quenching a large cooling
flow over the lifetime of a galaxy cluster could produce outsized
supermassive black holes in cDs, even if jet heating is efficient. NGC
1275 is one of the only cluster cooling flow cDs with a measured black
hole mass. Using molecular $\rm H_2$ emission line speeds,
\citet{wej05} measured a mass of $3 \times 10^8\ \msun$. Based on the
formula above, AGN feedback can quench a 150 -- $200\ \msunpy$ cooling
flow without producing an excessively massive black hole. This may
leave little room for an early buildup of the black hole and
subsequent long-term accretion from the cooling flow. It also requires
a relatively high efficiency for converting accreted mass to jet power
if the AGN is suppressing cooling. This level of growth could cause
significant departures from the established black hole mass versus
bulge mass (velocity dispersion) relation extrapolated to cD galaxy
luminosities \citep{lfr07}. Future measurements of black hole masses
in cD galaxies using very large aperture telescopes
\citep[e.g.,][]{lfr07} will be a sensitive probe of the accretion and
feedback history of clusters. 

\subsection{Accretion Mechanism}

The accretion mechanism is a critical piece of any operational
feedback loop. There are three broad categories: accretion from the
hot atmosphere surrounding the black hole through the Bondi mechanism
\citep[e.g.,][]{adf06}, accretion of cold clouds as the result of
stripping or cooling blobs of gas from a cooling flow
\citep[e.g.,][]{soker06}, and accretion of stars
\citep[e.g.,][]{wh05}. The actual mechanism must operate in the
sub-Eddington accretion regime, it must release most of its energy by 
mechanical winds or outflows and not by radiation, and it must be
responsive enough to prevent cooling and heating catastrophes.

Bondi accretion, with a rate that scales as $\nelec T^{-3/2}$, is
simple, natural, and in principle easy to regulate
\citep[e.g.,][]{csf02}. The local atmosphere responds to AGN heating
by expanding, which lowers the local gas density and the accretion
rate along with it. As the gas radiates away its energy, the
atmosphere contracts and compresses, and the accretion rate rises. For
$\gamma  = 5/3$, the Bondi accretion rate depends on gas properties
only through the entropy, which is affected directly by heating and 
cooling. The accretion rate depends on the square of the black hole
mass and properties of the atmosphere within the radius of influence
of the black hole, all of which are difficult or impossible to measure
with current instrumentation. Bondi accretion appears to be
energetically plausible in lower luminosity gE galaxies \citep{adf06},  
but is probably unable to power the largest outbursts in clusters of
galaxies \citep{rmn06}.

Ample supplies of cold gas are available in cDs to fuel the AGN. If
this gas were supplied by stripping or mergers, it would arrive
sporadically in a manner unrelated to the properties of the cooling
atmosphere. This makes the accretion rate difficult to tune, unless
the gas is stored in a disk, which is somehow regulated in the
vicinity of the black hole. Gas cooling out of the ICM would naturally
be subject to thermostatic control by periodic outbursts
\citep[e.g.,][]{soker06}. Even so, understanding how energy emerging
from a region smaller than the size of the solar system is able to
regulate flows on the vast scales of clusters is a monumental
challenge.

\subsection{Observational Constraints on Feedback Models}

We pointed out earlier that the conditions in the cores of clusters
lend themselves to detailed study of the rudiments of galaxy
formation. Because the entire cycle of heating, cooling, feedback, and
star formation can be explored there in great detail, cooling flows
provide a standard against which models of galaxy formation
\citep[e.g.,][]{ss06} can be tested. These models should satisfy the
following conditions:

\begin{enumerate}
\item A small fraction, $\lesssim 10\%$, of cDs at $z \sim 0$ are
  experiencing star formation perhaps as a consequence of the
  inability of AGN to balance radiative cooling at all times from
  their dense, $\sim 10^{-1}\ \pcm$, atmospheres with cooling times
  $\ll 10^9$ year. Infall of cold gas by mergers or
  stripping is another potential source of cold gas. However, it might
  be difficult to understand why stripping occurs preferentially in
  the cores of clusters with short cooling times.

\item In both gEs and cDs, AGN heating scales in proportion to the
  cooling luminosity, as expected in an operational feedback loop. It
  is not clear whether or how the level of feedback scales with black
  hole mass or halo mass. 

\item Bondi accretion may power the AGN in gEs \citep{adf06} but is
  probably unable to do so in more massive cD galaxies. 
Only the most powerful cluster outbursts require
  accretion rates approaching $\sim 1\ \msunpy$; typical rates are
  $\sim 10^{-2}\ \msunpy$, which is a small fraction of the Eddington
  accretion rate of a $10^8\ \msun$ black hole.

\item The jet model most directly associated with the Magorrian mode
  of \citet{ss06} and the radio mode of \citet{csw06} must account for
  the enormous range in radiative efficiency and large jet kinetic
  energy observed in cDs. It should also recover the observed
  temperature, density, and entropy profiles in the cores of clusters
  \citep{borgani,vd05}. 

\item The existence of very powerful AGN outbursts and the persistent
  energy demands of cooling imply substantial black hole growth at
  late times in cD galaxies.  There is tantalizing new evidence that
  cDs and their central black holes grow at an average rate that
  crudely follows the slope of the black hole mass versus bulge mass
  relation \citep{hr04} once star formation is taken into account
  \citep{rmn06}. However, bulges and black holes do not always grow in
  lockstep. Other lines of evidence suggest that the slope flattens in
  cD galaxies \citep{lfr07}.

\item Star formation parameterized with a Schmidt-Kennicutt law is
  probably a reasonable approximation in these systems
  \citep[e.g.,][]{emw06,mrb06}, although this issue is in need of
  further study. Disk formation, which is a staple of semianalytic
  galaxy formation models, is rare or short-lived
  \citep[e.g.,][]{hbb89,sce06}. Energy pumped into the hot gas by
  starburst winds, $\lesssim 10^{43}\ \ergps$, is negligible on the
  scale of the cooling flow \citep{mwm04,mrb06}, but may be important
  near the nucleus of the cD where fueling of the AGN is actively
  occurring.

\item Finally, the observed level and spatial distribution of chemical
  enrichment, and abundance ratios in the hot gas must constrain the
  history of star formation in the cores of clusters and the level of
  mixing generated by AGN outbursts and mergers.
\end{enumerate}

\section{CONCLUDING REMARKS}

The editors of Discover Magazine's Winter 2007 special issue ``Unseen 
Universe'' posed the following question to Martin Rees: ``Is there any
particular image that you saw recently that reminded you anew of just
how much progress we've made?''  Rees's response provides a succinct
history of this field. ``I've been especially impressed by the X-ray
images of galaxy clusters which are now becoming available from the
Chandra satellite and other instruments. We see gas being churned up
by explosions and huge black holes in the center of the cluster. We
see how it's cooling down and how the cooling is being balanced by
tremendous outbursts of jets and bubbles of hot gas. This is something
that most people didn't suspect was happening until these images
revealed it.'' 

If many people did not suspect this was happening before Chandra
revealed it, why not? The Rosat observatory had already established
that radio sources are interacting with the hot gas surrounding them
\citep{bvf93,cph94}, and it had been pointed out previously that AGN
\citep[e.g.,][]{pgd90,bo91,oe98,td97,bt95} and thermal conduction
\citep{rt89,m88} could offset cooling in some clusters. In our view,
these solutions were not widely embraced because, like the classical
cooling flow model itself, they lacked persuasive observational and 
theoretical support. As is often true in astronomy, the impasse was
broken by new, high-resolution instrumentation. XMM-Newton
spectroscopy revealed that hot gas in clusters cannot be cooling at
the classical rates. At about the same time, high resolution Chandra
images identified AGN feedback as the probable heating agent. But the
real situation is more complicated. The data show that AGN heating is
more subtle than early theoretical studies envisioned. Heating is not
a violent, local process. It is gentle and spatially dispersed. There
is scant evidence for constant density cores or central temperature
spikes, and entropy inversions as some nuclear heating models 
predicted. To our knowledge, no one anticipated cool rims surrounding
the cavities, ripples, ghost cavities, and quasar-like outbursts that
are barely audible in the radio and optically faint in the
nucleus. The data were inadequate and so was the physics. 

Much remains to be done. We do not understand how jets are powered and
what they are made of, how the putative feedback loop works, how
efficiently jets heat the gas, and we still cannot be sure that they
are the sole heating agent or even the principal one. These issues
will take time to resolve, but a great deal of progress has already
been made, notably in ever more realistic jet simulations (e.g.,
Figure \ref{fig:simulate}) that can now be tested against high quality
data. They must be resolved in order to understand the role of
supermassive black holes in galaxy and large-scale structure
formation. We conclude by listing several issues and avenues of
research that we believe will lead to substantial progress in this
field. 

\section*{FUTURE ISSUES}

\begin{enumerate}
\item XMM-Newton spectroscopy has excluded mass deposition at the
  classical rates in cooling flows. However, deposition at substantial
  levels comparable to observed star-formation rates have not been
  ruled out. A search for Fe XVII and other lines at levels that are
  consistent with observed star formation rates is within the grasp of
  XMM-Newton's reflection grating spectrometer, and will be within
  easy reach of a future Constellation-X. Combined with increasingly
  accurate star formation rates (item 5), this would constitute a
  strong test of new, feedback-based cooling flow models. 

\item It is difficult or impossible to detect cavities and shock
  fronts in distant clusters and groups beyond $z \sim 0.5$. In
  principle, radio observations, especially at low frequencies, can
  probe the history of feedback and heating at earlier times. However,
  the large variation in radio synchrotron efficiency must first be
  understood and calibrated. 

\item The time distribution of AGN jet power in galaxies and clusters,
  which is an essential part of a feedback model, is poorly
  understood. Existing X-ray cavity surveys are limited in size and
  suffer strong selection biases. A large, unbiased search for
  cavities and shocks in a flux or volume limited sample of groups
  and clusters is required to more accurately determine the average AGN heating rate. 

\item The environmental factors that trigger AGN outbursts,
  particularly the most powerful ones in cD galaxies ($> 10^{61}$
  erg), are poorly understood. High resolution imaging and
  spectroscopy of the stars and gas in nuclei of galaxies, made at a
  variety of wavelengths, will characterize the properties of nuclear
  gas disks, stellar cusps, and cores. This knowledge should provide a
  deeper understanding of the accretion and feedback process, and
  hopefully new insight as to why many systems are as powerful as
  quasars and yet they don't reveal themselves as such. Most
  importantly, measurements of black hole masses in cD galaxies using
  future large aperture telescopes will place restrictive limits on
  the history of AGN feedback in clusters. 

\item Limits on the rate of cooling in the cores of galaxies and
  clusters have become tight enough to warrant better measurements of
  star formation rates and histories. This will require precision
  photometry over a broad passband (UV to far IR) and careful
  accounting for dust and positive feedback from radio triggered star
  formation \citep[; see Figure
    \ref{fig:sf}]{deyoung95,obm04}. Combining this information
  with black hole growth rates estimated from cavities and shock
  fronts will yield new insight into how bulges and supermassive black
  holes grow at late times.

\item New high fidelity jet models \citep[e.g.,][~see Figure
  \ref{fig:simulate}]{hby06} combined with sensitive X-ray
  measurements of cavities and shock fronts will increase our
  understanding of jet dynamics, energetics, content, radiative
  efficiencies, and ultimately how jets form. The form of energy
  deposition by jets is a critical component of any feedback model. 

\item Models are needed for the fueling and triggering of AGN
  outbursts, including the part played by mergers. 

\item How cavity enthalpy and weak shock energy is dissipated, how
  efficiently it heats the gas, and where the heat is deposited are
  fundamentally important questions that have not been satisfactorily
  answered. 

\item The contribution of AGN outbursts to deviations from the
  expected scaling relationships between mass, temperature, and X-ray
  luminosity of clusters \citep[e.g.,][]{omb06,gmn07} is poorly
  understood. The energy required to quench a large cooling flow over
  its lifetime is comparable to the 1 keV per baryon necessary to
  preheat a cluster. AGN may contribute significantly to preheating.

\end{enumerate}

\section*{ACKNOWLEDGEMENTS}

We thank David Rafferty and Laura B\^{\i}rzan for helpful discussions
and for providing figures.  Michael Balogh read the draft in a
primitive state of development and offered helpful comments. Thanks to
Hans B\"ohringer, Larry David, Bill Forman, Dan Harris, Christine
Jones, Maxim Markevitch, and Alexey Vikhlinin for their advice. We
thank Chris O'Dea, Sebastian Heinz, and Marcus Br\"uggen for
permission to print figures from their work. We especially thank our
friend and collaborator Michael Wise, one of the pioneers of this
field, who has helped to shape our thoughts on this topic. B.R.M. was
supported in part by NASA Long Term Space Astrophysics grant
NAG5-11025.  P.E.J.N. acknowledges NASA grant NAS8-01130.


\begin{thebibliography}{}

\newcommand{\araa}{{\it Ann. Rev. Astron. Astrophys. }}
\newcommand{\apj}{{\it Astrophys. J. }}
\newcommand{\mnras}{{\it MNRAS }}
\newcommand{\aap}{{\it Astron. Astrophys. }}
\newcommand{\aj}{{\it Astron. J. }}
\newcommand{\nat}{{\it Nature }}
\newcommand{\an}{{\it Astron. Nachr. }}
\newcommand{\apjs}{{\it Astrophys. J. Suppl. }}
\newcommand{\apjl}{{\it ApJL }}
\newcommand{\physrep}{{\it Phys. Rep. }}
\newcommand{\etal}{{et al.}}

\bibitem[Allen et al.(2001)]{asf01} Allen SW, Schmidt RW, Fabian AC.
  2001. \mnras 328:L37-41 

\bibitem[Allen et al.(2004)]{ase04} Allen SW, Schmidt RW, Ebeling H,
  Fabian AC, van Speybroeck L. 2004. \mnras 353:457-467

\bibitem[Allen et al.(2006)]{adf06} Allen SW, Dunn RJH, Fabian AC,
  Taylor GB, Reynolds CS. 2006. \mnras 372:21-30

\bibitem[Arnaud(2005)]{a05} Arnaud M. 2005. In Background Microwave
    Radiation and Intracluster Cosmology, ed. F Melchiorri, Y
    Rephaeli, pp. 77-116. Netherlands: IOS Press 

\bibitem[Arnaud \& Evrard(1999)]{ae99} Arnaud M, Evrard
  AE. 1999. \mnras 305:631-640  

\bibitem[Arnaud et al.(1992)]{arb92} Arnaud M, Rothenflug R, Boulade
  O, Vigroux L, Vangioni-Flam E. 1992. \aap 254:49-64 

\bibitem[Balbus \& Soker(1989)]{bs89} Balbus SA, Soker N.  1989. \apj
341:611-630 

\bibitem[Babul et al.(2002)]{bab02} Babul, A., Balogh, M.~L., 
Lewis, G.~F., \& Poole, G.~B.\ 2002, \mnras, 330, 329


\bibitem[Baum \& O'Dea(1991)]{bo91} Baum SA, O'Dea CP. 1991. \mnras
  250:737-749 

\bibitem[Baumgartner et al.(2005)]{blh05} Baumgartner WH, Loewenstein
  M, Horner DJ, Mushotzky RF. 2005. \apj 620:680-696

\bibitem[Begelman(2001)]{b01} Begelman MC. 2001. In {\it ASP Conf. Proc.
240: Gas and Galaxy Evolution,} ed. JE Hibbard, M Rupen, JH van
Gorkom, p363.  San Francisco: ASP

\bibitem[Begelman(2004)]{b04} Begelman MC. 2004.  Coevolution of Black
Holes and Galaxies, from the Carnegie Observatories Centennial
Symposia, LC Ho ed. Carnegie Observatories Astrophysics Series:CUP p 374

\bibitem[Begelman et al.(1984)]{bbr84} Begelman MC, Blandford RD, Rees
MJ. 1984.  Reviews of Modern Physics 56:255-351 

%
\bibitem[Benson et al.(2003)]{bbf03} Benson AJ, Bower RG, Frenk CS,
  Lacey CG, Baugh CM, Cole S. 2003. \apj 599:38-49 

\bibitem[Best et al.(2006)]{bkh06} Best PN, Kaiser CR, Heckman TM,
  Kauffman G. 2006. \mnras 368:L67-70 

\bibitem[Best et al.(2007)]{bvk07} Best PN, von der Linden A,
  Kauffmann G, Heckman TM, Kaiser CR. 2007. \mnras in press


\bibitem[Binney \& Tabor(1995)]{bt95} Binney J, Tabor G. 1995. \mnras
  276:663-678 

\bibitem[Binney et al.(2007)]{bbo07} Binney J, Bibi A, Omma
  H. 2007. \mnras 377:142-146

\bibitem[B\^{\i}rzan et al.(2004)]{brm04} B\^{\i}rzan L, Rafferty DA,
  McNamara BR, Wise MW, Nulsen PEJ. 2004. \apj 607:800-809

\bibitem[Blandford(1999)]{b99} Blandford RD. 1999. In Galaxy Dynamics,
  ed. DR Merritt, M Valuri, JA Sellwood, pp. 87-95. San Francisco: ASP 

\bibitem[Blandford(2003)]{b03} Blandford RD. 2003. In High Energy
  Processes and Phenomena in Astrophysics, ed. XD Li, V Trimble,
  and ZR Wang, pp. 3-22. San Francisco: ASP 

\bibitem[Blandford \& Rees(1974)]{br74} Blandford RD, Rees
MJ. 1974. \mnras 169:395-415 

\bibitem[Blanton et al.(2001)]{bsm01} Blanton EL, Sarazin CL, McNamara
BR, Wise MW. 2001. \apj 558:L15-19  

\bibitem[Blanton et al.(2003)]{bsm03} Blanton EL, Sarazin CL, McNamara
  BR. 2003. \apj 585:227-243

%

\bibitem[B\"ohringer \& Morfill(1988)]{bm88} B\"ohringer H, Morfill
GE. 1988. \apj 330:609-619

\bibitem[B\"ohringer et al.(1993)]{bvf93} B\"ohringer H, Voges W,
Fabian AC, Edge AC, Neumann DM.  1993. \mnras 264:L25-L28 
 
\bibitem[B\"ohringer et al.(2002)]{bmc02} B\"ohringer H,
Matsushita K, Churazov E, Ikebe Y, Chen Y. 2002. \aap 382:804-820 
 
\bibitem[B\"ohringer et al.(2007)]{bsp07} B\"ohringer H, Schuecker
P, Pratt GW, Finoguenov A, eds. 2007. Proc. of Heating versus Cooling
in Galaxies and Clusters of Galaxies. Garching: Springer-Verlag 

\bibitem[Borgani(2004)]{borgani} Borgani S\ 2004, APS\&S, 294, 51

\bibitem[Braginskii(1965)]{b65} Braginskii SI. 1965. Rev. Plasma
  Phys. 1:205-311 

\bibitem[Branduardi-Raymont et al.(1981)]{bff81} 
Branduardi-Raymont G, Fabricant D, Feigelson E, Gorenstein P, 
Grindlay J, Soltan A, Zamorani G. 1981. \apj 248:55-60

\bibitem[Bregman \& David(1988)]{bd88}  Bregman JN, David LP.  1988.
\apj 326:639-644

\bibitem[Bregman et al.(2006)]{bfm06} Bregman JN, Fabian AC, Miller
ED, Irwin JA. 2006. \apj 642:746-51 

\bibitem[Brighenti \& Mathews(2003)]{bm03} Brighenti F, Mathews
  WG. 2003. \apj 587:580-588

\bibitem[Br\"uggen(2002)]{bruggen02} Br\"uggen M. 2002. \apj
  571:L13-L16

\bibitem[Br\"uggen \& Kaiser(2001)]{bk01} Br\"uggen M, Kaiser
CR. 2001. \mnras 325:676-684

\bibitem[Br\"uggen \& Kaiser(2002)]{bk02} Br\"uggen M, Kaiser
  CR. 2002. \nat 418:301-303 

\bibitem[Br\"uggen et al.(2005)]{brh05} Br\"uggen M, Ruszkowski M,
  Hallman E.  2005.  \apj 630:740-749 


\bibitem[Burbidge(1956)]{b56} Burbidge G. 1956. \apj 124:416-429

\bibitem[Burns(1990)]{burns90} Burns JO. 1990. \aj 99:14-30

\bibitem[Canizares et al.(1982)]{ccj82} Canizares CR, Clark GW,
Jernigan JG, Markert TH. 1982. \apj 262:33-43 

\bibitem[Canizares et al.(1988)]{cmd88} Canizares CR, Markert TH,
Donahue ME. 1988.  NATO ASIC Proc.~229: Cooling Flows in Clusters and
Galaxies, 63-72 

\bibitem[Carilli et al.(1994)]{cph94} Carilli CL, Perley RA, Harris
DE. 1994. \mnras 270:173-177 

\bibitem[Carilli \& Taylor(2002)]{ct02} Carilli CL, Taylor
  GB. 2002. \araa 40:319-348

\bibitem[Cavaliere \& Fusco-Femiano(1976)]{cf76} Cavaliere A,
  Fusco-Femiano R. 1976. \aap 49:137-144

\bibitem[Cavaliere \& Lapi(2006)]{cl06} Cavaliere A, Lapi
A. 2006. \apj 647:L5-8 


\bibitem[Churazov \& Inogamov(2004)]{ci04} Churazov E, Inogamov
N. 2004. \mnras 350:L52-L56

\bibitem[Churazov et al.(2001)]{cbk01} Churazov E, Br\"uggen M, Kaiser
CR, B\"ohringer H, Forman W. 2001. \apj 554:261-273

\bibitem[Churazov et al.(2002)]{csf02} Churazov E, Sunyaev R, Forman
  W, B\"ohringer H. 2002. \mnras 332:729-734

\bibitem[Churazov et al.(2004)]{cfj04} Churazov E, Forman W, Jones C, 
Sunyaev R, B\"ohringer H. 2004. \mnras 347:29-35

\bibitem[Churazov et al.(2005)]{css05} Churazov E, Sazonov S, Sunyaev
  R, Forman W, Jones C, Boehringer H. 2005. \mnras 363:L91-95

\bibitem[Ciotti \& Ostriker(2001)]{co01} Ciotti L, Ostriker
  JP. 2001. \apj 551:131-152 

\bibitem[Clarke et al.(2001)]{ckb01} Clarke TE, Kronberg PP,
  B\"ohringer H.  2001. \apj 547:L111-114

\bibitem[Clarke et al.(2004)]{cbs04} Clarke TE, Blanton EL, Sarazin
CL. 2004. \apj 616:178-191  

\bibitem[Clarke et al.(2005)]{csb05} Clarke TE, Sarazin CL, Blanton
EL, Neumann DM, Kassim NE. 2005. \apj 625:748-191 

\bibitem[Cowie \& Binney(1977)]{cb77} Cowie LL, Binney J. 1977. \apj
215:723-732  

\bibitem[Crawford et al.(1999)]{cae99} Crawford CS, Allen SW, Ebeling
  H, Edge AC, Fabian AC. 1999. \mnras 306:857-896 

\bibitem[Crawford et al.(2005)]{csf05} Crawford CS, Sanders JS, Fabian
  AC. 2005. \mnras 361:17-33

%
\bibitem[Croton et al.(2006)]{csw06} Croton DJ, et al. 2006. \mnras
  365:11-28 

\bibitem[Croston et al.(2005)]{chb05} Croston JH, Hardcastle MJ,
  Birkinshaw M. 2005. \mnras 357:279-294 

\bibitem[Croston et al.(2007)]{ckh07} Croston JH, Kraft RP, Hardcastle
MJ. 2007. \apj 660:191-199

\bibitem[Dalla Vecchia et al.(2004)]{dbt04} Dalla Vecchia C, Bower RG,
  Theuns T, Balogh ML, Mazzotta P, Frenk CS. 2004. \mnras 355:995-1004


\bibitem[David et al.(2001)]{dnm01}  David LP, Nulsen PEJ, McNamara
BR, Forman W, Jones C, Ponman T, Robertson B, Wise M. 2001. \apj
557:546-559 

\bibitem[Dennis \& Chandran(2005)]{dc05} Dennis TJ, Chandran
  BDG. 2005. \apj 622:205-216  

\bibitem[De Young(1995)]{deyoung95} De Young DS. 1995. \apj
  446:521-527

\bibitem[De Young(2001)]{deyoung01} De Young DS. 2001. The Physics of
Extragalactic Radio Sources. Chicago: Univ. Chicago Press. 569 pp.

\bibitem[De Young(2006)]{deyoung06} De Young DS. 2006. \apj 648:200-208

%

\bibitem[De Grandi \& Molendi(2001)]{dm01} De Grandi S, Molendi
  S. 2001. \apj 551:153-159 

\bibitem[De Grandi \& Molendi(2002)]{dm02} De Grandi S, Molendi
  S. 2002. \apj 567:163-177

\bibitem[De Young(2003)]{deyoung03} De Young DS. 2003. \mnras
343:719-24  

\bibitem[Diehl \& Statler(2007)]{ds07} Diehl S, Statler T.  2007. \apj
  in press (astro-ph/0606215)

\bibitem[Dolag et al.(2004)]{djs04} Dolag K, Jubelgas M, Springel V,
Borgani S, Rasia E.  2004.  \apjl 606:L97-L100

\bibitem[Donahue et al.(2006)]{dhc06} Donahue M, Horner DJ, Cavagnolo
  KW, Voit GM. 2006. \apj 643:730-750  

\bibitem[Dunn \& Fabian(2004)]{df04} Dunn RJH, Fabian AC. 2004. \mnras
355:862-873  

\bibitem[Dunn \& Fabian(2006)]{df06} Dunn RJH, Fabian AC. 2004. \mnras
373:959-971

\bibitem[Dunn et al.(2006)]{dfc06} Dunn RJH, Fabian 
AC, Celotti A\ 2006, \mnras, 372, 1741 

\bibitem[Dunn et al.(2005)]{dft05} Dunn RJH, Fabian AC, Taylor
GB. 2005. \mnras 364:1343-1353  

\bibitem[Dwarakanath \& Nath(2006)]{dn06} Dwarakanath KS, Nath
  BB. 2006. \apj 653:L9-12 

\bibitem[Edge(2001)]{edge01} Edge AC. 2001. \mnras 328:762-782 

\bibitem[Edge \& Stewart(1991)]{es91} Edge AC, Stewart GC.  1991.
  \mnras 252:414-427

\bibitem[Edwards et al.(2007)]{ehb07} Edwards LOV, Hudson MJ, Balogh
  ML, Smith RJ. 2007. \mnras 379:100-110

\bibitem[Egami et al.(2006)]{emw06} Egami E, Misselt KA,  Wise MW, et
  al. 2006. \apj 647:922-933

\bibitem[En{\ss}lin \& Heinz(2002)]{eh02} En{\ss}lin TA, Heinz
S. 2002. \aap 384:L27-L30  

\bibitem[En{\ss}lin \& Br\"uggen(2002)]{eb02}  En{\ss}lin TA,
Br\"uggen M.  2002.  \mnras 331:1011-1019

\bibitem[Ettori(2000)]{ettori00} Ettori S. 2000. \mnras 311:313-316 

\bibitem[Ettori \& Fabian(2000)]{ef00}  Ettori S, Fabian AC. 2000.
\mnras 317:L57-L59


\bibitem[Evrard et al.(1996)]{emn96} Evrard AE, Metzler CA, Navarro
  JF.  1996. \apj 469:494-507 

\bibitem[Ezawa et al.(1997)]{efm97} Ezawa H, Fukazawa Y, Makishima K,
  Ohashi T, Takahara F, Xu H, Yamasaki NY. 1997. \apjl 490:L33-L36


\bibitem[Fabian(1994)]{fab94} Fabian AC. 1994. \araa 32:277-318 

\bibitem[Fabian \& Nulsen(1977)]{fn77} Fabian AC, Nulsen PEJ. 1977
\mnras 180:479-484  

\bibitem[Fabian et al.(1981)]{fhc81} Fabian AC, Hu EM, Cowie LL,
  Grindlay J.  1981. \apj 248:47-54 

\bibitem[Fabian et al.(2000)]{fse00} Fabian AC, Sanders JS, Ettori S,
  Taylor GB, et al. 2000. \mnras 318:L65-L68

\bibitem[Fabian et al.(2001)]{fmn01} Fabian AC, Mushotzky RF, Nulsen
PEJ, Peterson JR. 2001. \mnras 321:L20-L24  

\bibitem[Fabian et al.(2002)]{fcb02} Fabian AC, Celotti A, Blundell
KM, Kassim NE, Perley RA. 2002. \mnras 331:369-375

\bibitem[Fabian et al.(2003a)]{fsa03} Fabian AC, Sanders JS, Allen SW,
Crawford CS, Iwasawa K, Johnstone RM, Schmidt RW, Taylor
GB. 2003. \mnras 344:L43-L47

\bibitem[Fabian et al.(2003b)]{fsc03} Fabian AC, Sanders JS, Crawford
CS, Conselice CJ, Gallagher JS, Wyse RFG. 2003. \mnras 344:L48-L52

\bibitem[Fabian et al.(2005)]{frt05}  Fabian AC, Reynolds CS, Taylor
GB, Dunn RJH.  2005.  \mnras 363:891-896

\bibitem[Fabian et al.(2006)]{fst06} Fabian AC, Sanders JS, Taylor GB,
Allen SW, Crawford CS, Johnstone RM, Iwasawa K. 2006. \mnras
366:417-428 

\bibitem[Finoguenov et al.(2000)]{fdp00} Finoguenov A, David LP,
  Ponman TJ.  2000. \apj 544:188-203 

\bibitem[Finoguenov \& Jones(2001)]{fj01} Finoguenov A, Jones
C. 2001. \apj 547:L107-L110  

\bibitem[Finoguenov et al.(2002)]{fmb02} Finoguenov A, Matsushita K,
  B\"ohringer H, Ikebe Y, Arnaud M. 2002 \aap 381:21-31

\bibitem[Forman et al.(1972)]{fkg72} Forman W, Kellogg E, Gursky H,
  Tananbaum H, Giacconi R. 1972. \apj 178:309-316 

\bibitem[Forman \& Jones(1982)]{fj82} Forman W, Jones C. 1982.
\araa 20:547-585 

\bibitem[Forman et al.(2005)]{fnh05} Forman W, Nulsen P, Heinz S, Owen
F, Eilek J, Vikhlinin A, Markevitch M, Kraft R, Churazov E, Jones
C. 2005. \apj 635:894-906

\bibitem[Forman et al.(2007)]{fcj07} Forman W, Churazov E, Jones C, et
al. 2007. \apj in press (astro-ph/0604583)

\bibitem[Fujita \& Reiprich(2004)]{fr04} Fujita Y, Reiprich
  TH. 2004. \apj 612:797-804 


\bibitem[Gallagher \& Ostriker(1972)]{go72} Gallagher JS, Ostriker
  JP. 1972. \apj 77:288-291

\bibitem[Giacconi et al.(1971)]{gkg71} Giacconi R, Kellogg E,
  Gorenstein P, Gursky H, Tananbaum H. 1971. \apj 165:L27-L35 

\bibitem[Gitti et al.(2006)]{gfs06} Gitti M, Feretti L, Schindler
  S. 2006. \aap 448:853-860 

\bibitem[Gitti et al.(2007)]{gmn07} Gitti M, McNamara BR, Nulsen PEJ,
Wise MW. 2007. \apj 660:1118-1136


\bibitem[Gomez et al.(2002)]{glr02} Gomez PL, Loken C, Roettiger K,
  Burns JO . 2002. \apj 569:122-133 

\bibitem[Govoni(2006)]{g06}  Govoni F.  2006.  \an 327:537-544

\bibitem[Govoni \& Feretti (2004)]{gf04} Govoni F, Feretti
  L. 2004. IJMod. Phys 13:1549-1594

\bibitem[Gu et al.(2007)]{gxg07} Gu J, Xu H, Gu L, An T, Wang Y, Zhang
  Z, Wu X-P. 2007 \apj 659:275-282

\bibitem[Gull \& Northover(1973)]{gn73} Gull SF, Northover
KJE. 1973. \nat 244:80-83

\bibitem[Gursky et al.(1971)]{gkm71} Gursky H, Kellogg E, Murray S,
  Leong C, Tananbaum H, Giacconi R.  1971. \apj 167:L81-L84

\bibitem[Gursky \& Schwartz(1977)]{gs77} Gursky H, Schwartz
  DA. 1977. \araa 15:541-568 

\bibitem[H\"aring \& Rix(2004)]{hr04}  H\"aring N, Rix H.  2004.  
  \apj 604:L89-L92 


\bibitem[Harris \& Krawczynski (2006)]{hk06} Harris DE, Krawczynski
H. 2006. \araa 44:463-506

\bibitem[Hatch et al.(2006)]{hcj06} Hatch NA, Crawford CS, Johnstone
  RM, Fabian AC. 2006. \mnras 367:433-448  

\bibitem[Heath et al.(2006)]{hka06} Heath D, Krause M, Alexander
  P. 2006. \mnras 374:787-792

\bibitem[Heckman(1981)]{heckman81} Heckman TM. 1981. \apj 250:L59-L63 

\bibitem[Heckman et al.(1989)]{hbb89} Heckman TM, Baum SA, van Breugel
  WJM, McCarthy P. 1989. \apj 338:48-77  


\bibitem[Heinz, Reynolds, \& Begelman(1998)]{hrb98} Heinz S, Reynolds
C, Begelman M. 1998. \apj 501:126-136

\bibitem[Heinz et al.(2002)]{hcr02} Heinz S, Choi Y-Y, Reynolds CS,
Begelman MC. 2002. \apj 569:L79-L82 

\bibitem[Heinz \& Churazov(2005)]{hc05}  Heinz S, Churazov E.  2005.
\apjl 634:L141-L144

\bibitem[Heinz et al.(2006)]{hby06} Heinz S, Bruggen M, Young A,
Levesque E. 2006. \mnras 373:L65-69 

\bibitem[Hicks \& Mushotzky(2005)]{hm05} Hicks AK, Mushotzky
  R. 2005. \apj 635:L9-L12  

\bibitem[Holtzman et al.(1992)]{hfs92} Holtzman JA, Faber SM, Shaya
  EJ, Lauer TR, Groth EJ, et al. 1992. \aj 103:691-702

%

\bibitem[Huang \& Sarazin(1998)]{hs98} Huang Z, Sarazin CL. 1998. \apj
496:728-736 

\bibitem[Ikebe et al.(1997)]{ime97} Ikebe Y, et al. 1997.
\apj 481:660-672 

\bibitem[Irwin \& Bregman(2000)]{ib00} Irwin JA Bregman JN. 2000. \apj
  538:543-554 

\bibitem[Irwin \& Bregman(2001)]{ib01} Irwin JA, Bregman
  JN. 2001. \apj 546:150-156  

\bibitem[Jaffe et al.(2005)]{jbb05} Jaffe W, Bremer MN, Baker
  K. 2005. \mnras 360:748-762 

\bibitem[Jetha et al.(2007)]{jph07} Jetha NN, Ponman TJ, Hardcastle
  MJ. Croston JH. 2007. \mnras 376:193-204

\bibitem[Johnstone et al.(1987)]{jfn87} Johnstone RM, Fabian AC,
Nulsen PEJ. 1987. \mnras 224:75-91 

\bibitem[Jones et al.(2002)]{jfv02} Jones C, Forman W, Vikhlinin A,
Markevitch M, David L, et al. 2002. \apj 567:L115-118

\bibitem[Jones et al.(2007)]{jfc07} Jones C, Forman W, Churazov E,
  Nulsen P, Kraft R, Murray S. 2007. See B\"ohringer et al.. 2007,
  pp. 143-152 

\bibitem[Jones \& de Young(2005)]{jd05}  Jones TW, de Young
DS. 2005. \apj 624:586-605

\bibitem[Juneau et al.(2005)]{jgc05} Juneau S, et al. 2005. \apjl
  619:L135-L138  

\bibitem[Kaiser et al.(2005)]{kpp05}  Kaiser CR, Pavlovski G, Pope ECD,
Fanghor H.  2005.  \mnras 359:493-503

\bibitem[Kempner et al.(2002)]{ksr02} Kempner JC, Sarazin CL, Ricker
PA.  2002. \apj 579:236-246

\bibitem[Kochanek et al.(2003)]{kwh03} Kochanek CS, White M, Huchra J,
  Macri L, Jarrett TH, Schneider SE, Mader J. 2003. \apj 585:161-181

\bibitem[Kraft et al.(2003)]{kvf03} Kraft RP, V{\'a}zquez SE, Forman
WR, Jones C, Murray SS, Hardcastle MJ, Worrall DM, Churazov
E. 2003. \apj 592:129-146  

\bibitem[Kravtsov et al.(2005)]{knv05} Kravtsov AV, Nagai D, Vikhlinin
  AV.  2005. \apj 625:588-595

\bibitem[Kronberg(2003)]{k03} Kronberg PP. 2003. Astronomical Society
  of the Pacific Conference Series, 301:169-183  

\bibitem[Krongold et al.(2007)]{kne07} Krongold Y, Nicastro F, Elvis
  M, Brickhouse N, Binette L, Mathur S, Jim{\'e}nez-Bail{\'o}n
  E. 2007. \apj  659:1022-1038

\bibitem[Landau \& Lifshitz(1987)]{ll87}  Landau LD, Lifshitz
EM. 1987.  {\it Course of Theoretical Physics, Vol.~6, Fluid
Mechanics.} London: Pergamon


\bibitem[Lauer et al.(2007)]{lfr07} Lauer TR, et al. 2007. \apj
  662:808-834

\bibitem[Lin et al.(2003)]{lms03} Lin YT, Mohr JJ, Stanford SA.
  2003. \apj 591:749-763


\bibitem[Lowenstein(2006)]{loew06} Loewenstein M. 2006. \apj
  648:230-249 

\bibitem[Lyutikov(2006)]{l06} Lyutikov M. 2006. \mnras 373:73-78 

\bibitem[Machachek et al.(2006)]{mnj06} Machacek M, Nulsen PEJ, Jones C,
Forman WR. 2006. \apj 648:947-955

\bibitem[Malagoli et al.(1987)]{mrb87} Malagoli A, Rosner R, Bodo
  G. 1987. \apj 319:632-636  

\bibitem[Markevitch(1998)]{markevitch98} Markevitch M. 1998. \apj
  504:27-34  

\bibitem[Markevitch et al.(2003)]{mmv03} Markevitch M, Mazzotta P,
Vikhlinin A, et al. 2003. \apj 586:L19-L23

\bibitem[Markevitch \& Vikhlinin(2007)]{mv07} Markevitch M, Vikhlinin
  A. 2007. Physics Reports 443:1-53

\bibitem[Markevitch et al.(1998)]{mfs98} Markevitch M, Forman WR,
  Sarazin CL, Vikhlinin A.  1998. \apj 503:77-96

\bibitem[Mathews \& Bregman(1978)]{mb78} Mathews WG, Bregman
JN. 1978. \apj 224:308-319 

\bibitem[Mathews \& Brighenti(2003)]{mb03} Mathews WG, Brighenti
F. 2003. \araa 41:191-239 

\bibitem[Mathews \& Brighenti(2007)]{mb07} Mathews WG, Brighenti
F. 2007. \apj 660:1137-1145

\bibitem[Mazzotta et al.(2002)]{mkp02} Mazzotta P, Kaastra JS, Paerels
FB, Ferrigno C, Colafrancesco S, et al. 2002. \apj 567:L37-40

\bibitem[Meiksin(1988)]{m88} Meiksin A. 1988. \apj 334:59-69 


\bibitem[McCarthy et al.(2004)]{mbb04} McCarthy IG, Balogh ML, Babul
  A, Poole GB, Horner DJ. 2004. \apj 613:811-830 

\bibitem[McNamara \& O'Connell(1989)]{mo89} McNamara BR, O'Connell
RW. 1989. \aj 98:2018-2043 

\bibitem[McNamara \& O'Connell(1993)]{mo93} McNamara BR, O'Connell
RW. 1993. McNamara BR, O'Connell RW. 1993. \aj 105:417-426

\bibitem[McNamara et al.(2000)]{mwn00} McNamara BR, et al. 2000. \apj
534:L135-138 

\bibitem[McNamara et al.(2001)]{mwn01} McNamara BR, Wise MW, Nulsen
PEJ, et al. 2001. \apj 562:L149-L12

\bibitem[McNamara et al.(2005)]{mnw05} McNamara BR, Nulsen PEJ, Wise
MW, Rafferty DA, Carilli C, Sarazin CL, Blanton EL. 2005. \nat
433:45-47 

\bibitem[McNamara et al.(2004)]{mwm04} McNamara BR, Wise MW, Murray
  SS. 2004. \apj 601:173-183  

\bibitem[McNamara et al.(2006)]{mrb06} McNamara BR, Rafferty DA,
  Birzan L, Steiner J, Wise MW, et al. 2006. \apj 648:164-175

\bibitem[Merritt(1985)]{m85} Merritt D. 1985. \apj 289:18-32

\bibitem[Mitchell et al.(1976)]{mcd76} Mitchell RJ, Culhane JL,
  Davison PJN, Ives JC.  1976. \mnras 175:29P-33P

\bibitem[Molendi \& Pizzolato(2001)]{mp01} Molendi S, Pizzolato
F. 2001. \apj 560:194-200 

\bibitem[Morita et al.(2006)]{miy06} Morita U, Ishisaki Y, Yamasaki
  NY, Ota N, Kawano N, et al. 2006. Publ. Astron. Soc. Jpn. 58:719-742 

\bibitem[Morris \& Fabian(2005)]{mf05} Morris RG, Fabian
AC. 2005. \mnras 358:585-600 


\bibitem[Mushotzky \& Loewenstein(1997)]{ml97} Mushotzky RF,
  Loewenstein M.  1997. \apj 481:L63-L66 

\bibitem[Mushotzky(2004)]{mushotzky04} Mushotzky RF.  2004. Clusters
  of Galaxies: Probes of Cosmological Structure and Galaxy Evolution,
  123-142

\bibitem[Nagai et al.(2007)]{nvk07} Nagai D, Vikhlinin A, Kravtsov
  AV. 2007. 655:98-108 

\bibitem[Nakazawa et al.(2007)]{nmf07} Nakazawa K, Makishima K,
  Fukazawa Y.  2007. Publ. Astron. Soc. Jpn. in press
  (astro-ph/0612753) 

\bibitem[Narayan \& Medvedev(2001)]{nm01} Narayan R, Medvedev
  MV. 2001. \apj 562:L129-L132 


\bibitem[Nipoti \& Binney(2004)]{nb04} Nipoti C, Binney
J. 2004. \mnras 349:1509-1515  

\bibitem[Nipoti \& Binney(2005)]{nb05} Nipoti C, Binney
  J. 2005. \mnras 361:428-436  


\bibitem[Nulsen(1986)]{nul86} Nulsen PEJ. 1986. \mnras 221:377-392 

\bibitem[Nulsen et al.(2002)]{ndm02} Nulsen PEJ, David LP, McNamara
  BR, Jones C, Forman WR, Wise M. 2002. \apj 568:163-173

\bibitem[Nulsen et al.(2005a)]{nhm05} Nulsen PEJ, Hambrick DC, McNamara
BR, Rafferty D, Birzan L, Wise MW, David LP. 2005. \apj 625:L9-L12

\bibitem[Nulsen et al.(2005b)]{nmw05} Nulsen PEJ, McNamara BR, Wise
MW, David LP. 2005. \apj 628:629-636 

\bibitem[Nulsen et al.(2007)]{njf07} Nulsen PEJ, Jones C, Forman WR,
David LP, McNamara BR, et al. 2007. See B\"ohringer et al. 2007,
pp. 223-28 

\bibitem[Nusser et al.(2006)]{nsb06} Nusser A, Silk J, Babul
  A. 2006. \mnras 373:739-746 

\bibitem[O'Dea et al.(2004)]{obm04} O'Dea CP, Baum SA, Mack J,
  Koekemoer AM, Laor A. 2004. \apj 612:131-151

\bibitem[Oegerle et al.(2001)]{ocd01} Oegerle WR, Cowie L, Davidsen A,
Hu E, Hutchings J, et al. 2001. \apj 560:187-193 

\bibitem[O'Hara et al.(2006)]{omb06} O'Hara TB, Mohr JJ, Bialek JJ,
  Evrard AE. 2006. \apj 639:64-80 

\bibitem[Omma \& Binney(2004)]{ob04} Omma H, Binney J. 2004. \mnras
  350:L13-L16  

\bibitem[Omma et al.(2004)]{obb04} Omma H, Binney J, Bryan G, Slyz
  A. 2004. \mnras 348:1105-1119 

\bibitem[Owen \& Eilek(1998)]{oe98} Owen FN, Eilek JA. 1998. \apj
493:73-80 

\bibitem[Peres et al.(1998)]{pfe98}  Peres CB, Fabian AC, Edge AC,
Allen SW, Johnstone RM, White DA. 1998. \mnras 298:416-423


\bibitem[Peterson et al.(2001)]{ppk01} Peterson JR, et al. 2001. \aap
365:L104-L109 

\bibitem[Peterson \& Fabian(2006)]{pf06} Peterson JR, Fabian
AC. 2006. \physrep 427:1-39  

\bibitem[Peterson et al.(2003)]{pkp03} Peterson JR, Kahn SM, Paerels
FBS, Kaastra JS, Tamura T, Bleeker JAM, Ferrigno C, Jernigan
JG. 2003. \apj 590:207-224  

\bibitem[Pedlar et al.(1990)]{pgd90} Pedlar A, Ghataure HS, Davies RD,
  Harrison BA, Perley R, Crane PC, Unger SW. 1990. \mnras 246:477-489 


\bibitem[Piffaretti et al.(2005)]{pjk05} Piffaretti R, Jetzer P,
  Kaastra JS, Tamura T. 2005. \aap 433:101-111  
 
\bibitem[Pfrommer et al.(2005)]{pes05} Pfrommer C, En{\ss}lin TA,
  Sarazin CL. 2005. \aap 430:799-810

\bibitem[Pizzolato \& Soker(2006)]{ps06} Pizzolato F, Soker N.  2006.
\mnras 371:1835-1848

\bibitem[Pointecouteau et al.(2004)]{pak04} Pointecouteau E, Arnaud M,
  Kaastra J, de Plaa J. 2004. \aap 423:33-47

\bibitem[Ponman et al.(1999)]{pcn99} Ponman TJ, Cannon DB, Navarro
  JF. 1999. \nat 397:135-137

\bibitem[Ponman et al.(2003)]{psf03} Ponman TJ, Sanderson AJR,
  Finoguenov A. 2003. \mnras 343:331-342


\bibitem[Pope et al.(2006)]{ppk06} Pope ECD, Pavlovski G, Kaiser CR,
  Fangohr H. 2006. \mnras 367:1121-1131 

\bibitem[Quilis et al.(2001)]{qbb01} Quilis V, Bower RG, Balogh
  ML. 2001. \mnras 328:1091-1097 

\bibitem[Rafferty et al.(2006)]{rmn06} Rafferty DA, McNamara BR,
Nulsen PEJ, Wise MW. 2006. \apj 652:216-231

\bibitem[Rebusco et al.(2005)]{rcb05} Rebusco P, Churazov E, Bohringer
  H, Forman W. 2005. MNRAS 359:1041-1048


\bibitem[Renzini(2004)]{renzini04} Renzini A. 2004. Clusters of 
Galaxies: Probes of Cosmological Structure and Galaxy Evolution, 260 

\bibitem[Reynolds et al.(2001)]{rhb01} Reynolds CS, Heinz S, Begelman
MC. 2001. \apj 549:L179-82 

\bibitem[Reynolds et al.(2002)]{rhb02} Reynolds CS, Heinz S, Begelman
  MC. 2002. \mnras 332:271-282 

\bibitem[Reynolds et al.(2005)]{rmf05}  Reynolds CS, McKernan B, Fabian
AC, Stone JM, Vernaleo JC.  2005.  \apj 357:242-250

\bibitem[Rizza et al.(2000)]{rlb00} Rizza E, Loken C, Bliton M,
Roettiger K, Burns JO, Owen FN. 2000. \aj 119:21-31 

\bibitem[Robinson et al.(2004)]{rdr04} Robinson K,  et al. 2004. \apj 
601:621-643

\bibitem[Roediger et al.(2006)]{rbr06} Roediger E, Br\"uggen M,
  Rebusco P, B\"ohringer H, Churazov E. 2006. \mnras 325:15-28

\bibitem[Rosner \& Tucker(1989)]{rt89}  Rosner R, Tucker
  WH. 1989. \apj 338:761-769 

\bibitem[Roychowdhury et al.(2004)]{rrn04}  Roychowdhury S, Ruszkowski
M, Nath BB, Begelman MC. 2004. \apj 615:681-688

\bibitem[Ruszkowski \& Begelman(2002)]{rb02}  Ruszkowski M, Begelman
  MC. 2002. \apj 581:223-228

\bibitem[Ruszkowski et al.(2004a)]{rbb04a}  Ruszkowski M, Br\"uggen M,
Begelman MC.  2004a.  \apj 611:158-163

\bibitem[Ruszkowski et al.(2004b)]{rbb04b}  Ruszkowski M, Br\"uggen M,
Begelman MC.  2004b.  \apj 615:675-680

\bibitem[Salom\'e et al.(2006)]{sce06}  Salom\'e P, Combes F, Edge AC,
  Crawford C, Erlund M, et al. 2006. \aa 454:437-445


\bibitem[Sanders et al.(2005)]{sfd05} Sanders JS, Fabian AC,
Dunn,RJH. 2005. \mnras 360:133-140  

\bibitem[Sanders et al.(2004)]{sfa04} Sanders JS, Fabian AC, Allen SW,
  Schmidt RW. 2004. \mnras 349:952-972
%

\bibitem[Sarazin(1988)]{s88} Sarazin CL. {\it X-ray Emission from
Clusters of Galaxies.} Cambridge: Cambridge University Press

\bibitem[Sazonov et al.(2005)]{soc05} Sazonov S Yu, Ostriker JP,
  Ciotti L, Sunyaev RA. 2005. \mnras 358:168-180 

\bibitem[Schindler et al.(2001)]{scd01} Schindler S, Castillo-Morales
A, De Filippis E, Schwope A, Wambsganss J. 2001. \aa 376:L27-30

\bibitem[Serlemitsos et al.(1977)]{ssb77} Serlemitsos PJ, Smith BW,
  Boldt EA, Holt SS, Swank JH. 1977. \apj 211:L63-L66 

\bibitem[Scheuer(1974)]{scheuer74} Scheuer PAG. 1974. \mnras
166:513-528

\bibitem[Siemiginowska et al.(2005)]{siem05} Siemiginowska, 
A., Cheung, C.~C., LaMassa, S., Burke, D.~J., Aldcroft, T.~L., Bechtold, 
J., Elvis, M., \& Worrall, D.~M.\ 2005, \apj, 632, 110

\bibitem[Sijacki \& Springel(2006)]{ss06} Sijacki D, Springel
  V. 2006. \mnras 366:397-416 

\bibitem[Silk(1976)]{silk76} Silk J. 1976. \apj 208:646-649 

%

\bibitem[Smith et al.(2002)]{swa02} Smith DA, Wilson AS, Arnaud KA,
Terashima Y, Young AJ. 2002. \apj 565:195-207 


\bibitem[Soker et al.(2002)]{sbs02} Soker N, Blanton EL, Sarazin CL. \
2002. \apj 573:533-541 

\bibitem[Soker(2006)]{soker06} Soker N. 2006. New Astron. 12:38-46

\bibitem[Spergel et al.(2007)]{sbd07} Spergel DN, Bean R, Dor\'e O,
  Nolta MR, Bennett CL, et al. 2007. \apjs 170:377-408

\bibitem[Spitzer(1962)]{s62} Spitzer L. 1962. {\it Physics of Fully
Ionized Gases.}  New York: Interscience

\bibitem[Springel et al.(2005)]{sdh05} Springel V, Di Matteo T,
  Hernquist L. 2005. \mnras 361:776-794 

\bibitem[Stewart et al.(1984)]{sfn84} Stewart GC, Fabian AC, Nulsen
  PEJ, Canizares CR. 1984. \apj 278:536-543  

\bibitem[Tamura et al.(2001)]{tkp01} Tamura T, et al. 2001. \aap
365:L87-L92  

\bibitem[Tamura et al.(2004)]{tkd04} Tamura T, Kaastra JS, den Herder
  JWA, Bleeker JAM, Peterson JR. 2004. \aap 420:135-146 


\bibitem[Tribble(1989)]{t89} Tribble PC. 1989. \mnras 238:1247-1260

\bibitem[Tucker \& David(1997)]{td97} Tucker W, David LP. 1997. \apj
  484:602-607 

\bibitem[Vall{\'e}e(2004)]{v04} Vall{\'e}e JP. 2004. New Astronomy
  Review 48:763-841 

\bibitem[Vernaleo \& Reynolds(2006)]{vr06} Vernaleo JC, Reynolds
  CS. 2006. \apj 645:83-94 

\bibitem[Vikhlinin et al.(2001a)]{vmf01} Vikhlinin A, Markevitch M,
Forman W, Jones C. 2001. \apj 555:L87-L90

\bibitem[Vikhlinin et al.(2001b)]{vmm01} Vikhlinin A, Markevitch M,
  Murray SS.  2001. \apjl 549:L47-L50

\bibitem[Vikhlinin et al.(2005)]{vmm05} Vikhlinin A, Markevitch M,
  Murray SS, Jones  C, Forman W, Van Speybroeck L. 2005. \apj
  628:655-672 

\bibitem[Vikhlinin et al.(2006)]{vkf06} Vikhlinin A, Kravtsov A,
  Forman W, Jones C, Markevitch M, Murray SS, Van Speybroeck L. 2006.
  \apj 640:691-709 

\bibitem[Vikhlinin et al.(2007)]{vbf07} Vikhlinin A, Burenin R, Forman
  W, Jones C, Hornstrup A, et al. 2007. See Bohringer et al. 2007,
  pp. 48-53 

\bibitem[Voigt \& Fabian(2004)]{vf04} Voigt LM, Fabian
  AC. 2004. \mnras 347:1130-1149

\bibitem[Voit(2005)]{voit05} Voit GM. 2005.  Reviews of Modern
  Physics, 77:207-258

\bibitem[Voit \& Donahue(2005)]{vd05} Voit GM, Donahue M. 2005. \apj
634:955-963 


\bibitem[Wang \& Hu(2005)]{wh05} Wang J-M, Hu C. 2005. \apj
  630:L125-L128 


\bibitem[Wilman et al.(2005)]{wej05} Wilman RJ, Edge AC, Johnstone
  RM. 2005. \mnras 359:755-764 

\bibitem[Wise et al.(2004)]{wmm04} Wise MW, McNamara BR, Murray
  SS. 2004. \apj  601:184-196 

\bibitem[Wise et al. (2007)]{wmn07} Wise MW, McNamara BR, Nulsen PEJ,
Houck JC,  David LP.  2007. \apj 659:1153-1158

\bibitem[Wu et al.(2000)]{wfn00} Wu KKS, Fabian AC, Nulsen
PEJ. 2000. \mnras 318:889-912  


\bibitem[Young et al.(2002)]{ywm02} Young AJ, Wilson AS, Mundell
CG. 2002. \apj 579:560-570 


\bibitem[Zakamska \& Narayan(2003)]{zn03} Zakamska NL, Narayan
   R. 2003. \apj 582:162-169



\end{thebibliography}
\end{document}